\documentclass[fleqn,usenatbib]{mnras}

\usepackage{newtxtext,newtxmath}
\usepackage{booktabs}
\newcommand{\mysplit}[1]{%
  \begin{tabular}{@{}c@{}}   
    #1
  \end{tabular}
  }

\usepackage[T1]{fontenc}
\usepackage[dvipsnames]{xcolor}

\DeclareRobustCommand{\VAN}[3]{#2}
\let\VANthebibliography\thebibliography
\def\thebibliography{\DeclareRobustCommand{\VAN}[3]{##3}\VANthebibliography}

\usepackage{graphicx}	
\usepackage{amsmath}	
\usepackage{soul}
\defcitealias{scott_2020_fornaxI}{Paper I}
\defcitealias{Eftekhari_2021_fornaxII}{Paper II}
\defcitealias{Smith2009_highalpha}{S09}
\defcitealias{McDermid2015}{M15}

\title[]{The SAMI - Fornax Dwarfs Survey III: Evolution of [$\alpha$/Fe] in dwarfs, from Galaxy Clusters to the Local Group.}

\author[]{
J. Romero-G\'omez,$^{1,2}$\thanks{E-mail: jorgerg658@gmail.com (IAC)},
Reynier F. Peletier$^{3}$,
J. A. L. Aguerri$^{1,2}$,
Steffen Mieske$^{4}$, 
Nicholas Scott$^{5,6}$,
\newauthor
Joss Bland-Hawthorn$^{5,6}$,
Julia J. Bryant$^{5,6}$,
Scott M. Croom$^{5,6}$,
F. Sara Eftekhari$^{3}$,
Jes\'us Falc\'on-Barroso$^{1,2}$,
\newauthor
Michael Hilker$^{7}$, 
Glenn van de Ven$^{8}$,
and Aku Venhola$^{9,3}$
\\
$^{1}$Instituto de Astrofísica de Canarias, Calle Vía Láctea S/N, La Laguna, Tenerife, Spain\\
$^{2}$Universidad de La Laguna Avda. Astrofísico Fco. Sánchez, E-38205 La Laguna, Tenerife, Spain\\
$^{3}$Kapteyn Institute, University of Groningen, Landleven 12, 9747, AD, Groningen, The Netherlands\\
$^{4}$European Southern Observatory, Alonso de Cordova 3107, 7630355 Vitacura, Santiago,Chile\\
$^{5}$Sydney Institute for Astronomy, School of Physics, A28, The University of Sydney, NSW, 2006, Australia\\
$^{6}$ARC Centre of Excellence for All Sky Astrophysics in 3 Dimensions (ASTRO 3D), Australia\\
$^{7}$ESO, European Southern Observatory, Karl-Schwarzschild-Str 2, D-85748 Garching bei München, Germany\\
$^{9}$Department of Astrophysics, University Vienna, Türkenschanzstrasse 17, A-1180 Wien, Austria\\
$^{9}$Space physics and astronomy research unit, University of Oulu, Pentti Kaiteran katu 1, FI-90014 Oulu, Finland
}

\date{Accepted XXX. Received YYY; in original form ZZZ}

\pubyear{2023}

\begin{document}
\label{firstpage}
\pagerange{\pageref{firstpage}--\pageref{lastpage}}
\maketitle

\begin{abstract}
Using very deep, high spectral resolution data from the SAMI Integral Field Spectrograph we study the stellar population properties of a sample of dwarf galaxies in the Fornax Cluster, down to a stellar mass of $10^{7}$ M$_{\odot}$, which has never been done outside the Local Group.
We use full spectral fitting to obtain stellar population parameters. Adding massive galaxies from the ATLAS$^{3D}$ project, which we re-analysed, and the satellite galaxies of the Milky Way, we obtained a galaxy sample that covers the stellar mass range $10^{4}$ to $10^{12} M_{\odot}$.  Using this large range we find that the mass – metallicity relation is not linear. We also find that the [$\alpha$/Fe]-stellar mass relation of the full sample shows a U-shape, with a minimum in [$\alpha$/Fe] for masses between $10^{9}-10^{10} M_{\odot}$. 
The relation between [$\alpha$/Fe] and stellar mass can be understood in the following way: When the faintest galaxies enter the cluster environment, a rapid burst of star formation is induced, after which the gas content is blown away by various quenching mechanisms. This fast star formation causes high [$\alpha$/Fe] values, like in the Galactic halo. 
More massive galaxies will manage to keep their gas longer and form several bursts of star formation, with lower [$\alpha$/Fe] as a result. For massive galaxies, stellar populations are regulated by internal processes, leading to [$\alpha$/Fe] increasing with mass. We confirm this model by showing that [$\alpha$/Fe] correlates with clustercentric distance in three nearby clusters, and also in the halo of the Milky Way.
\end{abstract}




\section{Introduction}\label{introduction}
Dwarf galaxies, the most abundant type of galaxies in the Universe, have been studied much less than their massive counterparts, because of their low surface brightness, making them difficult to study for telescopes. As an example, the \href{https://www.sdss.org/}{Sloan Digital Sky Survey} (SDSS) survey mostly concentrates on massive galaxies, since the exposure times are too short to get enough signal-to-noise (S/N) for all but the brightest and most nearby dwarfs. As a result, a large part of our knowledge of dwarf galaxies used to come from objects in the Local Group.

With the development of sensitive, wide-field CCD detectors in the last ~2-3 decades, however, the study of the low-surface brightness universe has received a boost and our knowledge of dwarf galaxies is slowly catching up. Dwarf galaxies come in many different types such as star-forming and quiescent dwarf galaxies \citep[][]{Sandage1984}, and each of them is subdivided into high and low surface brightness objects. In recent years some extreme classes have been added, like the small high-surface brightness Ultra Compact Dwarfs (UCDs) \citep[][]{Hilker1999, Drinkwater2000, Mieske2001, Wittmann2016, Saifollahi2021}, and the large, low surface brightness Ultra Diffuse Galaxies (UDGs) \citep[][]{Sandage1984, Dokkum2015}. Also, the ultra-faint class of dwarf galaxies (UFD) is a type of galaxy whose detected population have grown significantly in recent years thanks to the progress in imaging capabilities \citep{Simon2019}. In this paper, we will not study high surface brightness dwarfs, like UCDs  or compact ellipticals, but focus on the 'classical' low surface brightness dwarfs, mainly on the quiescent dwarf ellipticals (dEs)\citep[][]{Binggeli1988} and the dwarf spheroidals, but also on some star-forming dwarf irregular galaxies (dIrr).

Dwarf galaxies are usually defined to be fainter than ${\rm M}_B > -18$ mag. Although quiescent dwarfs look featureless at first view, deep imaging observations have shown that these galaxies can have complex sub-structures such as bars, spiral arms or disks \citep[][]{Jerjen2000_spiral, Barazza2002_spiral, Lisker2006_disks, Janz2014_bars, Michea2022}. The recent work of \citet{Michea2022} shows that only the brightest dwarfs contain bulges, disks and bars \citep[see also][]{Su2021}. 

The fact that the relative frequencies of quiescent and star-forming galaxies strongly depend on the environment \citep[][]{Binggeli-Sandange-Tammann1988} indicates that the role of the environment is large \citep[see also][]{Boselli2014_enrionment, Boselli2014_}. \citet{Binggeli1988} found a strong morphology-density relation for dwarfs: quiescent dwarfs are dominating in the Virgo Cluster, while blue, star-forming dwarfs are dominating outside it, in the outskirts. This shows the strong influence of the environment on the evolution of dwarf galaxies.

Physical mechanisms acting in high-density environments like ram-pressure stripping \citep{Gunn1972-rampressure} or strangulation \citep{Larson1980} can remove the gas from galaxies and stop their star formation \citep{Lisker2009}. This way they can change the morphology of the galaxies and transform star-forming into quiescent galaxies. In addition, other weaker processes like harassment \citep{Moore1998harasment,Aguerri-Garcia2009} or gravitational interactions with other galaxies or even the cluster \citep{Moore1999} may also affect the galaxy. If a galaxy has gone through some ram-pressure stripping, it can sometimes be detected in observations of neutral or ionized gas, detecting long tails being blown out of the in-falling galaxy \citep{Jaffe2018}. Additionally, this process can also alter the velocity field, since the potential well can be modified by the drag force of the expelled gas. This can be detected in the stellar kinematics, as a lower angular momentum of dwarf galaxies in clusters \citep{Boselli2022}. Given their low masses, dwarf galaxies are more susceptible to any physical process than giant galaxies.

Dwarf galaxies form continuous scaling relations with giant galaxies, for example, the Tully-Fisher relation \citep{Ponomareva2018}, the Fundamental Plane 
\citet{Eftekhari_2021_fornaxII}
\citepalias[hereinafter: ][]{Eftekhari_2021_fornaxII}, relations between morphological parameters, and between them, colours and line strengths \citep{Misgeld2011}. 

Apart from an external process, low-mass galaxies are also sensitive to an internal process that can quench or trigger star formation \citep[][]{Haines2007}. Studying the stellar populations, and their relation with internal and external properties of the galaxies, is a way to find out which one of these is more relevant.

Detailed knowledge of stellar populations of dwarf galaxies mostly comes from the Local Group, for example, star formation histories (SFH) and abundances of various elements \citep{Tolstoy2008}. In more distant galaxy clusters, knowledge about stellar populations in dwarfs is typically obtained by analysing their integrated properties.  Here we summarize some properties from the literature. In clusters of galaxies, quiescent dwarfs are usually old and metal-poor \citep{Koleva2009, Sybilska2017}, these systems might have formed at high redshift and evolved passively since then or could have formed from star-forming galaxies that fell from the surroundings into the cluster in early times and were quenched by the high-density environment \citep[][hereinafter:  \citetalias{Smith2009_highalpha}]{Smith2009_highalpha}. However, many dwarfs are young, e.g. star forming dwarfs, also called dwarf irregulars \citep{Michielsen2008_allages, Rys2015_allages}, while also having low metallicities. In fact, quiescent and star-forming dwarf galaxies lie on the same stellar mass - stellar metallicity relation \citep{Kirby2013}.

In terms of the abundance ratio [$\alpha$/Fe], dwarf galaxies usually have solar-like values \citep{Geha2003_solaralpha, Kaufer2004solaralpha, Seyda2018solaralpha}, although, in the core of a cluster, such as Coma, they could also be slightly more $\alpha$-enhanced than in the outskirts \citepalias{Smith2009_highalpha}. The abundance ratio [$\alpha$/Fe] is defined by the abundance of elements with $\alpha$ particles, like C, O or Mg among others. Abundance ratios are the effect of element enrichment in Supernovae Type Ia and II. Massive stars, which are the progenitors of SN Type II, live short lives and produce mostly $\alpha$-elements, like C, N, O, Mg etc. The other frequent type of SN, Type Ia, comes from binaries, of which one is a white dwarf. The material produced by these supernovae is richer in iron than in $\alpha$-elements. It also will take about 1 Gyr before such supernovae go off. To conclude, fast star formation causes high [$\alpha$/Fe] values, as is, for example, the case in the halo of our Milky Way. Slow star formation, like in the disk of our Milky Way, causes the enrichment to be dominated by SN Type Ia, so that [$\alpha$/Fe] values will be around zero. This means that the [$\alpha$/Fe] parameter is a useful parameter to measure how fast star formation proceeded \citep{Worthey1992, Reynier1989PhD}. We know that for giant galaxies there is a strong relation between [$\alpha$/Fe] and mass \citep{Traeger2000, McDermid2015, Watson2022}, which is thought to be due to a much faster enrichment in the more massive galaxies. Previous studies have shown that dwarf galaxies with stellar masses around $10^{9}$-$10^{10}M_{\odot}$ follow a relation of $\alpha$-enhancement vs mass similar to giant galaxies \citep{Smith2009_highalpha, Liu2016_linear_sigma, Sybilska2017}, however for even less massive dwarfs it is rather unknown what is the situation, and whether there continues to be a relation with galaxy mass or velocity dispersion, as one would expect from massive galaxies.

During the last decades, one of the most common methods to derive stellar population properties has been full spectral fitting (FSF) \citep{Vazdekis1999, Capellari2004, Cid2005, Ocvirk2006a,Ocvirk2006b, Conroy2014}, which makes use of the whole spectrum, fitting it with a combination of template spectra from a stellar model library, and obtaining the properties as a combination of the model properties. However, for this spectral fitting to work spectra with high S/N ratios are required. For this reason, stellar population properties derived using this method are only available for the brightest galaxies. For dwarfs, until recently, the only method available to obtain stellar population properties has been the measurement of some characteristics indices \citep{Burstein1984, Worthey1994, Trager1998}, the equivalent widths of those indices were then compared to some stellar models \citep{Thomas2003, Elodie-prugniel2007, Sanchez-Blazquez2006, Vazdekis2010, Falcon-Barroso2011}. Usually, Balmer lines like $H_{\beta}$ are used as tracers for age, and elements lines like $Mgb$ for metallicity. Nevertheless, in the last decade, good quality spectra have started to become available for less massive galaxies. Using data from the \href{https://www.sdss.org/}{Sloan Digital Sky Survey}(SDSS), \citet{Penny2016} were able to study galaxies using FSF. Their sample, however, was composed of fairly bright, M$_r$ > -19, and massive, $10^{9}M_{\odot} < M_{\star} < 5x10^{9}M_{\odot}$, galaxies, which we will not consider in this paper. Also \citet{Sybilska2017} (with 20 dwarfs of $10^{9}M_{\odot} < M_{\star} < 10^{10}M_{\odot}$) and \citet{Bidaran2022} (with 9 dwarf galaxies of $M_{\star} \sim 10^{9}M_{\odot}$) used FSF, indicating that this is a feasible method nowadays.
 
\subsection{The Fornax cluster}
Fornax is a galaxy cluster located at $\alpha$ (J2000) $= 3^h 38^m 30^s$; $\delta$ (J2000) = -35$^\circ$27'18", with the elliptical galaxy NGC 1399 at its centre. After Virgo, Fornax is the second nearest galaxy cluster to us at a distance of 20 Mpc \citep[][]{Blakeslee2009} and 1454$\pm$286 km/s mean recessional velocity \citep[][]{Maddox2019}. According to the mass and virial radius from \citet{Drinkwater2001}, $7\times10^{13} M_{\odot}$ and 0.7 Mpc respectively, Fornax is less massive than other clusters like Virgo or Coma, but with its small size is the densest mass aggregation in the Fornax-Eridanus filament \citep[][]{Nasonova2011}. Close in the sky, less than 5 deg from NGC 1399, is Fornax A, an adjacent galaxy group around the giant early-type radio galaxy NGC1316 \citep[][]{Ekers1983, Iodice2017}. This group has a similar order of magnitude mass as the main Fornax cluster \citep[][]{Maddox2019}, and a virial radius of 1.05 deg \citep{Drinkwater2001}, if it is in virial equilibrium. Inside, this group there are several dozens of possible members \citep{Venhola2019}, mostly dwarfs. It is expected that this group falls into the Fornax Cluster \citep[][]{Drinkwater2001}.

Together, both Fornax and Fornax A have about 1000 known galaxies \citep{Venhola2021}, but only a few of them are brighter than $M_B$ = -18 mag \citep[][]{Ferguson1989}. The cluster is, after Virgo, the closest galaxy cluster in the sky. This makes it very attractive to study spectroscopically, a unique environment to test the theories of dwarfs galaxy formation and evolution. The first survey to cover this region of the sky was the Fornax Cluster Catalogue (FCC) published by \citet{Ferguson1989}, where the membership of the cluster has been determined based on surface brightness and morphology. In the resulting catalogue, 340 dwarfs are classified as cluster members, while more than two thousand are not. More recently, and taking advantage of better telescopes, the Fornax Deep Survey (FDS) catalogue became available \citep[][]{Venhola_2018_sample_selection}. This recent survey covers the whole region of the Fornax and Fornax A group, making use of the Very Large Telescope Survey Telescope (VST) at Cerro Paranal, Chile. The observations were carried out between 2013 and 2017 with the \href{https://www.eso.org/sci/facilities/paranal/instruments/omegacam.html}{OmegaCAM}  instrument \citep[][]{Kuijken2002_omegacam}, acquiring deep imaging in u’, g’, r’, and i’-bands. In the FDS a final list of 564 Fornax dwarf galaxies is given, with 470 of them being early-types and 94 late-type galaxies. These two previous catalogues, FCC and FDS, were used to select a subsample with absolute magnitudes in the r-band fainter than -19 mag to obtain spectroscopic observations, as will be explained in the following section. Later on, a newer version of the FDS Dwarf Catalog (FDSDC) was published, including more low surface brightness dwarfs, such as UDGs \citep{Venhola2022}.
\begin{table*}
\caption{Table with general properties of the observed galaxies. The columns show FCC name, coordinates, galaxy type, whether the galaxy has ionized gas or not, and effective radius, all according to the FDS catalogue \citep{Venhola_2018_sample_selection, Venhola2019}. The last two columns show the stellar mass and the specific angular momentum from \citetalias{Eftekhari_2021_fornaxII} and \citetalias{scott_2020_fornaxI}, respectively.} 
\centering    
\begin{tabular}{cccccccc}     
\hline
FCC name & RA (deg) & DEC (deg) & Type(Venhola) & Ionized gas & R$_e$ (arcsec) & log(M$_{\star}$/M$_{\odot}$) & $\lambda_{R}$ \\

\noalign{\smallskip}
\hline\noalign{\smallskip}

\noalign{\smallskip}
FCC100 & 52.948479 & -35.051388 & e* & No & 19.77 & 8.42 & 0.28 \\

\noalign{\smallskip}
FCC106 & 53.198673 & -34.238728 & e(s) & No & 10.65 & 8.51 & 0.14 \\

\noalign{\smallskip}
FCC113 & 53.279419 & -34.805576 & l & Yes & 18.92 & 7.96 & - \\

\noalign{\smallskip}
FCC134 & 53.590393 & -34.592522 & e & No & 6.52 & 7.22 & - \\

\noalign{\smallskip}
FCC135 & 53.628445 & -34.297371 & e(s) & No & 14.72 & 8.36 & 0.31 \\

\noalign{\smallskip}
FCC136 & 53.622837 & -35.546459 & e & No & 17.5 & 8.76 & 0.15 \\

\noalign{\smallskip}
FCC143 & 53.746666 & -35.171089 & e(s) & No & 9.81 & 9.13 & 0.15 \\

\noalign{\smallskip}
FCC164 & 54.053589 & -36.166451 & e(s) & No & 9.95 & 7.95 & 0.08 \\

\noalign{\smallskip}
FCC178 & 54.202728 & -34.280102 & e & No & 11.26 & 7.62 & - \\

\noalign{\smallskip}
FCC181 & 54.2219 & -34.9384 & e* & No & 9.66 & 7.5 & - \\

\noalign{\smallskip}
FCC182 & 54.226295 & -35.374714 & e(s)* & No & 9.67 & 8.85 & 0.18 \\

\noalign{\smallskip}
FCC188 & 54.268906 & -35.590149 & e* & No & 12.2 & 8.12 & 0.2 \\

\noalign{\smallskip}
FCC195 & 54.347183 & -34.900108 & e & No & 12.78 & 7.74 & - \\

\noalign{\smallskip}
FCC202 & 54.527325 & -35.439911 & e* & No & 13.28 & 8.56 & 0.13 \\

\noalign{\smallskip}
FCC203 & 54.5382 & -34.518761 & e(s) & No & 16.04 & 8.41 & 0.33 \\

\noalign{\smallskip}
FCC207 & 54.580185 & -35.129124 & e & Yes & 9.59 & 8.18 & 0.31 \\

\noalign{\smallskip}
FCC211 & 54.589504 & -35.259689 & e* & No & 6.58 & 8.01 & 0.11 \\

\noalign{\smallskip}
FCC222 & 54.8055 & -35.37141 & e* & No & 16.1 & 8.43 & 0.32 \\

\noalign{\smallskip}
FCC223 & 54.8321 & -35.7247 & e* & No & 17.05 & 7.91 & - \\

\noalign{\smallskip}
FCC235 & 55.041069 & -35.629093 & l & Yes & 42.3 & 8.68 & - \\

\noalign{\smallskip}
FCC245 & 55.140991 & -35.022888 & e* & No & 14.52 & 8.21 & 0.19 \\

\noalign{\smallskip}
FCC250 & 55.184971 & -37.408268 & e & No & 9.22 & 7.64 & - \\

\noalign{\smallskip}
FCC252 & 55.209988 & -35.748455 & e* & No & 11.13 & 8.3 & 0.15 \\

\noalign{\smallskip}
FCC253 & 55.230301 & -37.837627 & e & No & 10.92 & 7.96 & 0.26 \\

\noalign{\smallskip}
FCC263 & 55.385574 & -34.888752 & l & Yes & 16.47 & 8.69 & 0.15 \\

\noalign{\smallskip}
FCC264 & 55.382313 & -35.58955 & e & No & 10.27 & 7.73 & - \\

\noalign{\smallskip}
FCC266 & 55.422161 & -35.170265 & e* & No & 6.91 & 8.14 & 0.17 \\

\noalign{\smallskip}
FCC274 & 55.571922 & -35.540737 & e* & No & 12.05 & 7.84 & - \\

\noalign{\smallskip}
FCC277 & 55.5949 & -35.1541 & e(s) & No & 10.1 & 9.3 & 0.27 \\

\noalign{\smallskip}
FCC285 & 55.760147 & -36.273357 & l & Yes & 32.65 & 8.47 & - \\

\noalign{\smallskip}
FCC298 & 56.18507 & -35.683716 & e* & No & 6.97 & 7.78 & 0.18 \\

\noalign{\smallskip}
FCC300 & 56.249588 & -36.319752 & e* & No & 20.82 & 8.25 & 0.28 \\

\noalign{\smallskip}
FCC301 & 56.2649 & -35.972668 & e(s) & No & 7.6 & 9.01 & 0.39 \\

\noalign{\smallskip}
FCC306 & 56.439095 & -36.3461 & l* & Yes & 7.26 & 7.52 & 0.52 \\

\noalign{\smallskip}
FCC033 & 51.243237 & -37.009613 & l & Yes & 16.89 & 8.77 & 0.46 \\

\noalign{\smallskip}
FCC037 & 51.289337 & -36.365185 & l & Yes & 33.89 & 8.7 & - \\

\noalign{\smallskip}
FCC046 & 51.604301 & -37.127785 & l & Yes & 8.51 & 7.91 & 0.38 \\

\noalign{\smallskip}
FCCB442 & 51.775901 & -36.635104 & e & No & 4.2 & 7.28 & - \\

\noalign{\smallskip}
FCCB904 & 53.484525 & -34.561623 & e & No & 5.12 & 7.55 & - \\

\hline\noalign{\smallskip}

\end{tabular}
             
\label{table_1_objects_info}
\end{table*}
The present work is part of a series of papers based on the analysis of IFU spectroscopic data of a subsample of dwarfs galaxies in the Fornax cluster. These very deep spectroscopic data present a unique opportunity to study the populations of less massive galaxies down to a level that has never been done before,  stellar mass below $10^{9.5} M_{\odot}$ or velocity dispersion lower than 40 km/s. On top of that, the high spectral resolution of a sample a few times larger than previous IFU studies allows us not only to study stellar populations properties but also to recover statistically significant analysis of the relation of these properties in dwarf galaxies with their environment and internal properties.

The paper is organized as follows: Section \ref{observs-reduct} summarizes the target selection, the spectroscopic observations and results from the previous papers of the series. In section \ref{Data_analysis} we explain the methodology used to analyse the data and obtain the results that we present in section \ref{results}. We then discuss the implication of our results in section \ref{discussion}, and finally, in section \ref{conclusions} we summarize our findings.

The properties of the Fornax cluster used in this work are stated in the previous paragraphs, and throughout this paper, we use magnitudes in the AB system and we adopt a $\Lambda$CDM cosmology with $\Omega_{m} = 0.3$, $\Omega_{\Lambda} = 0.7$ and $H_{0}$ = 70 km s$^{-1}$ Mpc$^{-1}$.
\begin{figure}
\centering
\includegraphics[scale=.55]{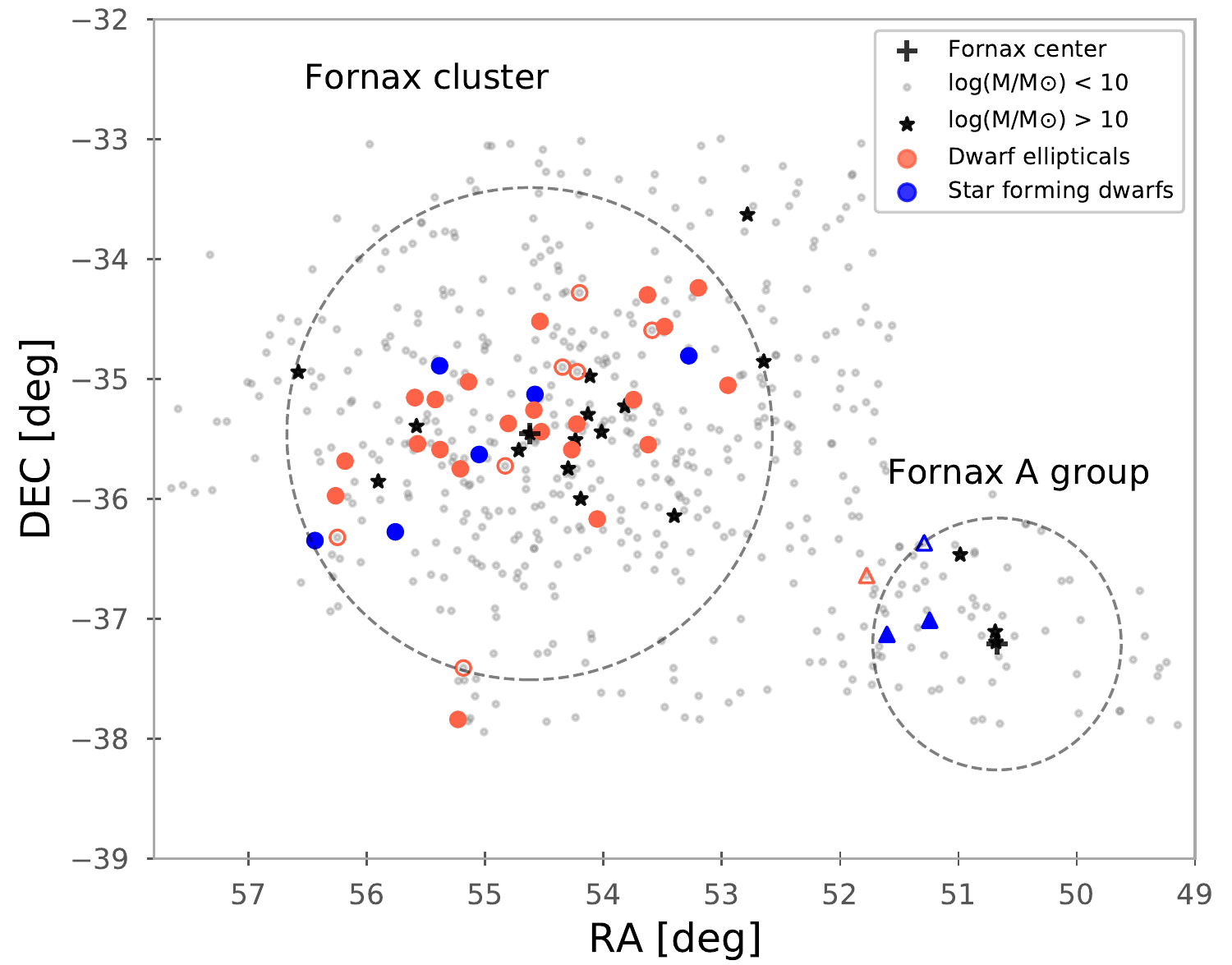}
\caption{Location in the Fornax cluster for all the galaxies in our sample (coloured symbols). Circles are for galaxies in the main Fornax cluster and triangles are for galaxies in the Fornax A group. Blue colours are for those galaxies with emission lines in their spectra and red is for the rest. The symbols are filled or not depending on the S/N, as will be later explained in section \ref{kinematics_subsection}. The black crosses represent the centre of the Fornax and Fornax A group. The background grey points are galaxies with a stellar mass lower than 10$^{10}$M$_{\odot}$ from a compilation catalogue of \citet{Su2021} \citep[see also][]{Venhola_2018_sample_selection, Iodice2019}, and black stars are for galaxies from that catalogue with a stellar mass higher than 10$^{10}$M$_{\odot}$.}
    \label{fig_catalogue}
\end{figure}
\section{Sample selection and spectroscopic observations}\label{observs-reduct}
In this work, we use the sample of dwarf galaxies from the Fornax cluster as defined by \citet{scott_2020_fornaxI} \citepalias[hereinafter: ][]{scott_2020_fornaxI}. Below we give a brief description of the process that ends up with this particular object list. In \citetalias{scott_2020_fornaxI} we not only give details about the selection of the sample, but also some details of the observations and data reduction, and more details about the data reduction are given in the second paper of our series \citepalias{Eftekhari_2021_fornaxII}. The distribution of our sample across the Fornax cluster can be seen in Figure \ref{fig_catalogue}. 

\subsection{Spectroscopic Observations and sample selection}\label{obs_seleciton_sec_2_1}
The observations were carried out at the Sydney-Australian Astronomical Observatory (AAO), using the Multi-Object Integral-Field Spectrograph called SAMI \citep{Croom2012}. Mounted on the 3.9 m Anglo-Australian Telescope (AAT), SAMI uses fibres to deploy 13 integral field units (IFUs), each of 15" diameter, across its 1-degree field of view. A total of 10 pointings of SAMI, covering a total of 118 dwarf galaxies in Fornax, were observed. Each observed field was integrated for a total exposure time of 7 hours, using a dither pattern to guarantee that the S/N is distributed uniformly for each IFU.

A first observing run was made in 2015 based on a selection from the FCC. The next observing runs, based on the FDS catalogue were in 2016 and 2018, and the selection criteria were slightly modified, based on the results of the 2015 run, to make sure that the S/N of the galaxies was large enough. Some other secondary targets like giant galaxies, UCDs or background galaxies were also observed.

The full spectroscopic sample consisted of 118 galaxies, of which 62 are classified as early-type, dE or dS0. The rest of the objects observed were UCDs, giant early-type and late-type cluster members, and background galaxies. From the primary targets, the kinematic analysis could be performed for 38 dwarfs. A list containing the basic properties of these dwarfs galaxies can be seen in Table \ref{table_1_objects_info}. Most objects are located inside the virial radius of the Fornax cluster, four others are associated with the Fornax A group. The majority of these objects are quiescent dwarfs, but the total sample also has spirals and irregular star-forming dwarfs. In Fig. \ref{fig_catalogue} we indicate in blue symbols those dwarf galaxies with emission lines (i.e. those with ionized gas). This group of galaxies are composed mainly of late-type dwarf irregulars and spirals, as classified by \citet{Venhola2019}, but it also contains one quiescent dwarf that has ionized gas, FCC207. These galaxies are characterized by strong emission lines in their spectra and will not be part of the main analysis in this paper, but will be briefly analysed and discussed in Section \ref{star_forming}.

\subsection{Results from the previous papers in this series}
In this paper, we analyse the stellar population properties of a sample of dwarfs galaxies in the Fornax cluster. The same spectroscopic data have already been examined in \citetalias{scott_2020_fornaxI}  and \citetalias{Eftekhari_2021_fornaxII}

In \citetalias{scott_2020_fornaxI} the target selection, integral field spectroscopy observations and data reduction of the sample are presented, along with a kinematic analysis of the data. Voronoi binning is used \citep[][]{Cappellari_Copin2003} to ensure a minimum S/N, then with FSF techniques spatially resolved maps of the line-of-sight velocity (V) and velocity dispersion ($\sigma$) were recovered and then used to acquire the specific angular momentum, $\lambda_R$ (see Table \ref{table_1_objects_info}). Based on these results and after comparison with more massive galaxies, we find that $\lambda_R$ is low for dwarf galaxies, only slightly higher than for massive ellipticals that have almost no rotation at all, but much lower than for galaxies of about $10^{10}M_{\odot}$. The paper implies that if quiescent dwarfs originate from star forming dwarfs falling into a cluster, these dwarf irregulars are not rotationally supported. If quiescent dwarfs originate from low-mass spirals, these spirals must lose about 90\% of their mass while falling into the cluster. Since these results are rather unexpected, more studies will be needed to be able to understand the results.

In \citetalias{Eftekhari_2021_fornaxII}, the galaxies are studied as a whole, collapsing all available spectral data into one single spectrum per galaxy. Then, with FSF the mean V and $\sigma$ are retrieved. With these and other galaxy parameters, \citetalias{Eftekhari_2021_fornaxII} studied scaling relations on the fundamental plane (FP) and stellar mass fundamental plane. Their results show that dwarfs with a mass between $10^{7}$ and $10^{8.5}$ $M_{\odot}$ deviate slightly from the FP as defined by all galaxies, indicating that the mass-to-light ratio of galaxies increases for lower masses. Also, from the relation between dynamical and stellar mass, they observed that low-mass galaxies have more dark matter than brighter dwarfs and giants, with the dwarfs in Fornax having similar dark matter ratios as dwarf galaxies of comparable stellar mass in the Local Group.
\section{Data analysis}\label{Data_analysis}
\subsection{Spectral fitting technique}\label{FSF_technique}

The AAOmega spectrograph \citep{Sharp2006} that is fed by the SAMI IFUs has a flexible range of resolutions and wavelengths. For our observations, we used a grating of 1500V and 1000R for the blue and red arm respectively, which corresponds to a resolution of about 1.05 and 1.6 \r{A} in FWHM. This translates into an instrumental resolution in sigma of about 26 km/s in the blue, allowing to study internal stellar kinematics down to internal dispersions of $\sim$10 km/s. In the two previous papers of the SAMI-Fornax dwarfs survey \citepalias[][]{scott_2020_fornaxI, Eftekhari_2021_fornaxII} all the analysis and results were obtained using only the blue part of the SAMI spectrograph, which covers the wavelength range between 4700 and 5400 \r{A}. In this work, we seek to extend that range to the red part of the spectrograph but to avoid contamination from sky noise we only include the wavelength range between the 6300 – 6800 \r{A} .
\begin{figure*}
\centering
\includegraphics[scale=0.85]{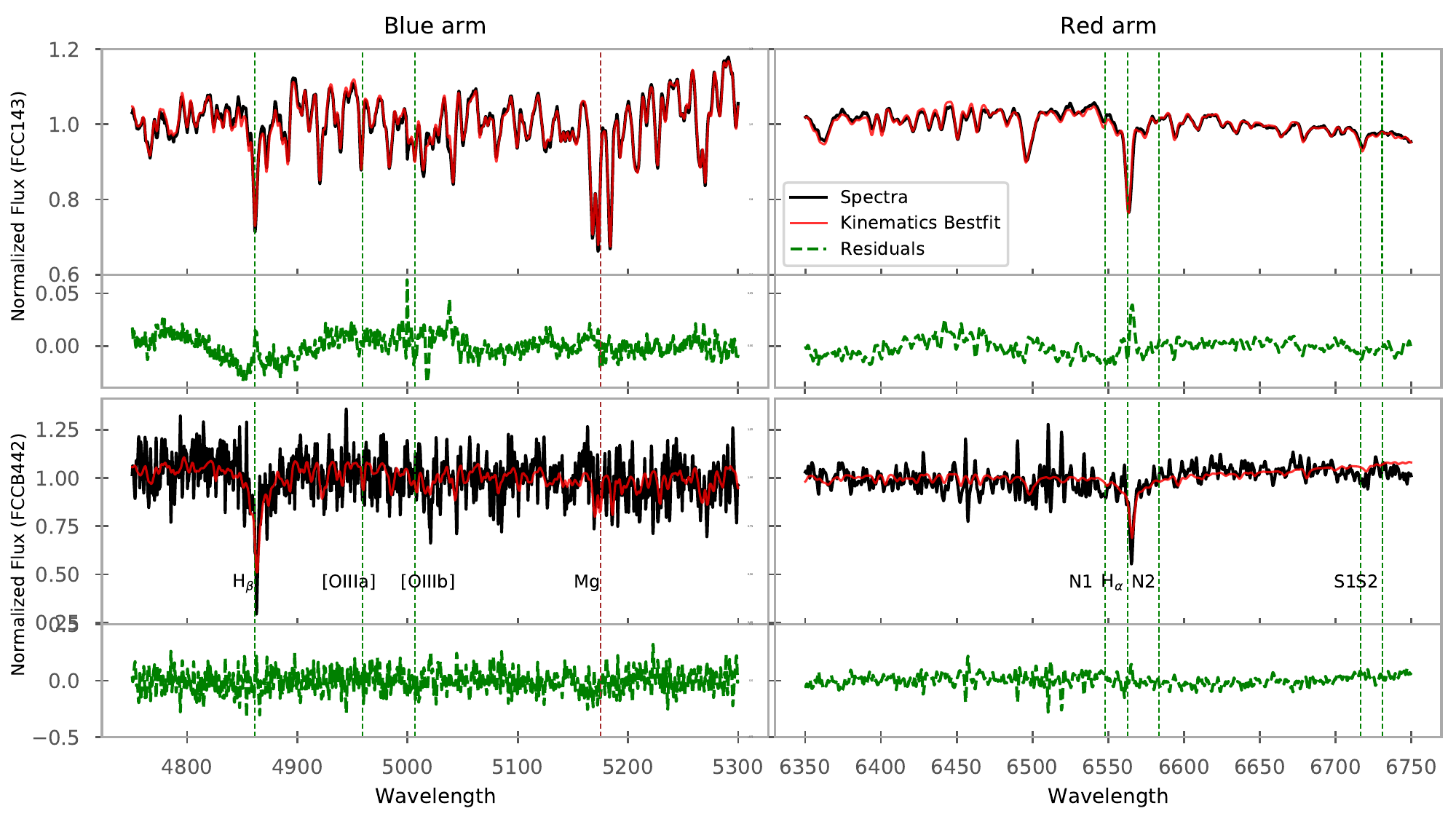}
\caption{Figure showing the best fit to a given spectrum, black line, by \href{https://www-astro.physics.ox.ac.uk/~mxc/software/}{pPXF} when fitting the stellar kinematics, red line, using only additive polynomials. Below each spectrum, the green line represents the residuals, calculated as the galaxy spectrum minus the best fit. The top two panels show the fits of one of the galaxies with the highest S/N and at the bottom one of the lowest S/N. In both cases, the absorption and emission spectral features are marked with vertical lines.}
    \label{fig_bestfit_pop_kin}
\end{figure*}
\begin{figure}
\centering
\includegraphics[scale=0.55]{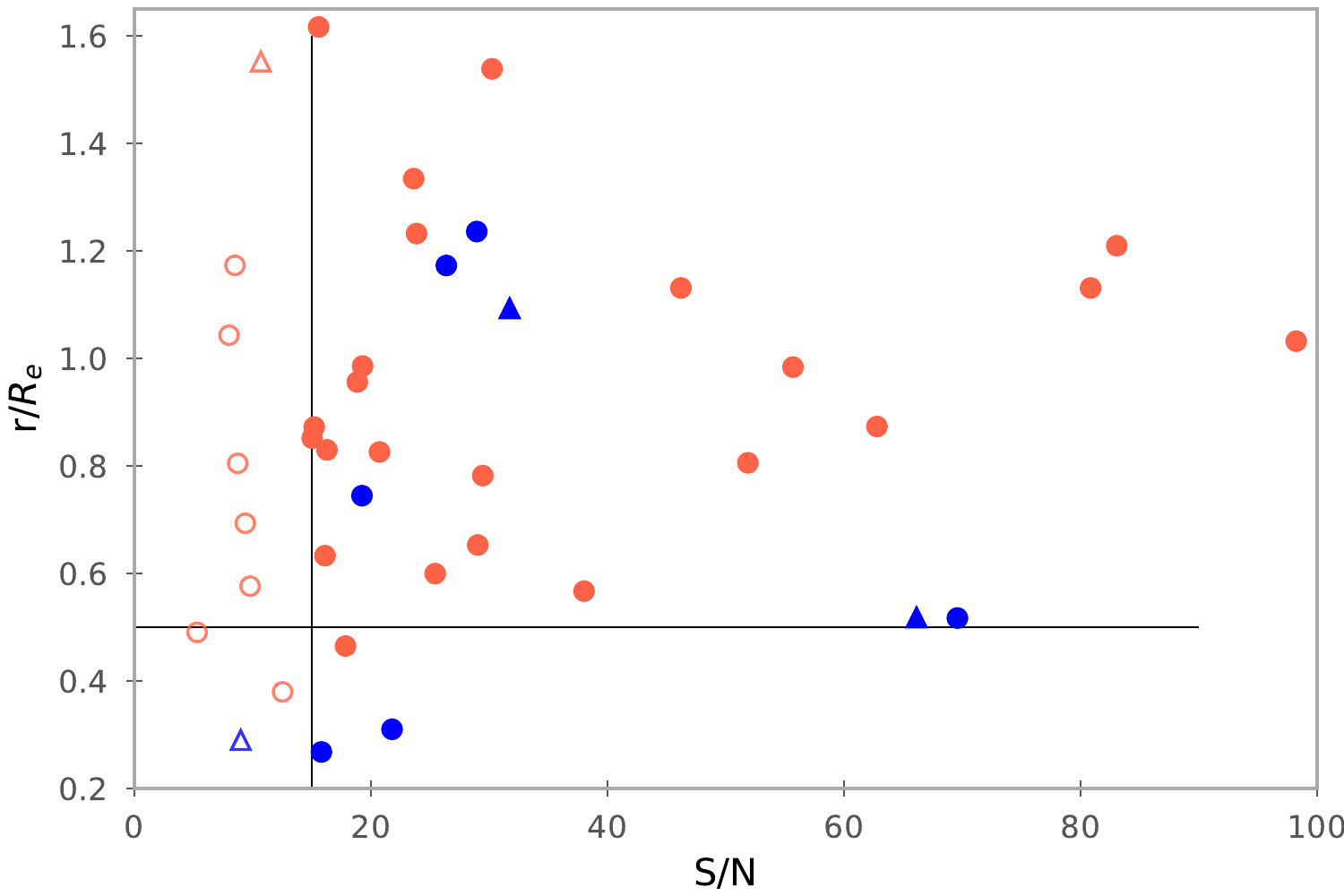}
\caption{Graph showing the relation between S/N and covered effective radius in our data. The vertical black line divides the galaxies in those with S/N higher than 15 or lower, which is where we separate our sample for more reliable results. The S/N is calculated from the ratio between the mean value of the flux in the spectra and the standard deviation of the residuals from the \href{https://www-astro.physics.ox.ac.uk/~mxc/software/}{pPXF} fitting. The horizontal black line points out for which galaxies we covered more than half effective radius during observations. The symbols are the same as in Fig. \ref{fig_catalogue}.}
    \label{fig_reff_vs_sn}
\end{figure}
\begin{figure*}
\centering
\includegraphics[scale=0.9]{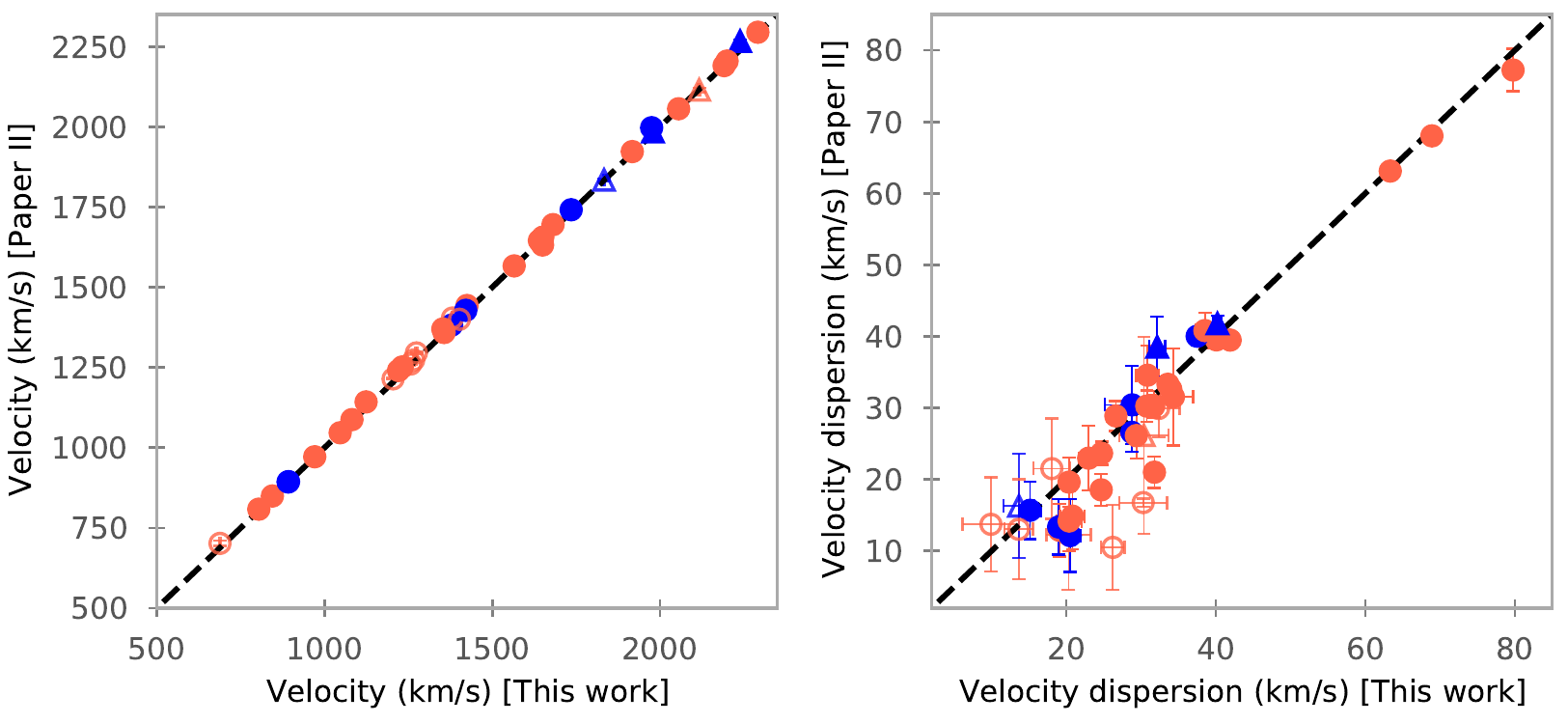}
\caption{\label{fig_kinematics}Comparison of the stellar kinematics with \citetalias{Eftekhari_2021_fornaxII}. The dotted black line indicates the 1:1 relation. The symbols are the same as in Fig. \ref{fig_catalogue}, circles or triangles depending if the galaxies are on Fornax or Fornax A, blue for those galaxies with emission lines in their spectra and red for the rest. The open symbols are galaxies with low S/N, as it is explained in section \ref{kinematics_subsection}.}
\end{figure*}
Examples of spectra of the galaxies analysed here can be seen in Fig. \ref{fig_bestfit_pop_kin}. To analyse the spectra from both arms at the same time we use the full spectral fitting algorithm of the penalized Pixel-Fitting Code (\href{https://www-astro.physics.ox.ac.uk/~mxc/software/}{pPXF}, \citeauthor{Capellari2004} \citeyear{Capellari2004} and \citeauthor{Capellari2017} \citeyear{Capellari2017}). This software has been widely used by the astrophysical community to extract information from spectra \citep{McDermid2015, Sybilska2017, Bidaran2020}. Using this tool we also define the S/N of a given galaxy as the ratio of the mean value of the flux in the spectrum and the standard deviation of the residuals from the \href{https://www-astro.physics.ox.ac.uk/~mxc/software/}{pPXF} fit. With this parameter, we will be able to test if merging the spectrum from both arms has any effect on the quality of the data, by comparing it with the results from the previous papers of this series. In \citetalias{Eftekhari_2021_fornaxII}, the line of sight velocity and velocity dispersion are determined by integrating the spectra inside the 15" diameter aperture of SAMI, which covers different effective radii fractions for each galaxy. In Fig. \ref{fig_reff_vs_sn} the distribution of S/N values as a function of the covered effective radius is shown, and for most galaxies, we cover at least half an effective radius. \citetalias{Eftekhari_2021_fornaxII} show that the velocity dispersions of these galaxies are constants as a function of radius. We could also assume that the stellar populations do not vary much either, since dwarfs usually have rather flat trends \citep{Koleva2011, denBrok2011, Rys2015}. This implies that for the conclusions of this paper, these varying radial cutoffs will not make much of a difference if we feed \href{https://www-astro.physics.ox.ac.uk/~mxc/software/}{pPXF} with a collapse spectrum like in \citetalias{Eftekhari_2021_fornaxII}.

The FSF technique fits the galaxy spectra with a combination of stellar templates using a maximum penalized likelihood technique, and to ensure that the best fit is appropriate there is a recommended procedure in the \href{https://www-astro.physics.ox.ac.uk/~mxc/software/}{pPXF} code \citep[][]{Capellari2017}. One should first use only additive Legendre polynomials to find the best solution for the velocity and the velocity dispersion of each galaxy, then fix the stellar kinematics and use only multiplicative polynomials to recover the stellar population properties, additive polynomials, on the other hand, are not allowed on this second fit. This is the best choice to ensure that the shape of the continuum is corrected but spectral features and line profiles are not affected.

In \citetalias{scott_2020_fornaxI} and \citetalias{Eftekhari_2021_fornaxII}, to study only the stellar kinematics, \href{https://www-astro.physics.ox.ac.uk/~mxc/software/}{pPXF} is applied by using as stellar models the \href{http://www.obs.u-bordeaux1.fr/m2a/soubiran/elodie_library.html}{ELODIE} library \citep{Elodie-prugniel2007}, because its resolution of 0.55 \r{A} in FWHM allowed them to reach the requirements for velocity dispersion previously mentioned.
In this current work, we are focusing on the study of stellar population properties, and for that purpose, we have opted for the stellar templates of the \href{http://miles.iac.es/}{Vazdekis/MILES} library \citep{Falcon-Barroso2011, Vazdekis2010, Sanchez-Blazquez2006}, which provide a resolution of 2.51 \r{A} (FWHM). 
We acknowledge that the library resolution is worse than our data, but the fact that \href{http://miles.iac.es/}{MILES} is a library with a much larger range of stellar parameters, like [$\alpha$/Fe], and is regularly being updated, makes it better for our scientific aims than \href{http://www.obs.u-bordeaux1.fr/m2a/soubiran/elodie_library.html}{ELODIE}. The resolution of \href{http://miles.iac.es/}{MILES} corresponds to about 60 km/s (FWHM), which is a reasonable compromise to study our sample of dwarfs.

\subsection{Calibration of the stellar kinematics}\label{kinematics_subsection}
A test is needed to ensure that with the complete spectra of both arms and convolved to a lower resolution, we can reproduce previous kinematics results. According to \citet{Toloba2012} \href{https://www-astro.physics.ox.ac.uk/~mxc/software/}{pPXF} is capable of obtaining results down to 0.4$\sigma_{resolution}$, in case of \href{http://miles.iac.es/}{MILES} this translate to $\sim$25 km/s, and 5-10 km/s for \href{http://www.obs.u-bordeaux1.fr/m2a/soubiran/elodie_library.html}{ELODIE}. The latter is comparable to the level that SAMI can reach, $\sim$10 km/s \citepalias{Eftekhari_2021_fornaxII}.

First, once the spectra of both arms are merged, we convolve the galaxy spectrum to the \href{http://miles.iac.es/}{MILES} resolution, since for correct use of \href{https://www-astro.physics.ox.ac.uk/~mxc/software/}{pPXF} both spectra and stellar templates have to be at the same resolution. Then, we follow the recommended \href{https://www-astro.physics.ox.ac.uk/~mxc/software/}{pPXF} procedure and using only additive polynomials of 6th order we fit the kinematics with the \href{http://www.obs.u-bordeaux1.fr/m2a/soubiran/elodie_library.html}{ELODIE} library. An example of these fits for different galaxies can be seen in Fig. \ref{fig_bestfit_pop_kin}, and the resulting kinematics are compared with previous results in Fig. \ref{fig_kinematics}. Later, with the stellar kinematics fixed and using \href{http://miles.iac.es/}{MILES} as templates, we used multiplicative polynomials of degree 6 to find the best stellar population parameters.

Comparing the velocity dispersions with \citetalias{Eftekhari_2021_fornaxII} (Fig. \ref{fig_kinematics}) we see that for most galaxies the agreement is excellent, except for galaxies for which our S/N is below 15. Throughout the results of this work, we prefer to give these low S/N galaxies different symbols, so that the reader can decide themselves whether to trust these results.
\section{Results}\label{results}
In this section, we present the resulting stellar population properties from \href{https://www-astro.physics.ox.ac.uk/~mxc/software/}{pPXF}, and show the relations with velocity dispersion, stellar mass, and the environment. As stated before, we limit this analysis to dwarf galaxies without emission lines.

In order to compare our results with more massive galaxies we  also include in our analysis galaxies of the \href{https://www-astro.physics.ox.ac.uk/atlas3d/}{ATLAS$^{3D}$ Project} \citep[][]{Cappellari2011-atlas3d}. The sample contains 260 massive early-type galaxies (ETGs), with a stellar mass between $10^{10}$ and $10^{12} M_{\odot}$, and located within a 42 Mpc distance. We re-analysed the spectra of each galaxy using the same methodology described in Section \ref{Data_analysis} and the same \href{http://miles.iac.es/}{MILES} grid (described in \ref{result_spp}) used for our SAMI-Fornax galaxies (see Fig. \ref{fig_atlas_populations} in appendix \ref{appendix_atlas} for a comparison of our results for the ETGs with the published data by the ATLAS$^{3D}$ Project). For the following figures and results, we removed from the ATLAS$^{3D}$ sample a few galaxies that were marked in \citet{McDermid2015} \citepalias[hereinafter: ][]{McDermid2015} as not having enough quality, due to low S/N, emission lines or some other problems that affect the spectrum analysis procedure. These galaxies are marked with a red X symbol in the figures from appendix \ref{appendix_atlas}. As for our SAMI-Fornax dwarfs, we have integrated all available spectra of each ATLAS$^{3D}$ galaxy, but for comparison with \citetalias{McDermid2015} we have also derived the properties inside 1.0 and 0.5 R$_e$ (see Fig. \ref{fig_atlas_populations_at_re}). After a careful inspection of these results, we decided to use in our study the total integrated spectrum of each galaxy, since there was only a small difference between these and the values at 1.0 R$_e$ and the results presented in this work are still maintained.

In \citet{Cappellari2011-atlas3d} the galaxies are also classified depending on whether they are in the Virgo cluster or not, but for a better comparison we divided galaxies into cluster and non-cluster using the density parameter from \citet{Cappellari2011-density}, taking as cluster members those with a local mean surface density of galaxies higher than log($\Sigma_{10}$) > 0.6 Mpc$^{-2}$. With this definition, 91$\%$ of Virgo galaxies are classified as cluster members. With all this we can compare our results with the ATLAS$^{3D}$ sample, obtaining a total mass range of 10$^{7}$ to 10$^{12}$ M$_{\odot}$. 

\subsection{Stellar population properties}\label{result_spp}

\begin{figure*}
\centering
\includegraphics[scale=0.8]{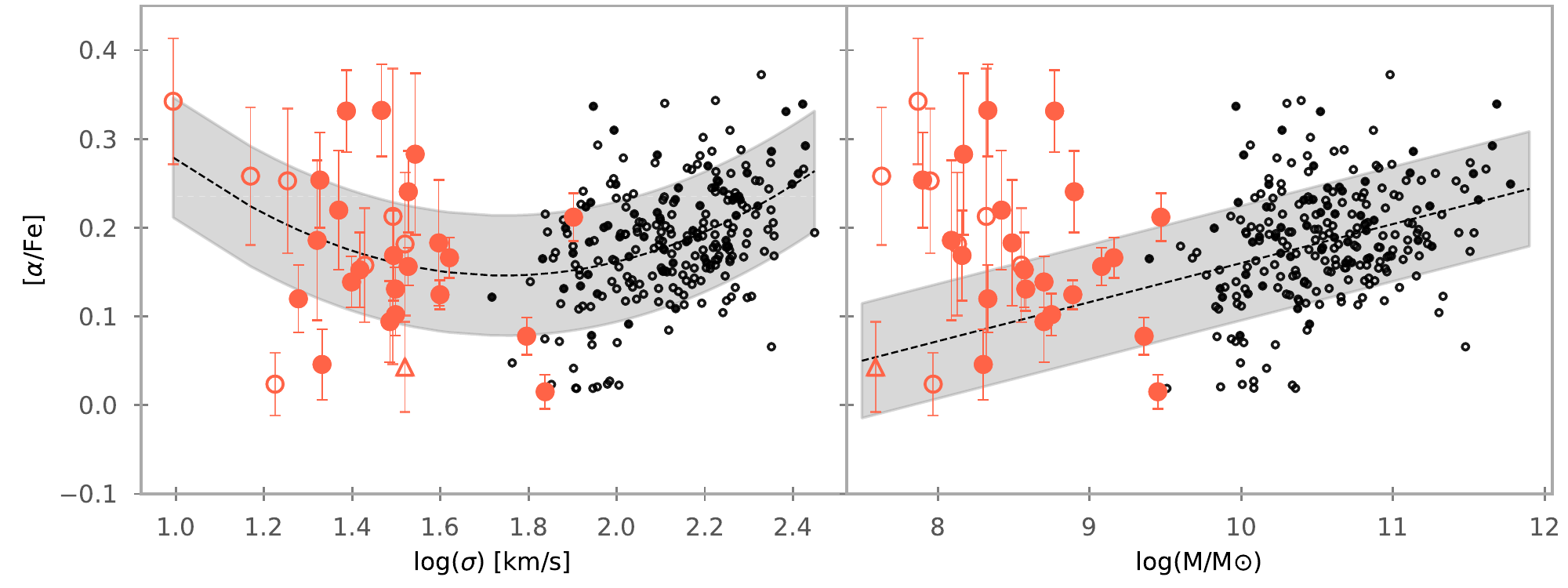}
\caption{\label{fig_alpha_vs_sigma}Abundance ratio [$\alpha$/Fe], obtained from FSF (Full Spectral Fitting) with \href{https://www-astro.physics.ox.ac.uk/~mxc/software/}{pPXF} and the \href{http://miles.iac.es/}{MILES} library, as a function of the logarithmic velocity dispersion and the stellar mass, in the left and right panel, respectively. The symbols for our dwarfs are the same as in Fig. \ref{fig_catalogue}. For ATLAS$^{3D}$ we used black-filled circles for cluster galaxies and non-filled black circles for the rest. The black dashed line and grey shaded area on the left represent a second-degree polynomial fit to all the data visible in the figure, making a U-shape.
In the right panel, however, we present the linear fit [$\alpha$/Fe]-log(M/$M_{\odot}$) taking into account only the galaxies from ATLAS$^{3D}$ project. Since stellar mass and velocity dispersion are tightly related, it is clear that this U-shape can also be applied to the [$\alpha$/Fe]-log(M/$M_{\odot}$) relation. The coefficients from both fits are in Tab. \ref{table_fit_coef}.}
\end{figure*}

\begin{figure*}
\centering
\includegraphics[scale=0.8]{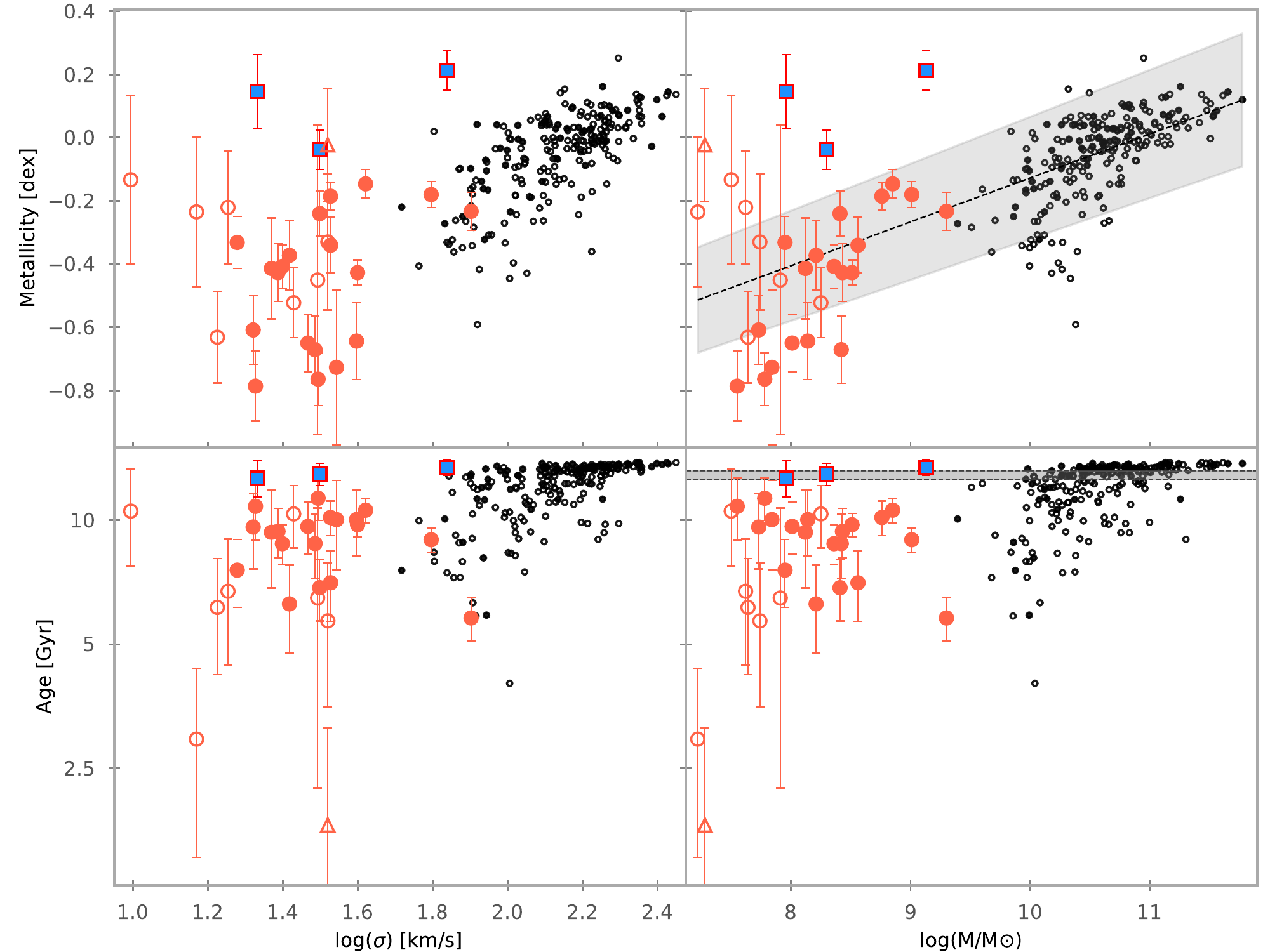}
\caption{\label{fig_all_vs_all} Relations between the stellar population parameter obtained from pPXF, as a function of the velocity dispersion and stellar mass. The top panels show the metallicity versus log($\sigma$) and log(M$_{\star}$/M$_{\odot}$) in the left and right panel, respectively. And in the bottom is the age versus log($\sigma$) and log(M$_{\star}$/M$_{\odot}$) in the left and right panel, respectively. The symbols are the same as in Fig. \ref{fig_alpha_vs_sigma}, except for the three old dwarf galaxies, which are outliers to the mass-metallicity relation, these have been highlighted with red edged squares filled with a clear blue colour (see also section \ref{discuss_origin}). These 3 outliers have also been included in the fits. In the top right panel, we used a black dashed line and grey shaded area to plot the linear fit of the mass-metallicity relation, considering the whole sample, and in the bottom-right panel, we also include a shaded region to indicate the re-ionization epoch between z$\sim$6 and z$\sim$14 \citep[][]{Fan2006_reionization}. The coefficients for the metallicity-mass relation are in Tab. \ref{table_fit_coef}. The three outlying dwarf galaxies in the top-right plot with high metallicities are discussed separately in subsection \ref{discuss_origin}.}
\end{figure*}

Having derived the kinematics from the fits of \href{https://www-astro.physics.ox.ac.uk/~mxc/software/}{pPXF}, we retrieved the age, metallicity and abundance ratio [$\alpha$/Fe] for each galaxy. The mean luminosity weighted values for every galaxy are shown in Fig. \ref{fig_alpha_vs_sigma} and \ref{fig_all_vs_all}. For this, we used \href{http://miles.iac.es/}{MILES} single stellar populations (SSP) models with the BaSTI isochrones and a bi-modal initial mass function (IMF) with a slope of 1.30 \citep[][]{Vazdekis2015}. The ages of the models from \href{http://miles.iac.es/}{MILES} range between 0.03 to 14 Gyr and metallicity from -2.27 to 0.4 dex, respectively, to optimise the time performance of the fitting, we limited the models to ages of 0.04, 0.5, 1, 3, 5, 7, 9, 11, 13, and 14 Gyr, and metallicities of -1.26, -0.66, -0.25, 0.06 and 0.26 dex. We chose these metallicities to be as consistent as possible with the range of the \href{http://www.obs.u-bordeaux1.fr/m2a/soubiran/elodie_library.html}{ELODIE} models used for the fits of the kinematics. We did several tests during the grid selection to ensure that the resulting stellar properties were not severely affected by this choice. As for the abundance ratio [$\alpha$/Fe], \href{http://miles.iac.es/}{MILES} only has models with solar values of 0.0 and super-solar 0.4 dex. More accurate values are determined by interpolation in the fit. Since the FSF technique does not offer a practical way to estimate the error of the fit, nor the errors in the obtained galaxy properties, after the first fit of the stellar population properties in \href{https://www-astro.physics.ox.ac.uk/~mxc/software/}{pPXF} we applied a Monte-Carlo (MC) method to retrieve the uncertainties of each parameter. For that purpose, we manipulate the best-fit spectra using the fit's residuals. For every pixel, the value of the best-fit is disrupted within the corresponding residual value using random normally-distributed numbers, and then, a new spectrum is obtained and fitted again with \href{https://www-astro.physics.ox.ac.uk/~mxc/software/}{pPXF}. All this process is repeated 100 times for each galaxy, and finally, the resulting parameters are the mean values of the distribution and the errors are the standard deviation. 
\subsubsection{$\alpha$-enhancement}\label{alpha_results}
In Fig. \ref{fig_alpha_vs_sigma} the inferred luminosity-weighted $\alpha$-enhancements are presented as a function of the logarithmic velocity dispersion, log($\sigma$), and the logarithmic stellar mass, log(M$_{\star}$/M$_{\odot}$). Individual values are listed in Tab. \ref{table_spp_ppxf}. Here we have not excluded galaxies with low S/N but we do use different symbols in every figure of this work to mark them.  Inside the sample of massive galaxies, we do not find any meaningful differences between the $\alpha$-enhancement values of the cluster and non-cluster members. For all the galaxies shown in Fig. \ref{fig_alpha_vs_sigma}, we see a large range of [$\alpha$/Fe] abundances, going from solar-like values up to $\sim$0.36 dex, for galaxies with a large range in velocity dispersion.

Our dwarf galaxies have a mean velocity dispersion of 32 km/s with a standard deviation of 14 km/s, while ATLAS$^{3D}$ galaxies have a mean of 136 km/s and a standard deviation of 49 km/s. There is essentially no overlap in log($\sigma$) between both samples, except for the three most massive dwarfs in our sample. For giant galaxies, strong relations between $\alpha$-enhancement and velocity dispersion, $\sigma$, are usually found, with [$\alpha$/Fe] increasing with higher velocity dispersion or stellar mass, since both properties are closely related. In the right panel of Fig. \ref{fig_alpha_vs_sigma} with a black dashed line and grey shadow, we plot the linear fit derived from the whole ATLAS$^{3D}$ sample, similar to the relation obtained by \citetalias{McDermid2015}.

The [$\alpha$/Fe] values of our SAMI-Fornax dwarf sample lie above the linear fit determined by the giants. This means that to fit both samples, covering a stellar mass range between 10$^{7.5}$ to 10$^{12}$ M$_{\odot}$, a second-order polynomial is better suited, creating a U-shape, as can be seen in the left panel of Fig. \ref{fig_alpha_vs_sigma} (the coefficients for the linear and quadratic fits can be seen in Tab. \ref{table_fit_coef}). The minimum of this parabola is located between 10$^{9}$ to 10$^{10}$ M$_{\odot}$, at about solar-like [$\alpha$/Fe] abundance. Analogously, a minimum in the [$\alpha$/Fe]-log($\sigma$) plane occurs at around log($\sigma$)=1.7 (50 km/s). Given that there is a linear relation for the giants with relatively little scatter, the way to keep the scatter limited when dwarfs are added is by adding another degree of freedom and fitting a parabola, the lowest order polynomial after the linear fit.

\subsubsection{Ages and metallicities}\label{ages_met_result}
In Fig. \ref{fig_all_vs_all} we present the ages and metallicities obtained from pPXF, as a function of log($\sigma$) and log(M$_{\star}$/M$_{\odot}$) (see individual values in Tab. \ref{table_spp_ppxf}).

In general, our SAMI-Fornax sample consists of intermediate-age to old galaxies with sub-solar metallicities, with mean luminosity-weighted values of 7.88 Gyr and -0.40 dex, respectively, while ATLAS$^{3D}$ galaxies are older and have solar-like metallicities, with mean values of 11.90 Gyr and -0.06 dex, respectively. For these massive galaxies, if we separate them into cluster or non-cluster galaxies we find that the mean age for field galaxies is slightly younger, $\sim$1 Gyr, although the mean metallicity remains the same.

These previous results are in agreement with stellar populations from other works on dwarfs spheroidal in galaxy clusters like Virgo \citep{Sybilska2017, Seyda2018solaralpha, Bidaran2022} or Coma \citep{Smith2009_highalpha}. Putting our sample and ATLAS$^{3D}$ together a clear mass-metallicity relation is visible. In the top right panel of Fig. \ref{fig_all_vs_all} we also present a linear fit to the [M/H]-M$_{\star}$ of both samples, with metallicity increasing for more massive galaxies with the exception of a few dwarfs, that have metallicities typical of giants (the parameters for this linear relation are in Tab. \ref{table_fit_coef}).

Additionally, when looking at the relation between stellar mass and age, we reproduce a similar distribution as in \citet[][Fig. 4]{Sybilska2017}. In the bottom right panel of Fig. \ref{fig_all_vs_all} we plot the ages of the galaxies against their stellar masses, and although we do not fit any linear nor quadratic function, it is clear that dwarfs are on average younger than giants.
We will discuss this topic more in a forthcoming paper (Romero-Gomez et al., in preparation).

\subsection{Environment dependencies}\label{result_environment}
\begin{figure}
\centering
\includegraphics[scale=0.67]{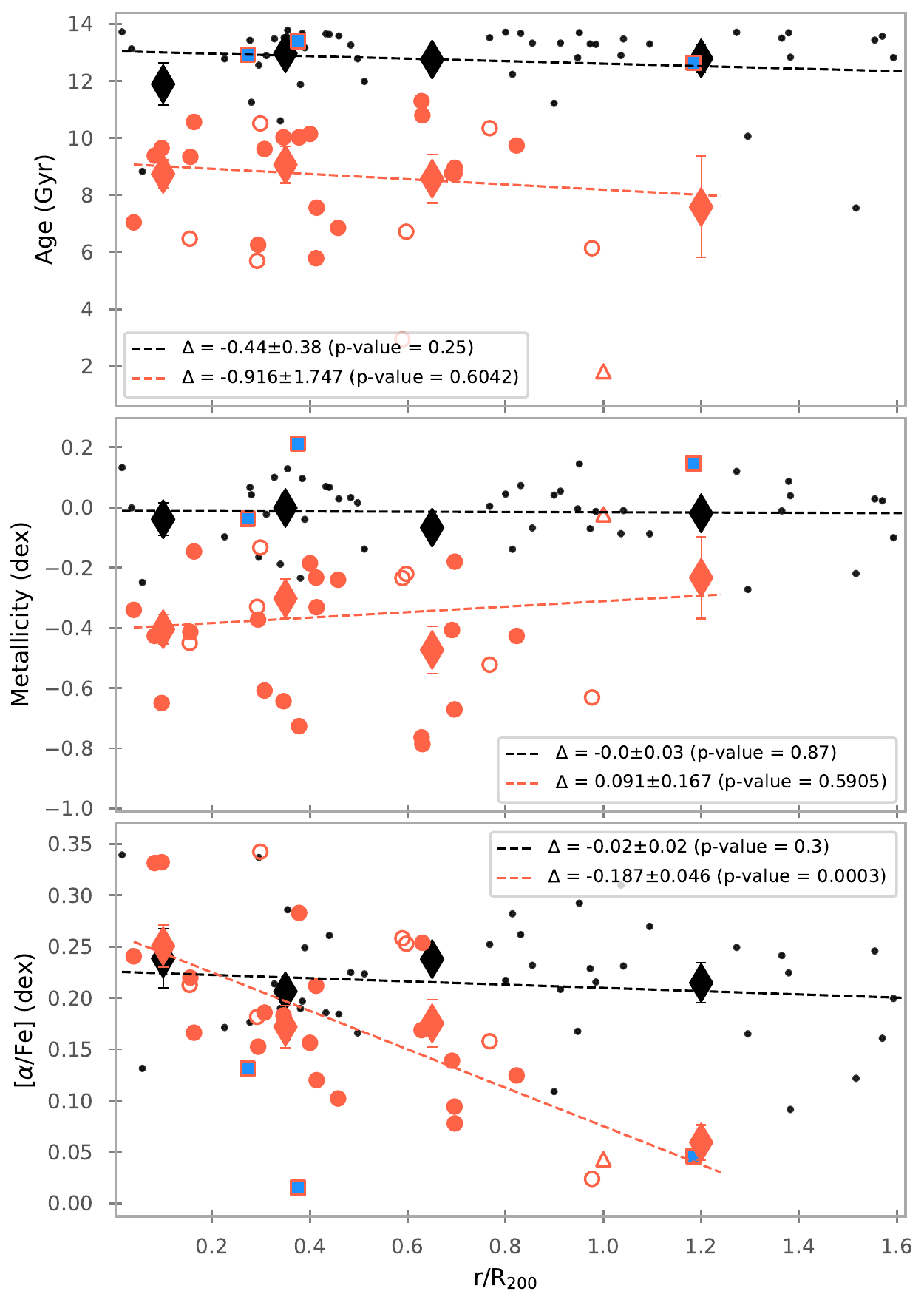}
\caption{Stellar population properties as a function of the environment. The three panels show, from top to bottom, the age, metallicity and [$\alpha$/Fe], respectively, versus projected distance to the centre Virgo, for ATLAS$^{3D}$ galaxies, the centre of Fornax for our dwarfs or to Fornax A for the galaxies belonging to this group. The dashed lines show a linear fit to the data, and the legend on each panel states the slope of the fit and its error, alongside the p-value of the fit. The latter value allows us the assessment of the null hypothesis, and when is lower than 0.05 indicates that the slope of the fit is statistically significant. For each of these panels, the coloured diamonds represent the mean value of the properties inside a given binned projected distance, which are included to guide the eye. We used 4 bins between 0-0.2, 0.2-0.5, 0.5-0.8, and 0.8-1.6 r/R$_{200}$.  The rest of the symbols are the same as in Fig. \ref{fig_all_vs_all}.}
    \label{fig_enviroment}
\end{figure}

One important key to understanding galaxy evolution is knowing what role the environment is playing in the evolution of dwarf galaxies. In Fig. \ref{fig_enviroment} we explore the possible relations of the stellar population properties with the projected distance to the centre of the cluster. In the case of the SAMI-Fornax sample, distances are taken with respect to the centre of the Fornax Cluster, or to the centre of the Fornax A group, if applicable. For ATLAS$^{3D}$ galaxies we determine them with respect to the centre of the Virgo Cluster. In both cases, the distances are expressed in R$_{200}$. For the massive galaxies, we have only included those that fulfil our cluster criteria (log($\Sigma_{10}$) > 0.6) and are classified as Virgo members in \citet{Cappellari2011-atlas3d}. 
For each cluster, we fitted the different properties as a function of the clustercentric distance with a linear relation, and we grouped the galaxies into 4 bins at different distances as a visualization aid. We have also computed the p-value of each linear fit, which represents a test of the null hypothesis that the slope is zero. Only when the p-value is below 0.05 we can confidently say that the property has a statistically significant relation with distance.
In the top panel of Fig. \ref{fig_enviroment} we see that the distribution of ages for the SAMI-Fornax dwarfs is on average rather constant as a function of clustercentric distance. The dependence on distance is not significant.
Something similar happens for the metallicity in the middle panel of Fig. \ref{fig_enviroment}, the binned values are mostly constant with distance. Here we can see again that the three outliers dwarfs to the metallicity-mass relation from Fig. \ref{fig_all_vs_all} are completely melted inside the ATLAS$^{3D}$ distribution. The absence of a correlation with projected distance for age and metallicity seems to be independent of stellar mass since the massive galaxies from Virgo in ATLAS$^{3D}$ do not show any visible relation either.

When looking at [$\alpha$/Fe], to the contrary, we see a clear relation for our SAMI-Fornax sample with clustercentric distance, where the most $\alpha$-enhanced galaxies are closer to the centre in projected distance while galaxies situated farther away tend to have lower $\alpha$-enhancement. This linear relation is statistically robust, as can be seen in the bottom panel of Fig. \ref{fig_enviroment}, the error of the slope is much smaller compared with the other fits and the p-value is the only one below the threshold of 0.05. Given the scatter of the [$\alpha$/Fe] values in the inner region of the Fornax cluster, this trend seemed to be dependent on the galaxies outside R$_{200}$. We checked that the linear fit without these galaxies is compatible with the relation presented in Fig. \ref{fig_enviroment} within 1$\sigma$ error. These numbers manifest the trend with distance for the SAMI-Fornax dwarfs, and while the two galaxies from the ATLAS$^{3D}$ sample with the highest [$\alpha$/Fe] ratio are both close to the centre in the projected distance, the rest of the Virgo sample does not experience such a strong relation between $\alpha$-enhancement and distance. 

\section{Discussion}\label{discussion}
Throughout this work, we have analysed the spectroscopic data of a sample of dwarf galaxies. In this section we discuss the resulting stellar population properties of our dwarfs and compare them with more massive galaxies from ATLAS$^{3D}$, with dwarfs from other galaxy clusters and from our Local Group, to put our results in a wider context.

\subsection{Age and metallicities of dwarfs galaxies}\label{discuss_ages_met}

Similar to us, \citet{Sybilska2017} studied stellar populations properties of a sample of dwarfs galaxies, although not for Fornax but for the Virgo Cluster, and compared them with the ATLAS$^{3D}$ results from \citetalias{McDermid2015}. Their sample is smaller and has a narrower mass range than ours, 9.0 < $log(M_{\star}/M_{\odot})$< 9.8, and instead of FSF to calculate the galaxy properties they used line index measurements to calculate age and metallicity: age from the $H\beta$ line and metallicity from Mgb and Fe5015. Because of this, the exact stellar population parameters of their dwarfs are not directly comparable to ours due to the different methodologies. Generally, there is an offset in age and metallicity between using FSF and index-fitting \citep[][]{Mentz2016, Bidaran2022} (see also the comparison with the populations derived from index-index diagrams for the ATLAS$^{3D}$ sample in Fig. \ref{fig_atlas_populations}). Overall, they measured similar differences between dwarfs and giants as we do. Both dwarf samples have intermediate-old ages and low metal content, with more massive dwarfs having the same metallicity as the intermediate mass galaxies from ATLAS$^{3D}$. As can be seen in Fig. \ref{fig_all_vs_all} and Fig. \ref{fig_enviroment}, between our least massive dwarfs and the most massive galaxy on ATLAS$^{3D}$ there is almost a difference of 1.0 dex in metallicity and around 2-3 Gyr in age. \citet{Sybilska2017} also presented a tight relation between [M/H] and log($\sigma$), similar to what we obtain in Fig. \ref{fig_all_vs_all}. 

We decided to expand the relation to less massive objects by including a sample of dwarf galaxies consisting of all quiescent satellite dwarfs of the Milky Way. For these types of objects, spectroscopic studies of the galaxies as a whole do not exist, instead, chemical abundance studies are made for individual stars. For this reason, after going through the literature and selecting a list of reliable references, for each galaxy we have taken the mean value of all its available stars and used the standard deviation as the error. For the metallicity, we use the [Fe/H] abundance, since iron is a common feature present in spectra. We are aware that most observations in the Local Group come from metal-poor stars. For that reason, we want to highlight that the values presented here are just the mean of the compiled sample of stars. Although for some Local Group galaxies metallicities are available for hundreds of stars, for some only measurements of 2 or 3 stars are available. One should keep in mind that these could be peculiar and that therefore the metallicities of the smallest galaxies could be biased. The full table with stellar population properties and a more detailed list of references are given in the appendix \ref{appendix_local}. In Fig. \ref{fig_local_group_alpha-sigma} (top) we present the relation between metallicity and velocity dispersion, a proxy of the dynamical mass of galaxies. It shows that the ATLAS$^{3D}$ galaxies and the SAMI-Fornax dwarfs together can be fitted well with a linear relation (purple line). The Local Group galaxies can also be fitted with a linear relation (grey line), the relation of \citet{Kirby2013}. However, to fit all samples together, a higher order relation, for example of second order, as is plotted, is needed (green line).

\begin{figure}
\centering
\includegraphics[scale=0.06]{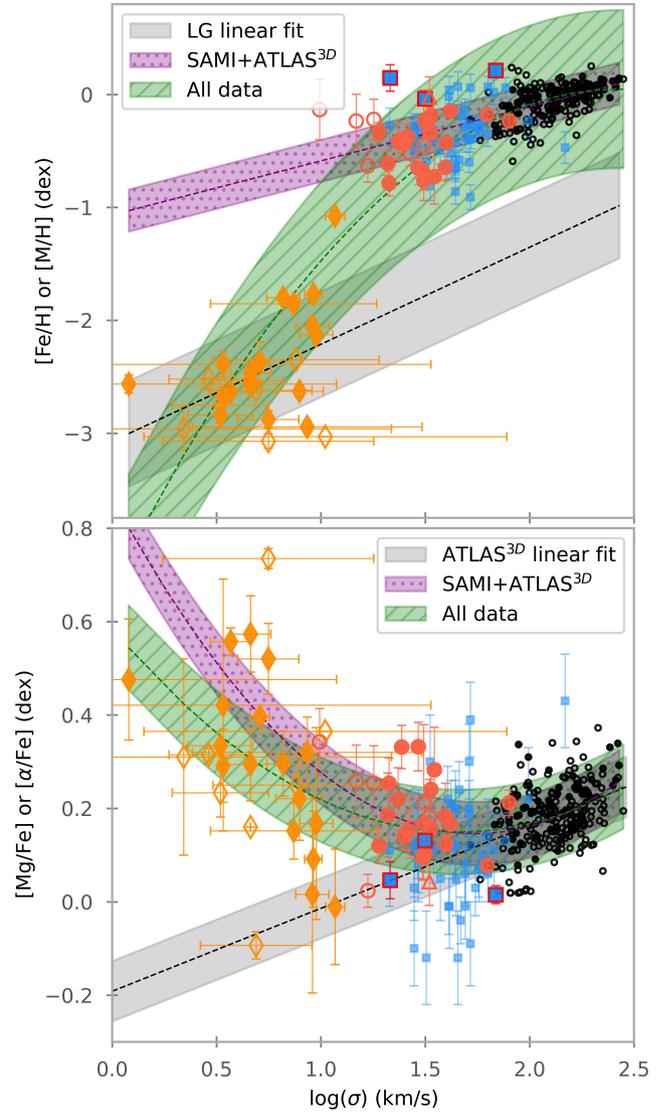}
\caption{Relation between metallicity and [$\alpha$/Fe] abundance ratio and stellar velocity dispersion for dwarf galaxies. Apart from the objects of this paper we also include a sample of dwarfs in the Coma cluster \citetalias{Smith2009_highalpha} and measurements of Local Group dwarfs from different works in the literature. For most works of the Local Group dwarfs, the metallicity or $\alpha$-enhancement is given as chemical abundances for individual stars, and here we show the mean value of all the values we could compile (see Tab. \ref{table_apx_lg_2} in appendix \ref{appendix_local}). For our SAMI-Fornax dwarfs and ATLAS$^{3D}$ galaxies, the symbols are the same as in Fig. \ref{fig_enviroment}. Light blue squares are for Coma's dwarfs from \citetalias{Smith2009_highalpha}, the orange diamonds are for Local Group dwarfs and the empty orange diamonds are for those objects with 3 or fewer stars in \ref{table_apx_lg_2}. The different lines and shadows in the figure are polynomial fits and their corresponding uncertainties to the quantities represented in both panels. In the top panel, we present the metallicity as a function of the logarithmic velocity dispersion. The purple dashed line and shadow are the linear fit to metallicity-log($\sigma$) for SAMI-Fornax and ATLAS$^{3D}$ samples, like in Fig. \ref{fig_all_vs_all}, the black dashed line and gray shadow is a linear fit to the Local Group objects and the green dashed line and shadow is a second-degree polynomial fit to metallicity-log($\sigma$) for all the objects in the figure. In the bottom, the abundance ratio [$\alpha$/Fe] is plotted as a function of log($\sigma$).  With a black dashed line and grey shadow, we show the linear relation [$\alpha$/Fe]-log($\sigma$) only for the massive galaxies from ATLAS$^{3D}$. Purple dashed line and shadow represent the U-shape fit for our SAMI-Fornax and ATLAS$^{3D}$ data, same as in Fig. \ref{fig_alpha_vs_sigma}, and the green dashed line and shadow represent a second-degree fit to all the data visible in the plot. All the coefficients of the fits presented in this figure are in Tab. \ref{table_fit_coef}.}
\label{fig_local_group_alpha-sigma}
\end{figure}
\subsection{$\alpha$-enhancement relations}\label{discuss_alpha}
\subsubsection{The U-shape}\label{U-shape}

\citet{Traeger2000}; \citetalias{McDermid2015} and \citet{Watson2022} studied different samples of giant galaxies and found strong relations between $\alpha$-enhancement and velocity dispersion, or stellar mass, like the one we present in Fig. \ref{fig_alpha_vs_sigma}. Such analysis, however, is difficult to perform in low-mass galaxies because of their low surface brightness and therefore objects with $\sigma$ <30 km/s have rarely been studied up to now outside the Local Group \citep[\citetalias{Smith2009_highalpha},][]{Sybilska2017, Seyda2018solaralpha, Bidaran2022}. Our SAMI-Fornax dwarfs are the first sample of more than a few galaxies to reach as low as 10 km/s.
Studies comparing with galaxies of higher stellar mass, $M_{\star} > 10^{10}M_{\odot}$, usually find a linear relation [$\alpha$/Fe]-log($\sigma$) similar to the one presented in Fig. \ref{fig_alpha_vs_sigma} for ATLAS$^{3D}$ ETGs. However, fitting our whole sample of galaxies reveals a U-shape that hints at a minimum in $\alpha$-enhancement for galaxies with stellar masses between $10^{9}-10^{10} M_{\odot}$. 

To further inspect this U-shape and its consequences we again include dwarf galaxies within our Local Group, to extend the range in velocity dispersion. The abundance ratio [Mg/Fe] is usually used as an approximation for the overall $\alpha$-enhancement \citep{Traeger2000, Vazdekis2015}, although this clearly is an approximation, as can be seen by detailed studies of Local Group galaxies \citep[e.g.][]{Tolstoy2009}. It is important to notice that since these are chemical abundance values for one element, the values are not directly comparable with our FSF results, which give one, effective, abundance ratio. Also, in order to visually fill the small velocity dispersion gap between our SAMI-Fornax dwarfs and ATLAS$^{3D}$ galaxies, we include the results for a sample of Coma dwarfs from \citetalias{Smith2009_highalpha}. These results are obtained using SSPs, but using index measurements instead of FSF.

In Fig. \ref{fig_local_group_alpha-sigma} (lower panel) we present all these values of $\alpha$-enhancement as a function of log($\sigma$). Note that the same relation, as a function of stellar mass, is given in Appendix \ref{appendix_ppxf_table} (see Fig. \ref{fig_alpha_local_vs_mass}). In Fig. \ref{fig_local_group_alpha-sigma} we have added the linear relation for ATLAS$^{3D}$ galaxies, a second order polynomial fit for the ATLAS$^{3D}$ and SAMI-Fornax sample together, and another second-order fit for all samples. It is clear that a U-shape is a good fit for all samples together. We see that there is a continuous sequence from giants to dwarfs from 10$^7$-10$^9$ M$_\odot$ to the faintest Local Group dwarfs.
There is the exception of one Local Group object that is closer to the linear relation, Horologium I. For this ultra-faint dwarf, we were only able to obtain abundances for three stars, which could mean that our statistical value is an underestimate. To support this idea we refer to the work of \citet{Jerjen2018}, who obtained [$\alpha$/Fe] = 0.2$\pm$0.1 dex, <[Fe/H]> = -2.4$^{+0.10}_{-0.35}$ dex and an age of 13.7$^{+0.3}_{-0.8}$ Gyr from deep photometric data. With this value, Horologium I would fall directly inside the U-shape and would agree with the idea that all intermediate-old galaxies follow the U-shape.

As a whole, the U-shape relation points towards the fact that both very massive and very faint galaxies have high $\alpha$-enhancement, with the intermediate-sized galaxies in between having solar-like values. The general picture is that Mg, or generally speaking $\alpha$-elements are mostly made in SN type II, while Fe is predominantly made in SN type Ia. Since SN type II have a much shorter timescale than SN type Ia, high [$\alpha$/Fe] is generally interpreted as indicating the fast formation of the galaxy, while solar-type [$\alpha$/Fe] is associated with slow formation \citep[e.g.][]{Arnone2005}. Our result implies that the very massive and the very faint galaxies have been forming fast, and that star-formation was halted/quenched quickly due to AGN and stellar (SNe) feedback in the very massive and very faint galaxies respectively, while this is not the case for the objects of around 10$^9$ M$_\odot$. 
The abundance ratios do not say {\it when} the galaxies were formed. For our sample, the star formation histories will be studied in a forthcoming paper 
(Romero-Gomez et al., in prep.).

\subsubsection{Environment}\label{alpha-environmet}
To understand better how galaxies are enriched in $\alpha$-elements, we now study the distribution of $\alpha$-enhancement for dwarf galaxies as a function of clustercentric radius, and, for the Local Group, as a function of distance to the Milky Way. We use here our SAMI-Fornax dwarfs, the
Coma dwarfs of \citetalias{Smith2009_highalpha}, the Local Group dwarfs that are satellites of the Milky Way and the dwarfs in the Virgo Cluster from \citet{Liu2016_distance}, \citet{Sybilska2017} and \citet{Bidaran2022}. In the Milky Way we use [Mg/Fe], and for objects in the Fornax A group we use the distance to Fornax A. Note that the [$\alpha$/Fe] values of \citet{Liu2016_distance} are systematically lower than all other samples, by about 0.2, because of the different way that is used in that paper to calculate these ratios.

\begin{figure}
\centering
\includegraphics[scale=0.85]{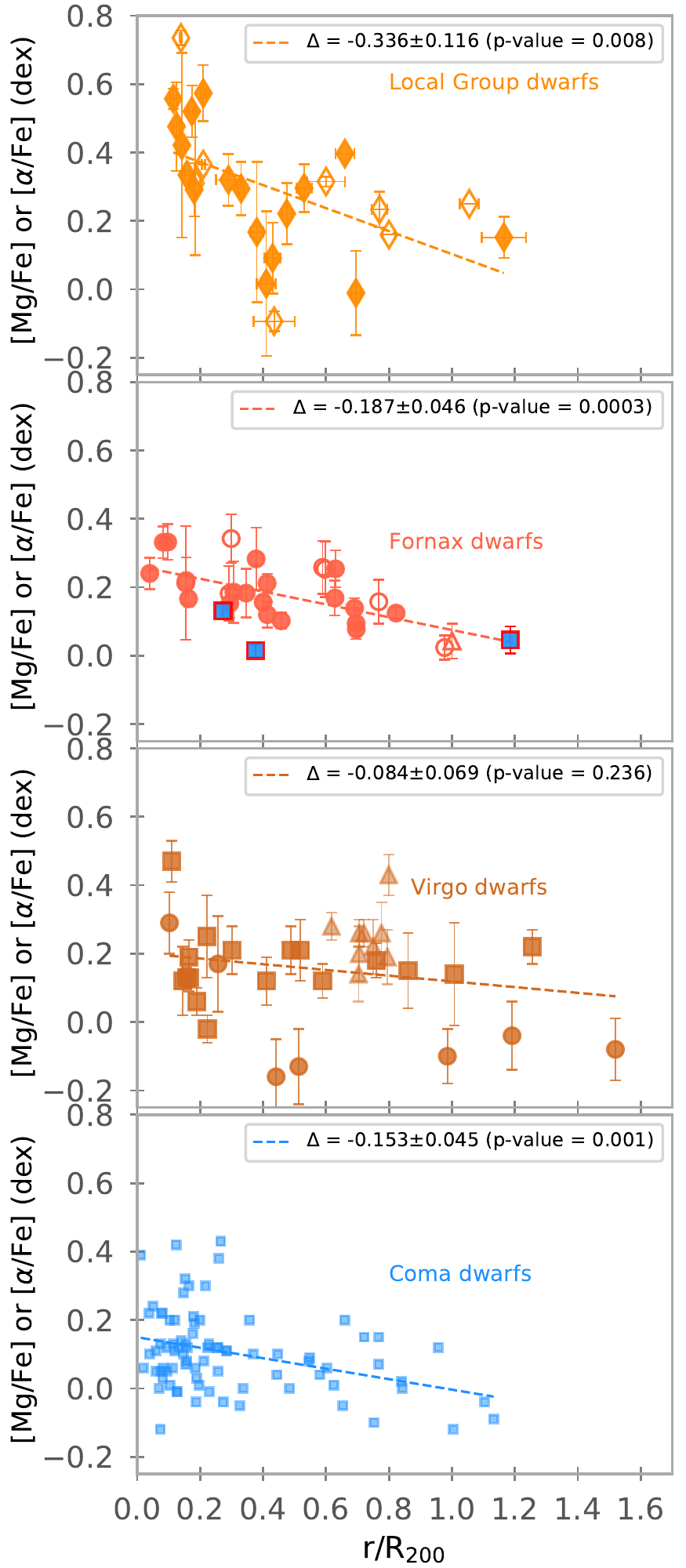}
\caption{$\alpha$-enhancement as a function of the cluster-centric distance for dwarf galaxies in different environments. Comparison of our results with those of \citetalias{Smith2009_highalpha} for Coma dwarfs, and Local Group dwarfs from different works of the literature (see Tab. \ref{table_apx_lg_2}).  For the Virgo dwarfs we include data from \citet{Sybilska2017} (brown squares), \citet{Bidaran2022} (brown triangles) and \citet{Liu2016_distance} (brown circles). The dashed lines show a linear fit to all the points on each panel, and the legends state the slope of the fit and its error, alongside the p-value of the fit. The rest of the symbols are the same as in Fig. \ref{fig_local_group_alpha-sigma}. All the coefficients of the fits presented in this figure are in Tab. \ref{table_fit_coef}.}
\label{fig_local_group_alpha-distance}
\end{figure}

\begin{figure}
\centering
\includegraphics[scale=0.85]{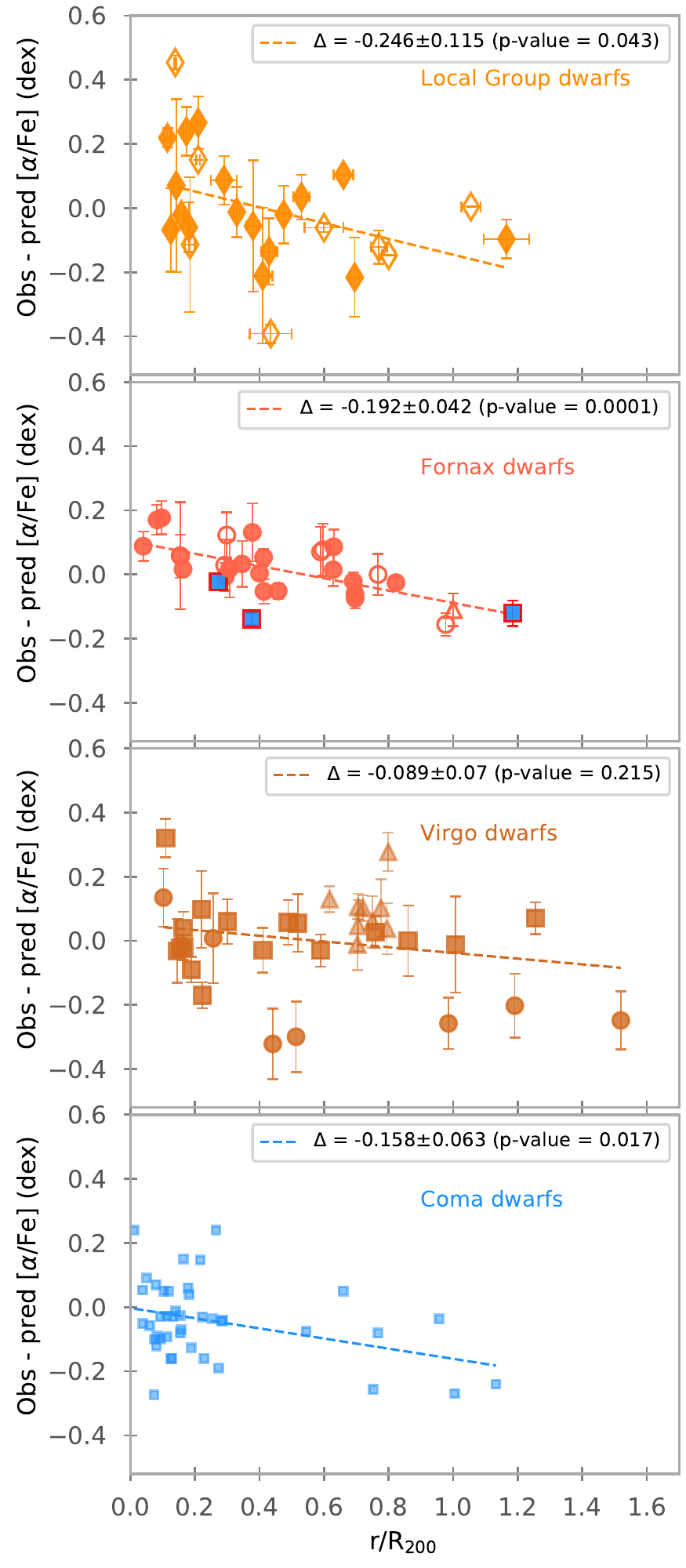}
\caption{$\alpha$-enhancement corrected for mass-dependence. Here we plot the difference between the observed $\alpha$-enhancement and the prediction based on the U-shape, as a function of the cluster-centric distance for dwarf galaxies in different environments. Similar to Fig. \ref{fig_local_group_alpha-distance}.}
\label{fig_delta_alpha_distance}
\end{figure}

In Fig. \ref{fig_local_group_alpha-distance} we show that the most $\alpha$-enhanced galaxies in every cluster are close to the centre, and that there is a negative trend in [$\alpha$/Fe] in each case as well, the smallest in the Virgo Cluster. So, can we deduce that the cluster environment is responsible for this relation, the fact that [$\alpha$/Fe] is enhanced in the centre of a cluster, or group, in the case of the Milky Way? To find out, we now remove the mass dependency of the [$\alpha$/Fe] values by subtracting the [$\alpha$/Fe] value that is expected for a galaxy of a certain mass following the U-shape of Fig. \ref{fig_local_group_alpha-sigma}. This difference is shown in Fig.\ref{fig_delta_alpha_distance}. It shows that galaxy mass does not play an important role in this relation. One can explain it in the following way: when galaxies fall into the cluster at some point an interaction will occur with the intra-cluster medium. Through ram-pressure stripping a strong star formation burst is induced, and the rest of the gas is expelled, causing the high [$\alpha$/Fe] ratios for low mass galaxies \citep[see e.g.][]{Liu2016_distance, Bidaran2022}. Since these events are more likely to occur near the cluster centre, a trend in [$\alpha$/Fe] as a function of radius is created. It is made stronger by the fact that in the outer parts of the cluster these bursts were not so strong, so that star formation is still going on and [$\alpha$/Fe] is lower as a consequence. A similar conclusion is given in \citet{Smith2008} and  \citetalias{Smith2009_highalpha}, who also say that {\it in this scenario, it is qualitatively
expected that a trend to younger ages would be accompanied by a trend towards lower Mg/Fe.} We will study the age dependence of our samples in the next paper (Romero-Gomez et al., in preparation), where we will study the star formation histories in detail.

With all this in mind and after looking at our results, it is clear that whatever process is at play while galaxies are falling into the cluster, dwarfs are much more affected by the environment than massive galaxies.

The statistics shown in Fig. \ref{fig_delta_alpha_distance} indicate that there is a strong relation between the [$\alpha$/Fe] abundance of the dwarfs and the distance in Fornax, Coma and the Local Group. These relations are important considering that we are comparing values from galaxy clusters and a group, environments with a large range in mass and dynamically distinct.

While in clusters, a dense intra-cluster medium can affect dwarf galaxies through ram-pressure stripping, there is the circumgalactic medium in the halo of our Milky Way that can play a similar role, quenching satellites as they orbit around our Galaxy \citep[][]{Putman2021, Akins2021-simulations}. We notice a similarity between the galaxy clusters and the Local Group. The dwarfs with low $\alpha$-enhancement,  the ones that are close to the linear relation with log($\sigma$) in the lower panel of Fig. \ref{fig_local_group_alpha-sigma}, are galaxies that far away from the centre of the cluster or group.
Those objects did not fall into the cluster, therefore did not have a burst of star formation increasing their [$\alpha$/Fe]-ratio, but are still classified as quiescent dwarfs. It is expected that they might still contain some gas. 

We note that the [$\alpha$/Fe] trend in the Virgo Cluster is much smaller than in the other clusters and in the Milky Way. This is possibly the case because Virgo is a less relaxed cluster than Coma and Fornax \citep{Choque-Chapalla2022_thesis}, with the fraction of late-type galaxies larger than in Coma and Fornax. The data in Virgo come from \citet{Liu2016_distance}, \citet{Sybilska2017} and \citet{Bidaran2022}. In this latter paper galaxies are studied from a group falling into the Virgo Cluster. Since these galaxies are of intermediate age, the last burst, leading to the high [$\alpha$/Fe] must have happened in the group as so-called pre-processing. 
The dependence of the slope of the [$\alpha$/Fe] vs. distance diagram on cluster evolution offers interesting possibilities for future investigations.

\subsubsection{Concluding remarks about the relation between [$\alpha$/Fe] and stellar mass}\label{alpha-final-remarks}

As a result of our analysis of the relation between [$\alpha$/Fe] and log($\sigma$), we find a U-shape that seems to be dictated by a combination of internal properties and environmental processes. So, for which galaxies did the internal processes dominate, and for which the external ones? For massive galaxies like in ATLAS$^{3D}$ their stronger potential wells allow them to be less affected by the environment, and they are quenched early due to internal processes that are mass-sensitive, the more massive galaxies are quenched faster and we see higher ratios of [$\alpha$/Fe] abundances. This is confirmed by \citet{Zheng2019}, who studied a sample of MaNGA galaxies and found the usual correlation between [$\alpha$/Fe] and velocity dispersion for massive galaxies, down to a velocity dispersion of about $\sigma$ $\sim$ 80 km/s. They concluded that overall values depend mostly on internal properties and to a lesser degree on environmental effects, confirmed by the fact that we do not see much difference in the stellar populations of ATLAS$^{3D}$ galaxies as a function of clustercentric distance in Virgo. When reaching the massive dwarf regime, 10$^{8}$-10$^{9}$ M$_{\odot}$, we see a mix of $\alpha$-enhanced values that start to be dominated by the environment, deviating from the linear [$\alpha$/Fe]-log($\sigma$) relation.

In clusters like Fornax, as dwarf galaxies with mass smaller than 10$^8$ M$_\odot$ fall into the cluster environment, ram-pressure stripping could remove all the gas and incite a burst of star formation, increasing the [$\alpha$/Fe] ratio. More massive dwarfs are less affected and can retain gas in their inner parts, and for new generations of stars, prolonging star formation in these regions and lowering the [$\alpha$/Fe] ratio. When objects are even more massive, the effect of the environment becomes less strong.

For very low-mass dwarfs, like the ones in our Local Group sample, we see how satellites closer to our Milky Way are more enhanced,  showing a much more pronounced U-shape. This is clearly a strong effect of the environment as we can see in Fig. \ref{fig_local_group_alpha-distance} and \ref{fig_delta_alpha_distance}. This quenching relation with the environment or distance to our Galaxy has been already pointed out by other studies. \citet{Putman2021} showed that most of these dwarfs have not been detected in gas and only in those farther away than the virial radius of the Milky Way HI gas has been detected. These objects tend to be classified as star forming dwarfs (dIrr) and are therefore not included in our sample. Also in \citet{Naidu2022}, studying recently discovered disrupted dwarfs, galaxies whose debris form the stellar halo, they found that disrupted dwarfs are more $\alpha$-enhanced and metal-poor than other surviving dwarfs. This means that the gaseous halo of our Milky Way could produce a quenching process similar to that of the intra-cluster medium, quenching mostly the least massive galaxies and giving us the observed higher [$\alpha$/Fe] abundances. 

All of this is consistent with the simulations of dwarf satellite galaxies in Milky Way-mass halos made by \citet{Akins2021-simulations}. They found that dwarfs with stellar masses between 10$^{6}$-10$^{8}$ M$_{\odot}$ quickly quench after infall, although others with the same order of magnitude in mass but with a larger gas fraction can carry on forming stars a few more Gyr after infall. This explains not only high $\alpha$-enhancement values in the U-shape, but also the small branch of galaxies with solar-like abundance ratios that lie on the linear relation with velocity dispersion (see section \ref{star_forming}), and may have fallen later in their cluster or group and still retain some gas. \citet{Akins2021-simulations} also found a threshold around 10$^{8}$M$_{\odot}$ in stellar mass where the simulated galaxies with larger masses start to be less quenched by the environment in short timescales, and the quenching efficiency changes. Their proposed stellar mass threshold coincides with the low part of our U-shape, where we find the lower $\alpha$-enhancement values.

\subsection{Possible origin of the three outliers}\label{discuss_origin}
As mentioned before, from the age and metallicity distribution, three particular galaxies call our attention, FCC143, FCC252 and FCC253, the oldest galaxies in our sample of dwarfs (see Fig. \ref{fig_enviroment}). Looking at the [M/H]-M$_{\star}$ relation in the top right panel of Fig. \ref{fig_all_vs_all} they have metallicities compatible with galaxies of 10$^{10}$-10$^{11}$ M$_{\odot}$. But because of their low masses, they should be metal-poor in this scenario. The resemblance of these three galaxies to more massive galaxies indicates that their progenitors were probably from a different, more massive, morphological type, and that they were stripped by some environmental, probably tidal, process down to their current sizes, transforming them into quiescent dwarfs \citep[][]{Paudel2010}. Given the age of these galaxies, their quenching could have happened during or even before the re-ionization epoch. A more detailed study of their star formation histories is needed to see if the time at which 90\% of their mass was formed is set during re-ionization. It is unlikely that re-ionization alone was the reason why these galaxies were quenched. Different studies based on cosmological simulations suggest that only less massive dwarfs, $M_{\star} < 10^{6}M_{\odot}$, are quenched by cosmic re-ionization \citep{Simon2019, Rey_2022, Pereira_2023}. Since these galaxies live in a high-density environment, like a cluster, they have more chances of maintaining this slow accreting rate. For this, we will study their star formation histories in more detail (Romero-Gomez et al., in preparation).

For FCC143, based on its surface brightness, effective radius and S\'ersic index \citepalias{Eftekhari_2021_fornaxII}, we noticed that it seems to be a compact elliptical. Systems like this in environments of high density are expected to have older and more metal-rich populations than similar objects with lower masses \citep{Guerou2015}. Also, judging by its position, FCC143 could have been a more extended galaxy that transformed into a compact dwarf after falling through the cluster. FCC252 does not have compact-like properties, and it could be a primordial galaxy that evolved alongside Fornax. On the other hand, the location of FCC253 just outside the virial radius suggests other possibilities. A simple explanation would be that it suffered some pre-processing in another environment \citep{Bidaran2022}. Another possibility is that it might be a backsplash galaxy \citep{Sales2007}. Between typical distances of one or two virial radii, backsplash galaxies are objects that fell into the cluster a long time ago but after passing near the core their orbits create a slingshot effect that throws these galaxies to the outskirts of the cluster. According to \citet{Gill2005}, in this manoeuvre galaxies could lose up to 40\% of their mass, which would reinforce the possibility that the progenitor of FCC253 is a more massive galaxy. Compared with the phase-space presented by \citet{Gill2005}, it even has kinematics compatible with a backsplash galaxy. Unfortunately, with observations is hard to distinguish these galaxies from those that are falling to the environment \citep{Pimbblet2011_blacksplash}, and their study is more common in $\Lambda$CDM cosmological simulations of clusters \citep{Haggar2020}.
\begin{figure}
\centering
\includegraphics[scale=0.06]{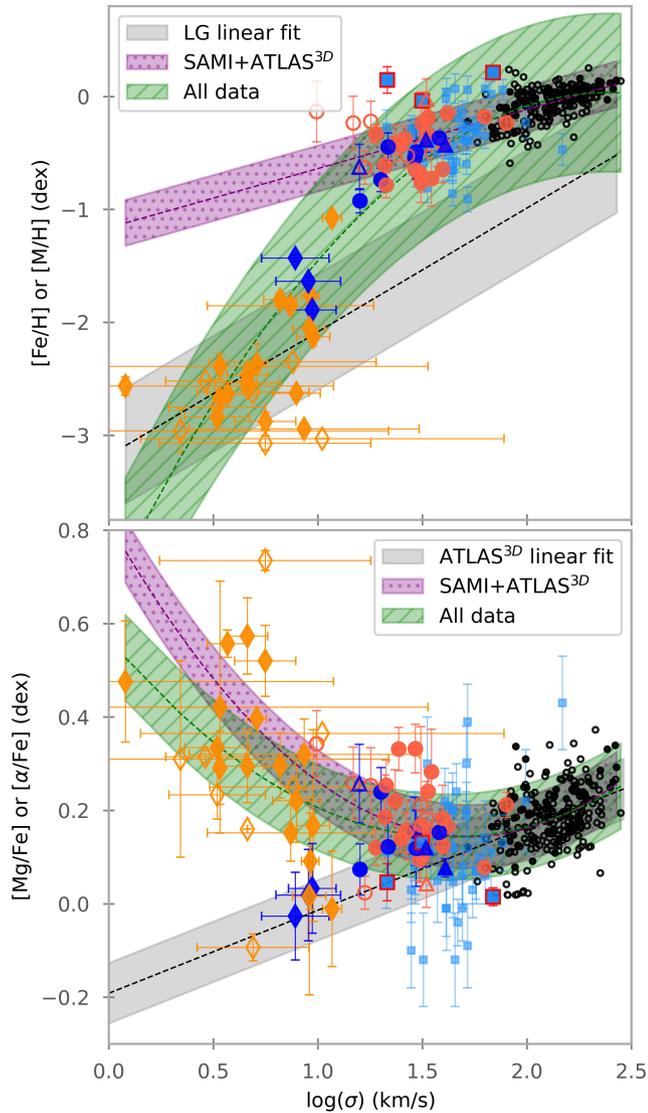}
\caption{Same as Fig. \ref{fig_local_group_alpha-sigma} but with the star forming galaxies, in blue. And also the three dIrr from the Local Group are shown with filled blue diamonds.}
\label{fig_local_group_alpha-sigma_WE}
\end{figure}
\subsection{Star forming galaxies}\label{star_forming}
In all our previous analyses and discussions we have excluded star forming dwarfs (dIrr) galaxies from our SAMI-Fornax sample. However, several of them were observed when the sample was taken. Analyzing the spectra in the same way, excluding the emission lines, we find that these galaxies are younger and less metal-rich than the quiescent dwarfs. With a mean luminosity-weighted age and metallicity of 5.22$\pm$2.16 Gyr and -0.55$\pm$0.18 dex, respectively.
Adding these objects to Fig. \ref{fig_all_vs_all} does not show anything new, the age and metallicity relations with stellar mass are quite similar. The same thing happens with the distribution, as a function of clustercentric distance, that the stellar populations of star forming dwarfs do not imply any kind of relation or trend.

In Fig. \ref{fig_local_group_alpha-sigma_WE} we replicate Fig. \ref{fig_local_group_alpha-sigma} but include the star forming dwarfs from our SAMI-Fornax sample and also from the Local Group. For the metallicity, we notice that these objects fit perfectly into the second-degree [M/H]-log($\sigma$) relation. For the $\alpha$-enhancement, these young objects appear to have in general lower $\alpha$-enhancement values, placing them closer to the linear relation. This indicates that star-forming galaxies sit on a different line than non-star-forming ones. Given our previous interpretation of the relation between [$\alpha$/Fe] and stellar mass, these low-mass objects have probably not fallen into the cluster environment yet. The dIrr we have in the Local Group sample are galaxies positioned far away from the Milky Way. There are, however, studies, like the SAGA survey, that indicate that there should be more star-forming satellites at closer distances \citep{Geha_2017}. This study looked for nearby Milky Way analogues and found a completely different fraction of quiescent satellites than was expected. While their radial distribution agrees well with simulations \citep{Samuel_2020}, the quiescent fraction is only reproduced at larger masses \citep{Samuel_2022}. In any case, it would be interesting to know if these star-forming galaxies that are closer to their hosts can be found on the linear relation, as we show in Fig. \ref{fig_local_group_alpha-sigma_WE}, to study in detail whether the environmental effects that we are inferring from the Local Group sample are representative for other groups of galaxies.

\section{Summary and Conclusions}\label{conclusions}
We present the spectral analysis of 31 quiescent dwarfs in the Fornax galaxy cluster with a stellar mass between $10^{7}$-$10^{9.5} M_{\odot}$, a sample one order of magnitude less massive and a factor 5 larger in number than previous IFU studies. Also for comparison, we include spectra from the 260 ETGs of the ATLAS$^{3D}$ project in our sample. Using FSF techniques we derive stellar populations properties and look for possible relations of these with stellar mass, velocity dispersion or environment, from which we draw the following conclusions:
\begin{enumerate}
    \item {\bf{$\alpha$-enhancement relations:}}
    \begin{itemize}
  \item The linear relation between [$\alpha$/Fe] and log($\sigma$), which is most commonly used in the literature, does not fit our dwarf galaxies. Instead, a second-degree polynomial is a better fit for the whole sample, confirming a U-shape. Here galaxies with masses of $10^9$-$10^{10}$ M$_{\odot}$ have solar abundance ratios, and then $\alpha$-enhancement values increase for both massive and less massive galaxies. For less massive dwarfs this relation connects with the satellite galaxies of the Milky Way. 
  \item When looking at the relation with the environment we find that $\alpha$-enhancement decreases with the projected distance to the Fornax cluster centre. This relation is also present in dwarfs in other environments like Coma or the Local Group, but to a lesser degree in Virgo. This is caused by the effect that the Virgo Cluster is not fully relaxed and that the relation of abundance ratio with clustercentric distance is not fully in place yet. The relation between $\alpha$-enhancement and clustercentric distance is not affected by galaxy mass.
  \item We can explain both  the mass-metallicity and the mass-$\alpha$-enhancement relations using a plausible combination of internal and environmental factors that affect galaxies of different mass in different ways.
\end{itemize}
\item {\bf{Stellar populations properties:}}
\begin{itemize}
  \item When comparing with more massive ATLAS$^{3D}$ galaxies we show that our dwarfs follow the known metallicity-stellar mass linear relation. After including Local Group dwarfs, we notice that it is not feasible to fit all samples using the same linear relation, as was done, for example, by \citet{Kirby2013}.  To do so a second-degree polynomial is needed.
  \item In our sample we identify three outliers to the mass-metallicity and mass-age relation. FCC143, 252 and 253 have solar-like metallicities commonly found only in the most massive giants. Their current positions and structural properties favour a scenario where their progenitors could have been massive galaxies that through various mechanisms suffered mass loss. FCC143 is a compact elliptical, which is expected to have these extreme properties in high-density environments, FCC252 could possibly be a primordial galaxy in the Fornax cluster and FCC253 could have undergone some pre-processing or be a backsplash galaxy.
  \item In general, the stellar population parameters for our dwarf galaxies are consistent with the typical values reported in previous works of the literature, intermediate-old ages with considerable scatter, and sub-solar metallicities. The same is true for the abundance ratio [$\alpha$/Fe], values close to solar but with some galaxies a little bit enhanced.
\end{itemize}
\end{enumerate}

As we have seen there is quite a debate in the literature about the possible relation of the abundance ratio [$\alpha$/Fe] with the environment and formation time scales. To research deeper into the latter, in the next papers of this series, we are obtaining the SFH of every galaxy in our sample, and studying the possible relations with the stellar population properties or environment.

\section*{Data availability}
The reduced data underlying this article will be made available through the CDS. The raw data are publicly available in the AAT data archive.

\section*{Acknowledgements}
This research has been supported by the Spanish Ministry of Education, Culture and Sports under grant AYA2017-83204-P.

The Sydney-AAO Multi-object Integral field spectrograph (SAMI) was developed jointly by the University of Sydney and the Australian Astronomical Observatory.

Parts of this research were supported by the Australian Research Council Centre of Excellence for All Sky Astrophysics in 3 Dimensions (ASTRO 3D), through project number CE170100013.

GvdV acknowledges funding from the European Research Council (ERC) under the European Union's Horizon 2020 research and innovation programme under grant agreement No. 724857 (Consolidator Grant ArcheoDyn).

RFP acknowledges the financial support from the European Union’s Horizon 2020 research and innovation program under the Marie Sklodowska-Curie grant agreement No. 721463 to the SUNDIAL ITN network.

JF-B acknowledges support through the RAVET project by the grant PID2019-107427GB-C32 from the Spanish Ministry of Science, Innovation and Universities (MCIU), and through the IAC project TRACES which is partially supported through the state budget and the regional budget of the Consejería de Economía, Industria, Comercio y Conocimiento of the Canary Islands Autonomous Community.

For the analysis we have used Python \href{http://www.python.org}{http://www.python.org}; Matplotlib \citep{Hunter2007}, a suite of open source python modules that provide a framework for creating scientific plots; and Astropy, a community-developed core Python package for Astronomy \citep{Astropy2013}.



\bibliographystyle{mnras}
\bibliography{References} 

\begin{thebibliography}{}
\makeatletter
\relax
\def\mn@urlcharsother{\let\do\@makeother \do\$\do\&\do\#\do\^\do\_\do\%\do\~}
\def\mn@doi{\begingroup\mn@urlcharsother \@ifnextchar [ {\mn@doi@}
  {\mn@doi@[]}}
\def\mn@doi@[#1]#2{\def\@tempa{#1}\ifx\@tempa\@empty \href
  {http://dx.doi.org/#2} {doi:#2}\else \href {http://dx.doi.org/#2} {#1}\fi
  \endgroup}
\def\mn@eprint#1#2{\mn@eprint@#1:#2::\@nil}
\def\mn@eprint@arXiv#1{\href {http://arxiv.org/abs/#1} {{\tt arXiv:#1}}}
\def\mn@eprint@dblp#1{\href {http://dblp.uni-trier.de/rec/bibtex/#1.xml}
  {dblp:#1}}
\def\mn@eprint@#1:#2:#3:#4\@nil{\def\@tempa {#1}\def\@tempb {#2}\def\@tempc
  {#3}\ifx \@tempc \@empty \let \@tempc \@tempb \let \@tempb \@tempa \fi \ifx
  \@tempb \@empty \def\@tempb {arXiv}\fi \@ifundefined
  {mn@eprint@\@tempb}{\@tempb:\@tempc}{\expandafter \expandafter \csname
  mn@eprint@\@tempb\endcsname \expandafter{\@tempc}}}

\bibitem[\protect\citeauthoryear{{Aguerri} \&
  {Gonz{\'a}lez-Garc{\'\i}a}}{{Aguerri} \&
  {Gonz{\'a}lez-Garc{\'\i}a}}{2009}]{Aguerri-Garcia2009}
{Aguerri} J.~A.~L.,  {Gonz{\'a}lez-Garc{\'\i}a} A.~C.,  2009, \mn@doi [\aap]
  {10.1051/0004-6361:200810339}, \href
  {https://ui.adsabs.harvard.edu/abs/2009A&A...494..891A} {494, 891}

\bibitem[\protect\citeauthoryear{{Akins}, {Christensen}, {Brooks}, {Munshi},
  {Applebaum}, {Engelhardt}  \& {Chamberland}}{{Akins}
  et~al.}{2021}]{Akins2021-simulations}
{Akins} H.~B.,  {Christensen} C.~R.,  {Brooks} A.~M.,  {Munshi} F.,
  {Applebaum} E.,  {Engelhardt} A.,   {Chamberland} L.,  2021, \mn@doi [\apj]
  {10.3847/1538-4357/abe2ab}, \href
  {https://ui.adsabs.harvard.edu/abs/2021ApJ...909..139A} {909, 139}

\bibitem[\protect\citeauthoryear{{Arnone}, {Ryan}, {Argast}, {Norris}  \&
  {Beers}}{{Arnone} et~al.}{2005}]{Arnone2005}
{Arnone} E.,  {Ryan} S.~G.,  {Argast} D.,  {Norris} J.~E.,   {Beers} T.~C.,
  2005, \mn@doi [\aap] {10.1051/0004-6361:20041034}, \href
  {https://ui.adsabs.harvard.edu/abs/2005A&A...430..507A} {430, 507}

\bibitem[\protect\citeauthoryear{{Astropy Collaboration} et~al.,}{{Astropy
  Collaboration} et~al.}{2013}]{Astropy2013}
{Astropy Collaboration} et~al., 2013, \mn@doi [\aap]
  {10.1051/0004-6361/201322068}, \href
  {https://ui.adsabs.harvard.edu/abs/2013A&A...558A..33A} {558, A33}

\bibitem[\protect\citeauthoryear{{Barazza}, {Binggeli}  \& {Jerjen}}{{Barazza}
  et~al.}{2002}]{Barazza2002_spiral}
{Barazza} F.~D.,  {Binggeli} B.,   {Jerjen} H.,  2002, \mn@doi [\aap]
  {10.1051/0004-6361:20020875}, \href
  {https://ui.adsabs.harvard.edu/abs/2002A&A...391..823B} {391, 823}

\bibitem[\protect\citeauthoryear{{Bidaran} et~al.,}{{Bidaran}
  et~al.}{2020}]{Bidaran2020}
{Bidaran} B.,  et~al., 2020, \mn@doi [\mnras] {10.1093/mnras/staa2097}, \href
  {https://ui.adsabs.harvard.edu/abs/2020MNRAS.497.1904B} {497, 1904}

\bibitem[\protect\citeauthoryear{{Bidaran} et~al.,}{{Bidaran}
  et~al.}{2022}]{Bidaran2022}
{Bidaran} B.,  et~al., 2022, arXiv e-prints, \href
  {https://ui.adsabs.harvard.edu/abs/2022arXiv220706977B} {p. arXiv:2207.06977}

\bibitem[\protect\citeauthoryear{{Binggeli}, {Sandage}  \&
  {Tammann}}{{Binggeli} et~al.}{1988a}]{Binggeli1988}
{Binggeli} B.,  {Sandage} A.,   {Tammann} G.~A.,  1988a, \mn@doi [\araa]
  {10.1146/annurev.aa.26.090188.002453}, \href
  {https://ui.adsabs.harvard.edu/abs/1988ARA&A..26..509B} {26, 509}

\bibitem[\protect\citeauthoryear{{Binggeli}, {Sandage}  \&
  {Tammann}}{{Binggeli} et~al.}{1988b}]{Binggeli-Sandange-Tammann1988}
{Binggeli} B.,  {Sandage} A.,   {Tammann} G.~A.,  1988b, \mn@doi [\araa]
  {10.1146/annurev.aa.26.090188.002453}, \href
  {https://ui.adsabs.harvard.edu/abs/1988ARA&A..26..509B} {26, 509}

\bibitem[\protect\citeauthoryear{{Blakeslee} et~al.,}{{Blakeslee}
  et~al.}{2009}]{Blakeslee2009}
{Blakeslee} J.~P.,  et~al., 2009, \mn@doi [\apj] {10.1088/0004-637X/694/1/556},
  \href {https://ui.adsabs.harvard.edu/abs/2009ApJ...694..556B} {694, 556}

\bibitem[\protect\citeauthoryear{{Boselli} \& {Gavazzi}}{{Boselli} \&
  {Gavazzi}}{2014}]{Boselli2014_}
{Boselli} A.,  {Gavazzi} G.,  2014, \mn@doi [\aapr]
  {10.1007/s00159-014-0074-y}, \href
  {https://ui.adsabs.harvard.edu/abs/2014A&ARv..22...74B} {22, 74}

\bibitem[\protect\citeauthoryear{{Boselli} et~al.,}{{Boselli}
  et~al.}{2014}]{Boselli2014_enrionment}
{Boselli} A.,  et~al., 2014, \mn@doi [\aap] {10.1051/0004-6361/201424419},
  \href {https://ui.adsabs.harvard.edu/abs/2014A&A...570A..69B} {570, A69}

\bibitem[\protect\citeauthoryear{{Boselli}, {Fossati}  \& {Sun}}{{Boselli}
  et~al.}{2022}]{Boselli2022}
{Boselli} A.,  {Fossati} M.,   {Sun} M.,  2022, \mn@doi [\aapr]
  {10.1007/s00159-022-00140-3}, \href
  {https://ui.adsabs.harvard.edu/abs/2022A&ARv..30....3B} {30, 3}

\bibitem[\protect\citeauthoryear{{Burstein}, {Faber}, {Gaskell}  \&
  {Krumm}}{{Burstein} et~al.}{1984}]{Burstein1984}
{Burstein} D.,  {Faber} S.~M.,  {Gaskell} C.~M.,   {Krumm} N.,  1984, \mn@doi
  [\apj] {10.1086/162718}, \href
  {https://ui.adsabs.harvard.edu/abs/1984ApJ...287..586B} {287, 586}

\bibitem[\protect\citeauthoryear{{Cappellari}}{{Cappellari}}{2017}]{Capellari2017}
{Cappellari} M.,  2017, \mn@doi [\mnras] {10.1093/mnras/stw3020}, \href
  {https://ui.adsabs.harvard.edu/abs/2017MNRAS.466..798C} {466, 798}

\bibitem[\protect\citeauthoryear{{Cappellari} \& {Copin}}{{Cappellari} \&
  {Copin}}{2003}]{Cappellari_Copin2003}
{Cappellari} M.,  {Copin} Y.,  2003, \mn@doi [\mnras]
  {10.1046/j.1365-8711.2003.06541.x}, \href
  {https://ui.adsabs.harvard.edu/abs/2003MNRAS.342..345C} {342, 345}

\bibitem[\protect\citeauthoryear{{Cappellari} \& {Emsellem}}{{Cappellari} \&
  {Emsellem}}{2004}]{Capellari2004}
{Cappellari} M.,  {Emsellem} E.,  2004, \mn@doi [\pasp] {10.1086/381875}, \href
  {https://ui.adsabs.harvard.edu/abs/2004PASP..116..138C} {116, 138}

\bibitem[\protect\citeauthoryear{{Cappellari} et~al.,}{{Cappellari}
  et~al.}{2011a}]{Cappellari2011-atlas3d}
{Cappellari} M.,  et~al., 2011a, \mn@doi [\mnras]
  {10.1111/j.1365-2966.2010.18174.x}, \href
  {https://ui.adsabs.harvard.edu/abs/2011MNRAS.413..813C} {413, 813}

\bibitem[\protect\citeauthoryear{{Cappellari} et~al.,}{{Cappellari}
  et~al.}{2011b}]{Cappellari2011-density}
{Cappellari} M.,  et~al., 2011b, \mn@doi [\mnras]
  {10.1111/j.1365-2966.2011.18600.x}, \href
  {https://ui.adsabs.harvard.edu/abs/2011MNRAS.416.1680C} {416, 1680}

\bibitem[\protect\citeauthoryear{{Cappellari} et~al.,}{{Cappellari}
  et~al.}{2013}]{Cappellari2013a}
{Cappellari} M.,  et~al., 2013, \mn@doi [\mnras] {10.1093/mnras/stt562}, \href
  {https://ui.adsabs.harvard.edu/abs/2013MNRAS.432.1709C} {432, 1709}

\bibitem[\protect\citeauthoryear{{Chiti}, {Frebel}, {Ji}, {Jerjen}, {Kim}  \&
  {Norris}}{{Chiti} et~al.}{2018}]{Chiti2018}
{Chiti} A.,  {Frebel} A.,  {Ji} A.~P.,  {Jerjen} H.,  {Kim} D.,   {Norris}
  J.~E.,  2018, \mn@doi [\apj] {10.3847/1538-4357/aab4fc}, \href
  {https://ui.adsabs.harvard.edu/abs/2018ApJ...857...74C} {857, 74}

\bibitem[\protect\citeauthoryear{{Choque Challapa}}{{Choque
  Challapa}}{2022}]{Choque-Chapalla2022_thesis}
{Choque Challapa} N.,  2022, PhD thesis, University of Groningen,
  \mn@doi{10.33612/diss.224335388}

\bibitem[\protect\citeauthoryear{{Cid Fernandes}, {Mateus}, {Sodr{\'e}},
  {Stasi{\'n}ska}  \& {Gomes}}{{Cid Fernandes} et~al.}{2005}]{Cid2005}
{Cid Fernandes} R.,  {Mateus} A.,  {Sodr{\'e}} L.,  {Stasi{\'n}ska} G.,
  {Gomes} J.~M.,  2005, \mn@doi [\mnras] {10.1111/j.1365-2966.2005.08752.x},
  \href {https://ui.adsabs.harvard.edu/abs/2005MNRAS.358..363C} {358, 363}

\bibitem[\protect\citeauthoryear{{Conroy}, {Graves}  \& {van Dokkum}}{{Conroy}
  et~al.}{2014}]{Conroy2014}
{Conroy} C.,  {Graves} G.~J.,   {van Dokkum} P.~G.,  2014, \mn@doi [\apj]
  {10.1088/0004-637X/780/1/33}, \href
  {https://ui.adsabs.harvard.edu/abs/2014ApJ...780...33C} {780, 33}

\bibitem[\protect\citeauthoryear{{Croom} et~al.,}{{Croom}
  et~al.}{2012}]{Croom2012}
{Croom} S.~M.,  et~al., 2012, \mn@doi [\mnras]
  {10.1111/j.1365-2966.2011.20365.x}, \href
  {https://ui.adsabs.harvard.edu/abs/2012MNRAS.421..872C} {421, 872}

\bibitem[\protect\citeauthoryear{{Drinkwater}, {Jones}, {Gregg}  \&
  {Phillipps}}{{Drinkwater} et~al.}{2000}]{Drinkwater2000}
{Drinkwater} M.~J.,  {Jones} J.~B.,  {Gregg} M.~D.,   {Phillipps} S.,  2000,
  \mn@doi [\pasa] {10.1071/AS00034}, \href
  {https://ui.adsabs.harvard.edu/abs/2000PASA...17..227D} {17, 227}

\bibitem[\protect\citeauthoryear{{Drinkwater}, {Gregg}  \&
  {Colless}}{{Drinkwater} et~al.}{2001}]{Drinkwater2001}
{Drinkwater} M.~J.,  {Gregg} M.~D.,   {Colless} M.,  2001, \mn@doi [\apjl]
  {10.1086/319113}, \href
  {https://ui.adsabs.harvard.edu/abs/2001ApJ...548L.139D} {548, L139}

\bibitem[\protect\citeauthoryear{{Eftekhari} et~al.,}{{Eftekhari}
  et~al.}{2022}]{Eftekhari_2021_fornaxII}
{Eftekhari} F.~S.,  et~al., 2022, arXiv e-prints, \href
  {https://ui.adsabs.harvard.edu/abs/2022arXiv220905525E} {p. arXiv:2209.05525}

\bibitem[\protect\citeauthoryear{{Ekers}, {Goss}, {Wellington}, {Bosma},
  {Smith}  \& {Schweizer}}{{Ekers} et~al.}{1983}]{Ekers1983}
{Ekers} R.~D.,  {Goss} W.~M.,  {Wellington} K.~J.,  {Bosma} A.,  {Smith} R.~M.,
    {Schweizer} F.,  1983, \aap, \href
  {https://ui.adsabs.harvard.edu/abs/1983A&A...127..361E} {127, 361}

\bibitem[\protect\citeauthoryear{{Fabrizio} et~al.,}{{Fabrizio}
  et~al.}{2015}]{Fabrizio2015}
{Fabrizio} M.,  et~al., 2015, \mn@doi [\aap] {10.1051/0004-6361/201525753},
  \href {https://ui.adsabs.harvard.edu/abs/2015A&A...580A..18F} {580, A18}

\bibitem[\protect\citeauthoryear{{Falc{\'o}n-Barroso},
  {S{\'a}nchez-Bl{\'a}zquez}, {Vazdekis}, {Ricciardelli}, {Cardiel}, {Cenarro},
  {Gorgas}  \& {Peletier}}{{Falc{\'o}n-Barroso}
  et~al.}{2011}]{Falcon-Barroso2011}
{Falc{\'o}n-Barroso} J.,  {S{\'a}nchez-Bl{\'a}zquez} P.,  {Vazdekis} A.,
  {Ricciardelli} E.,  {Cardiel} N.,  {Cenarro} A.~J.,  {Gorgas} J.,
  {Peletier} R.~F.,  2011, \mn@doi [\aap] {10.1051/0004-6361/201116842}, \href
  {https://ui.adsabs.harvard.edu/abs/2011A&A...532A..95F} {532, A95}

\bibitem[\protect\citeauthoryear{{Fan}, {Carilli}  \& {Keating}}{{Fan}
  et~al.}{2006}]{Fan2006_reionization}
{Fan} X.,  {Carilli} C.~L.,   {Keating} B.,  2006, \mn@doi [\araa]
  {10.1146/annurev.astro.44.051905.092514}, \href
  {https://ui.adsabs.harvard.edu/abs/2006ARA&A..44..415F} {44, 415}

\bibitem[\protect\citeauthoryear{{Ferguson}}{{Ferguson}}{1989}]{Ferguson1989}
{Ferguson} H.~C.,  1989, \mn@doi [\aj] {10.1086/115152}, \href
  {https://ui.adsabs.harvard.edu/abs/1989AJ.....98..367F} {98, 367}

\bibitem[\protect\citeauthoryear{{Fran{\c{c}}ois}, {Monaco}, {Bonifacio}, {Moni
  Bidin}, {Geisler}  \& {Sbordone}}{{Fran{\c{c}}ois}
  et~al.}{2016}]{Francois2016}
{Fran{\c{c}}ois} P.,  {Monaco} L.,  {Bonifacio} P.,  {Moni Bidin} C.,
  {Geisler} D.,   {Sbordone} L.,  2016, \mn@doi [\aap]
  {10.1051/0004-6361/201527181}, \href
  {https://ui.adsabs.harvard.edu/abs/2016A&A...588A...7F} {588, A7}

\bibitem[\protect\citeauthoryear{{Frebel}, {Simon}, {Geha}  \&
  {Willman}}{{Frebel} et~al.}{2010}]{Frebel2010}
{Frebel} A.,  {Simon} J.~D.,  {Geha} M.,   {Willman} B.,  2010, \mn@doi [\apj]
  {10.1088/0004-637X/708/1/560}, \href
  {https://ui.adsabs.harvard.edu/abs/2010ApJ...708..560F} {708, 560}

\bibitem[\protect\citeauthoryear{{Frebel}, {Simon}  \& {Kirby}}{{Frebel}
  et~al.}{2014}]{Frebel2014}
{Frebel} A.,  {Simon} J.~D.,   {Kirby} E.~N.,  2014, \mn@doi [\apj]
  {10.1088/0004-637X/786/1/74}, \href
  {https://ui.adsabs.harvard.edu/abs/2014ApJ...786...74F} {786, 74}

\bibitem[\protect\citeauthoryear{{Frebel}, {Norris}, {Gilmore}  \&
  {Wyse}}{{Frebel} et~al.}{2016}]{Frebel2016}
{Frebel} A.,  {Norris} J.~E.,  {Gilmore} G.,   {Wyse} R. F.~G.,  2016, \mn@doi
  [\apj] {10.3847/0004-637X/826/2/110}, \href
  {https://ui.adsabs.harvard.edu/abs/2016ApJ...826..110F} {826, 110}

\bibitem[\protect\citeauthoryear{{Geha}, {Guhathakurta}  \& {van der
  Marel}}{{Geha} et~al.}{2003}]{Geha2003_solaralpha}
{Geha} M.,  {Guhathakurta} P.,   {van der Marel} R.~P.,  2003, \mn@doi [\aj]
  {10.1086/377624}, \href
  {https://ui.adsabs.harvard.edu/abs/2003AJ....126.1794G} {126, 1794}

\bibitem[\protect\citeauthoryear{{Geha} et~al.,}{{Geha}
  et~al.}{2017}]{Geha_2017}
{Geha} M.,  et~al., 2017, \mn@doi [\apj] {10.3847/1538-4357/aa8626}, \href
  {https://ui.adsabs.harvard.edu/abs/2017ApJ...847....4G} {847, 4}

\bibitem[\protect\citeauthoryear{{Gill}, {Knebe}  \& {Gibson}}{{Gill}
  et~al.}{2005}]{Gill2005}
{Gill} S. P.~D.,  {Knebe} A.,   {Gibson} B.~K.,  2005, \mn@doi [\mnras]
  {10.1111/j.1365-2966.2004.08562.x}, \href
  {https://ui.adsabs.harvard.edu/abs/2005MNRAS.356.1327G} {356, 1327}

\bibitem[\protect\citeauthoryear{{Gilmore}, {Norris}, {Monaco}, {Yong}, {Wyse}
  \& {Geisler}}{{Gilmore} et~al.}{2013}]{Gilmore2013}
{Gilmore} G.,  {Norris} J.~E.,  {Monaco} L.,  {Yong} D.,  {Wyse} R. F.~G.,
  {Geisler} D.,  2013, \mn@doi [\apj] {10.1088/0004-637X/763/1/61}, \href
  {https://ui.adsabs.harvard.edu/abs/2013ApJ...763...61G} {763, 61}

\bibitem[\protect\citeauthoryear{{Gu{\'e}rou} et~al.,}{{Gu{\'e}rou}
  et~al.}{2015}]{Guerou2015}
{Gu{\'e}rou} A.,  et~al., 2015, \mn@doi [\apj] {10.1088/0004-637X/804/1/70},
  \href {https://ui.adsabs.harvard.edu/abs/2015ApJ...804...70G} {804, 70}

\bibitem[\protect\citeauthoryear{{Gunn} \& {Gott}}{{Gunn} \&
  {Gott}}{1972}]{Gunn1972-rampressure}
{Gunn} J.~E.,  {Gott} J.~Richard I.,  1972, \mn@doi [\apj] {10.1086/151605},
  \href {https://ui.adsabs.harvard.edu/abs/1972ApJ...176....1G} {176, 1}

\bibitem[\protect\citeauthoryear{{Haggar}, {Gray}, {Pearce}, {Knebe}, {Cui},
  {Mostoghiu}  \& {Yepes}}{{Haggar} et~al.}{2020}]{Haggar2020}
{Haggar} R.,  {Gray} M.~E.,  {Pearce} F.~R.,  {Knebe} A.,  {Cui} W.,
  {Mostoghiu} R.,   {Yepes} G.,  2020, \mn@doi [\mnras]
  {10.1093/mnras/staa273}, \href
  {https://ui.adsabs.harvard.edu/abs/2020MNRAS.492.6074H} {492, 6074}

\bibitem[\protect\citeauthoryear{{Haines}, {Gargiulo}, {La Barbera},
  {Mercurio}, {Merluzzi}  \& {Busarello}}{{Haines} et~al.}{2007}]{Haines2007}
{Haines} C.~P.,  {Gargiulo} A.,  {La Barbera} F.,  {Mercurio} A.,  {Merluzzi}
  P.,   {Busarello} G.,  2007, \mn@doi [\mnras]
  {10.1111/j.1365-2966.2007.12189.x}, \href
  {https://ui.adsabs.harvard.edu/abs/2007MNRAS.381....7H} {381, 7}

\bibitem[\protect\citeauthoryear{{Hansen} et~al.,}{{Hansen}
  et~al.}{2017}]{Hansen2017}
{Hansen} T.~T.,  et~al., 2017, \mn@doi [\apj] {10.3847/1538-4357/aa634a}, \href
  {https://ui.adsabs.harvard.edu/abs/2017ApJ...838...44H} {838, 44}

\bibitem[\protect\citeauthoryear{{Hilker}, {Infante}, {Vieira}, {Kissler-Patig}
   \& {Richtler}}{{Hilker} et~al.}{1999}]{Hilker1999}
{Hilker} M.,  {Infante} L.,  {Vieira} G.,  {Kissler-Patig} M.,   {Richtler} T.,
   1999, \mn@doi [\aaps] {10.1051/aas:1999434}, \href
  {https://ui.adsabs.harvard.edu/abs/1999A&AS..134...75H} {134, 75}

\bibitem[\protect\citeauthoryear{{Hunter}}{{Hunter}}{2007}]{Hunter2007}
{Hunter} J.~D.,  2007, \mn@doi [Computing in Science and Engineering]
  {10.1109/MCSE.2007.55}, \href
  {https://ui.adsabs.harvard.edu/abs/2007CSE.....9...90H} {9, 90}

\bibitem[\protect\citeauthoryear{{Iodice} et~al.,}{{Iodice}
  et~al.}{2017}]{Iodice2017}
{Iodice} E.,  et~al., 2017, \mn@doi [\apj] {10.3847/1538-4357/aa6846}, \href
  {https://ui.adsabs.harvard.edu/abs/2017ApJ...839...21I} {839, 21}

\bibitem[\protect\citeauthoryear{{Iodice} et~al.,}{{Iodice}
  et~al.}{2019}]{Iodice2019}
{Iodice} E.,  et~al., 2019, \mn@doi [\aap] {10.1051/0004-6361/201833741}, \href
  {https://ui.adsabs.harvard.edu/abs/2019A&A...623A...1I} {623, A1}

\bibitem[\protect\citeauthoryear{{Ishigaki}, {Aoki}, {Arimoto}  \&
  {Okamoto}}{{Ishigaki} et~al.}{2014}]{Ishigaki2014}
{Ishigaki} M.~N.,  {Aoki} W.,  {Arimoto} N.,   {Okamoto} S.,  2014, \mn@doi
  [\aap] {10.1051/0004-6361/201322796}, \href
  {https://ui.adsabs.harvard.edu/abs/2014A&A...562A.146I} {562, A146}

\bibitem[\protect\citeauthoryear{{Jablonka} et~al.,}{{Jablonka}
  et~al.}{2015}]{Jablonka2015}
{Jablonka} P.,  et~al., 2015, \mn@doi [\aap] {10.1051/0004-6361/201525661},
  \href {https://ui.adsabs.harvard.edu/abs/2015A&A...583A..67J} {583, A67}

\bibitem[\protect\citeauthoryear{{Jaff{\'e}} et~al.,}{{Jaff{\'e}}
  et~al.}{2018}]{Jaffe2018}
{Jaff{\'e}} Y.~L.,  et~al., 2018, \mn@doi [\mnras] {10.1093/mnras/sty500},
  \href {https://ui.adsabs.harvard.edu/abs/2018MNRAS.476.4753J} {476, 4753}

\bibitem[\protect\citeauthoryear{{Janz} et~al.,}{{Janz}
  et~al.}{2014}]{Janz2014_bars}
{Janz} J.,  et~al., 2014, \mn@doi [\apj] {10.1088/0004-637X/786/2/105}, \href
  {https://ui.adsabs.harvard.edu/abs/2014ApJ...786..105J} {786, 105}

\bibitem[\protect\citeauthoryear{{Jerjen}, {Kalnajs}  \& {Binggeli}}{{Jerjen}
  et~al.}{2000}]{Jerjen2000_spiral}
{Jerjen} H.,  {Kalnajs} A.,   {Binggeli} B.,  2000, \aap, \href
  {https://ui.adsabs.harvard.edu/abs/2000A&A...358..845J} {358, 845}

\bibitem[\protect\citeauthoryear{{Jerjen}, {Conn}, {Kim}  \&
  {Schirmer}}{{Jerjen} et~al.}{2018}]{Jerjen2018}
{Jerjen} H.,  {Conn} B.,  {Kim} D.,   {Schirmer} M.,  2018, arXiv e-prints,
  \href {https://ui.adsabs.harvard.edu/abs/2018arXiv180902259J} {p.
  arXiv:1809.02259}

\bibitem[\protect\citeauthoryear{{Ji}, {Frebel}, {Simon}  \& {Chiti}}{{Ji}
  et~al.}{2016}]{Ji2016}
{Ji} A.~P.,  {Frebel} A.,  {Simon} J.~D.,   {Chiti} A.,  2016, \mn@doi [\apj]
  {10.3847/0004-637X/830/2/93}, \href
  {https://ui.adsabs.harvard.edu/abs/2016ApJ...830...93J} {830, 93}

\bibitem[\protect\citeauthoryear{{Ji}, {Simon}, {Frebel}, {Venn}  \&
  {Hansen}}{{Ji} et~al.}{2019}]{Ji2019}
{Ji} A.~P.,  {Simon} J.~D.,  {Frebel} A.,  {Venn} K.~A.,   {Hansen} T.~T.,
  2019, \mn@doi [\apj] {10.3847/1538-4357/aaf3bb}, \href
  {https://ui.adsabs.harvard.edu/abs/2019ApJ...870...83J} {870, 83}

\bibitem[\protect\citeauthoryear{{Ji} et~al.,}{{Ji} et~al.}{2020}]{Ji2020}
{Ji} A.~P.,  et~al., 2020, \mn@doi [\apj] {10.3847/1538-4357/ab6213}, \href
  {https://ui.adsabs.harvard.edu/abs/2020ApJ...889...27J} {889, 27}

\bibitem[\protect\citeauthoryear{{Kaufer}, {Venn}, {Tolstoy}, {Pinte}  \&
  {Kudritzki}}{{Kaufer} et~al.}{2004}]{Kaufer2004solaralpha}
{Kaufer} A.,  {Venn} K.~A.,  {Tolstoy} E.,  {Pinte} C.,   {Kudritzki} R.-P.,
  2004, \mn@doi [\aj] {10.1086/383209}, \href
  {https://ui.adsabs.harvard.edu/abs/2004AJ....127.2723K} {127, 2723}

\bibitem[\protect\citeauthoryear{{Kirby}, {Cohen}, {Guhathakurta}, {Cheng},
  {Bullock}  \& {Gallazzi}}{{Kirby} et~al.}{2013}]{Kirby2013}
{Kirby} E.~N.,  {Cohen} J.~G.,  {Guhathakurta} P.,  {Cheng} L.,  {Bullock}
  J.~S.,   {Gallazzi} A.,  2013, \mn@doi [\apj] {10.1088/0004-637X/779/2/102},
  \href {https://ui.adsabs.harvard.edu/abs/2013ApJ...779..102K} {779, 102}

\bibitem[\protect\citeauthoryear{{Kirby} et~al.,}{{Kirby}
  et~al.}{2015}]{Kirby2015}
{Kirby} E.~N.,  et~al., 2015, \mn@doi [\apj] {10.1088/0004-637X/801/2/125},
  \href {https://ui.adsabs.harvard.edu/abs/2015ApJ...801..125K} {801, 125}

\bibitem[\protect\citeauthoryear{{Kirby}, {Cohen}, {Simon}, {Guhathakurta},
  {Thygesen}  \& {Duggan}}{{Kirby} et~al.}{2017}]{Kirby2017}
{Kirby} E.~N.,  {Cohen} J.~G.,  {Simon} J.~D.,  {Guhathakurta} P.,  {Thygesen}
  A.~O.,   {Duggan} G.~E.,  2017, \mn@doi [\apj] {10.3847/1538-4357/aa6570},
  \href {https://ui.adsabs.harvard.edu/abs/2017ApJ...838...83K} {838, 83}

\bibitem[\protect\citeauthoryear{{Koch}, {McWilliam}, {Grebel}, {Zucker}  \&
  {Belokurov}}{{Koch} et~al.}{2008}]{Koch2008}
{Koch} A.,  {McWilliam} A.,  {Grebel} E.~K.,  {Zucker} D.~B.,   {Belokurov} V.,
   2008, \mn@doi [\apjl] {10.1086/595001}, \href
  {https://ui.adsabs.harvard.edu/abs/2008ApJ...688L..13K} {688, L13}

\bibitem[\protect\citeauthoryear{{Koleva}, {de Rijcke}, {Prugniel}, {Zeilinger}
   \& {Michielsen}}{{Koleva} et~al.}{2009}]{Koleva2009}
{Koleva} M.,  {de Rijcke} S.,  {Prugniel} P.,  {Zeilinger} W.~W.,
  {Michielsen} D.,  2009, \mn@doi [\mnras] {10.1111/j.1365-2966.2009.14820.x},
  \href {https://ui.adsabs.harvard.edu/abs/2009MNRAS.396.2133K} {396, 2133}

\bibitem[\protect\citeauthoryear{{Koleva}, {Prugniel}, {De Rijcke}  \&
  {Zeilinger}}{{Koleva} et~al.}{2011}]{Koleva2011}
{Koleva} M.,  {Prugniel} P.,  {De Rijcke} S.,   {Zeilinger} W.~W.,  2011,
  \mn@doi [\mnras] {10.1111/j.1365-2966.2011.19057.x}, \href
  {https://ui.adsabs.harvard.edu/abs/2011MNRAS.417.1643K} {417, 1643}

\bibitem[\protect\citeauthoryear{{Kuijken} et~al.,}{{Kuijken}
  et~al.}{2002}]{Kuijken2002_omegacam}
{Kuijken} K.,  et~al., 2002, The Messenger, \href
  {https://ui.adsabs.harvard.edu/abs/2002Msngr.110...15K} {110, 15}

\bibitem[\protect\citeauthoryear{{Lai}, {Lee}, {Bolte}, {Lucatello}, {Beers},
  {Johnson}, {Sivarani}  \& {Rockosi}}{{Lai} et~al.}{2011}]{Lai2011}
{Lai} D.~K.,  {Lee} Y.~S.,  {Bolte} M.,  {Lucatello} S.,  {Beers} T.~C.,
  {Johnson} J.~A.,  {Sivarani} T.,   {Rockosi} C.~M.,  2011, \mn@doi [\apj]
  {10.1088/0004-637X/738/1/51}, \href
  {https://ui.adsabs.harvard.edu/abs/2011ApJ...738...51L} {738, 51}

\bibitem[\protect\citeauthoryear{{Larson}, {Tinsley}  \& {Caldwell}}{{Larson}
  et~al.}{1980}]{Larson1980}
{Larson} R.~B.,  {Tinsley} B.~M.,   {Caldwell} C.~N.,  1980, \mn@doi [\apj]
  {10.1086/157917}, \href
  {https://ui.adsabs.harvard.edu/abs/1980ApJ...237..692L} {237, 692}

\bibitem[\protect\citeauthoryear{{Lisker}}{{Lisker}}{2009}]{Lisker2009}
{Lisker} T.,  2009, \mn@doi [Astronomische Nachrichten]
  {10.1002/asna.200911291}, \href
  {https://ui.adsabs.harvard.edu/abs/2009AN....330.1043L} {330, 1043}

\bibitem[\protect\citeauthoryear{{Lisker}, {Grebel}  \& {Binggeli}}{{Lisker}
  et~al.}{2006}]{Lisker2006_disks}
{Lisker} T.,  {Grebel} E.~K.,   {Binggeli} B.,  2006, \mn@doi [\aj]
  {10.1086/505045}, \href
  {https://ui.adsabs.harvard.edu/abs/2006AJ....132..497L} {132, 497}

\bibitem[\protect\citeauthoryear{{Liu} et~al.,}{{Liu}
  et~al.}{2016a}]{Liu2016_distance}
{Liu} Y.,  et~al., 2016a, \mn@doi [\apj] {10.3847/0004-637X/818/2/179}, \href
  {https://ui.adsabs.harvard.edu/abs/2016ApJ...818..179L} {818, 179}

\bibitem[\protect\citeauthoryear{{Liu}, {Ho}  \& {Peng}}{{Liu}
  et~al.}{2016b}]{Liu2016_linear_sigma}
{Liu} Y.,  {Ho} L.~C.,   {Peng} E.,  2016b, \mn@doi [\apjl]
  {10.3847/2041-8205/829/2/L26}, \href
  {https://ui.adsabs.harvard.edu/abs/2016ApJ...829L..26L} {829, L26}

\bibitem[\protect\citeauthoryear{{Maddox}, {Serra}, {Venhola}, {Peletier},
  {Loubser}  \& {Iodice}}{{Maddox} et~al.}{2019}]{Maddox2019}
{Maddox} N.,  {Serra} P.,  {Venhola} A.,  {Peletier} R.,  {Loubser} I.,
  {Iodice} E.,  2019, \mn@doi [\mnras] {10.1093/mnras/stz2530}, \href
  {https://ui.adsabs.harvard.edu/abs/2019MNRAS.490.1666M} {490, 1666}

\bibitem[\protect\citeauthoryear{{Marshall} et~al.,}{{Marshall}
  et~al.}{2019}]{Marshall2019}
{Marshall} J.~L.,  et~al., 2019, \mn@doi [\apj] {10.3847/1538-4357/ab3653},
  \href {https://ui.adsabs.harvard.edu/abs/2019ApJ...882..177M} {882, 177}

\bibitem[\protect\citeauthoryear{{Mashonkina}, {Pakhomov}, {Sitnova},
  {Jablonka}, {Yakovleva}  \& {Belyaev}}{{Mashonkina}
  et~al.}{2022}]{Mashonkina2022}
{Mashonkina} L.,  {Pakhomov} Y.~V.,  {Sitnova} T.,  {Jablonka} P.,  {Yakovleva}
  S.~A.,   {Belyaev} A.~K.,  2022, \mn@doi [\mnras] {10.1093/mnras/stab3189},
  \href {https://ui.adsabs.harvard.edu/abs/2022MNRAS.509.3626M} {509, 3626}

\bibitem[\protect\citeauthoryear{{McConnachie}}{{McConnachie}}{2012}]{McConnachie2012}
{McConnachie} A.~W.,  2012, \mn@doi [\aj] {10.1088/0004-6256/144/1/4}, \href
  {https://ui.adsabs.harvard.edu/abs/2012AJ....144....4M} {144, 4}

\bibitem[\protect\citeauthoryear{{McConnachie} \& {Venn}}{{McConnachie} \&
  {Venn}}{2020}]{McConnachie2020}
{McConnachie} A.~W.,  {Venn} K.~A.,  2020, \mn@doi [\aj]
  {10.3847/1538-3881/aba4ab}, \href
  {https://ui.adsabs.harvard.edu/abs/2020AJ....160..124M} {160, 124}

\bibitem[\protect\citeauthoryear{{McDermid} et~al.,}{{McDermid}
  et~al.}{2015}]{McDermid2015}
{McDermid} R.~M.,  et~al., 2015, \mn@doi [\mnras] {10.1093/mnras/stv105}, \href
  {https://ui.adsabs.harvard.edu/abs/2015MNRAS.448.3484M} {448, 3484}

\bibitem[\protect\citeauthoryear{{Mentz} et~al.,}{{Mentz}
  et~al.}{2016}]{Mentz2016}
{Mentz} J.~J.,  et~al., 2016, \mn@doi [\mnras] {10.1093/mnras/stw2129}, \href
  {https://ui.adsabs.harvard.edu/abs/2016MNRAS.463.2819M} {463, 2819}

\bibitem[\protect\citeauthoryear{{Michea}, {Pasquali}, {Smith},
  {Calder{\'o}n-Castillo}, {Grebel}  \& {Peletier}}{{Michea}
  et~al.}{2022}]{Michea2022}
{Michea} J.,  {Pasquali} A.,  {Smith} R.,  {Calder{\'o}n-Castillo} P.,
  {Grebel} E.~K.,   {Peletier} R.~F.,  2022, arXiv e-prints, \href
  {https://ui.adsabs.harvard.edu/abs/2022arXiv220506281M} {p. arXiv:2205.06281}

\bibitem[\protect\citeauthoryear{{Michielsen} et~al.,}{{Michielsen}
  et~al.}{2008}]{Michielsen2008_allages}
{Michielsen} D.,  et~al., 2008, in {Knapen} J.~H.,  {Mahoney} T.~J.,
  {Vazdekis} A.,  eds,  Astronomical Society of the Pacific Conference Series
  Vol. 390, Pathways Through an Eclectic Universe. p.~308

\bibitem[\protect\citeauthoryear{{Mieske} \& {Hilker}}{{Mieske} \&
  {Hilker}}{2001}]{Mieske2001}
{Mieske} S.,  {Hilker} M.,  2001, in Astronomische Gesellschaft Meeting
  Abstracts. p.~P161

\bibitem[\protect\citeauthoryear{{Misgeld} \& {Hilker}}{{Misgeld} \&
  {Hilker}}{2011}]{Misgeld2011}
{Misgeld} I.,  {Hilker} M.,  2011, \mn@doi [\mnras]
  {10.1111/j.1365-2966.2011.18669.x}, \href
  {https://ui.adsabs.harvard.edu/abs/2011MNRAS.414.3699M} {414, 3699}

\bibitem[\protect\citeauthoryear{{Moore}, {Lake}  \& {Katz}}{{Moore}
  et~al.}{1998}]{Moore1998harasment}
{Moore} B.,  {Lake} G.,   {Katz} N.,  1998, \mn@doi [\apj] {10.1086/305264},
  \href {https://ui.adsabs.harvard.edu/abs/1998ApJ...495..139M} {495, 139}

\bibitem[\protect\citeauthoryear{{Moore}, {Lake}, {Quinn}  \& {Stadel}}{{Moore}
  et~al.}{1999}]{Moore1999}
{Moore} B.,  {Lake} G.,  {Quinn} T.,   {Stadel} J.,  1999, \mn@doi [\mnras]
  {10.1046/j.1365-8711.1999.02345.x}, \href
  {https://ui.adsabs.harvard.edu/abs/1999MNRAS.304..465M} {304, 465}

\bibitem[\protect\citeauthoryear{{Nagasawa} et~al.,}{{Nagasawa}
  et~al.}{2018}]{Nagasawa2018}
{Nagasawa} D.~Q.,  et~al., 2018, \mn@doi [\apj] {10.3847/1538-4357/aaa01d},
  \href {https://ui.adsabs.harvard.edu/abs/2018ApJ...852...99N} {852, 99}

\bibitem[\protect\citeauthoryear{{Naidu} et~al.,}{{Naidu}
  et~al.}{2022}]{Naidu2022}
{Naidu} R.~P.,  et~al., 2022, arXiv e-prints, \href
  {https://ui.adsabs.harvard.edu/abs/2022arXiv220409057N} {p. arXiv:2204.09057}

\bibitem[\protect\citeauthoryear{{Nasonova}, {de Freitas Pacheco}  \&
  {Karachentsev}}{{Nasonova} et~al.}{2011}]{Nasonova2011}
{Nasonova} O.~G.,  {de Freitas Pacheco} J.~A.,   {Karachentsev} I.~D.,  2011,
  \mn@doi [\aap] {10.1051/0004-6361/201016004}, \href
  {https://ui.adsabs.harvard.edu/abs/2011A&A...532A.104N} {532, A104}

\bibitem[\protect\citeauthoryear{{Norris}, {Yong}, {Gilmore}  \&
  {Wyse}}{{Norris} et~al.}{2010a}]{Norris2010a}
{Norris} J.~E.,  {Yong} D.,  {Gilmore} G.,   {Wyse} R. F.~G.,  2010a, \mn@doi
  [\apj] {10.1088/0004-637X/711/1/350}, \href
  {https://ui.adsabs.harvard.edu/abs/2010ApJ...711..350N} {711, 350}

\bibitem[\protect\citeauthoryear{{Norris}, {Gilmore}, {Wyse}, {Yong}  \&
  {Frebel}}{{Norris} et~al.}{2010b}]{Norris2010b}
{Norris} J.~E.,  {Gilmore} G.,  {Wyse} R. F.~G.,  {Yong} D.,   {Frebel} A.,
  2010b, \mn@doi [\apjl] {10.1088/2041-8205/722/1/L104}, \href
  {https://ui.adsabs.harvard.edu/abs/2010ApJ...722L.104N} {722, L104}

\bibitem[\protect\citeauthoryear{{Norris}, {Yong}, {Venn}, {Gilmore},
  {Casagrande}  \& {Dotter}}{{Norris} et~al.}{2017}]{Norris2017}
{Norris} J.~E.,  {Yong} D.,  {Venn} K.~A.,  {Gilmore} G.,  {Casagrande} L.,
  {Dotter} A.,  2017, \mn@doi [\apjs] {10.3847/1538-4365/aa755e}, \href
  {https://ui.adsabs.harvard.edu/abs/2017ApJS..230...28N} {230, 28}

\bibitem[\protect\citeauthoryear{{Ocvirk}, {Pichon}, {Lan{\c{c}}on}  \&
  {Thi{\'e}baut}}{{Ocvirk} et~al.}{2006a}]{Ocvirk2006a}
{Ocvirk} P.,  {Pichon} C.,  {Lan{\c{c}}on} A.,   {Thi{\'e}baut} E.,  2006a,
  \mn@doi [\mnras] {10.1111/j.1365-2966.2005.09182.x}, \href
  {https://ui.adsabs.harvard.edu/abs/2006MNRAS.365...46O} {365, 46}

\bibitem[\protect\citeauthoryear{{Ocvirk}, {Pichon}, {Lan{\c{c}}on}  \&
  {Thi{\'e}baut}}{{Ocvirk} et~al.}{2006b}]{Ocvirk2006b}
{Ocvirk} P.,  {Pichon} C.,  {Lan{\c{c}}on} A.,   {Thi{\'e}baut} E.,  2006b,
  \mn@doi [\mnras] {10.1111/j.1365-2966.2005.09323.x}, \href
  {https://ui.adsabs.harvard.edu/abs/2006MNRAS.365...74O} {365, 74}

\bibitem[\protect\citeauthoryear{{Paudel}, {Lisker}, {Kuntschner}, {Grebel}  \&
  {Glatt}}{{Paudel} et~al.}{2010}]{Paudel2010}
{Paudel} S.,  {Lisker} T.,  {Kuntschner} H.,  {Grebel} E.~K.,   {Glatt} K.,
  2010, \mn@doi [\mnras] {10.1111/j.1365-2966.2010.16507.x}, \href
  {https://ui.adsabs.harvard.edu/abs/2010MNRAS.405..800P} {405, 800}

\bibitem[\protect\citeauthoryear{{Peletier}}{{Peletier}}{1989}]{Reynier1989PhD}
{Peletier} R.~F.,  1989, PhD thesis, -

\bibitem[\protect\citeauthoryear{{Penny} et~al.,}{{Penny}
  et~al.}{2016}]{Penny2016}
{Penny} S.~J.,  et~al., 2016, \mn@doi [\mnras] {10.1093/mnras/stw1913}, \href
  {https://ui.adsabs.harvard.edu/abs/2016MNRAS.462.3955P} {462, 3955}

\bibitem[\protect\citeauthoryear{{Pereira-Wilson}, {Navarro},
  {Ben{\'\i}tez-Llambay}  \& {Santos-Santos}}{{Pereira-Wilson}
  et~al.}{2023}]{Pereira_2023}
{Pereira-Wilson} M.,  {Navarro} J.~F.,  {Ben{\'\i}tez-Llambay} A.,
  {Santos-Santos} I.,  2023, \mn@doi [\mnras] {10.1093/mnras/stac3633}, \href
  {https://ui.adsabs.harvard.edu/abs/2023MNRAS.519.1425P} {519, 1425}

\bibitem[\protect\citeauthoryear{{Pimbblet}}{{Pimbblet}}{2011}]{Pimbblet2011_blacksplash}
{Pimbblet} K.~A.,  2011, \mn@doi [\mnras] {10.1111/j.1365-2966.2010.17869.x},
  \href {https://ui.adsabs.harvard.edu/abs/2011MNRAS.411.2637P} {411, 2637}

\bibitem[\protect\citeauthoryear{{Planck Collaboration} et~al.,}{{Planck
  Collaboration} et~al.}{2014}]{Planck2014}
{Planck Collaboration} et~al., 2014, \mn@doi [\aap]
  {10.1051/0004-6361/201321529}, \href
  {https://ui.adsabs.harvard.edu/abs/2014A&A...571A...1P} {571, A1}

\bibitem[\protect\citeauthoryear{{Ponomareva}, {Verheijen}, {Papastergis},
  {Bosma}  \& {Peletier}}{{Ponomareva} et~al.}{2018}]{Ponomareva2018}
{Ponomareva} A.~A.,  {Verheijen} M. A.~W.,  {Papastergis} E.,  {Bosma} A.,
  {Peletier} R.~F.,  2018, \mn@doi [\mnras] {10.1093/mnras/stx3066}, \href
  {https://ui.adsabs.harvard.edu/abs/2018MNRAS.474.4366P} {474, 4366}

\bibitem[\protect\citeauthoryear{{Prugniel}, {Soubiran}, {Koleva}  \& {Le
  Borgne}}{{Prugniel} et~al.}{2007}]{Elodie-prugniel2007}
{Prugniel} P.,  {Soubiran} C.,  {Koleva} M.,   {Le Borgne} D.,  2007, arXiv
  e-prints, \href {https://ui.adsabs.harvard.edu/abs/2007astro.ph..3658P} {pp
  astro--ph/0703658}

\bibitem[\protect\citeauthoryear{{Putman}, {Zheng}, {Price-Whelan}, {Grcevich},
  {Johnson}, {Tollerud}  \& {Peek}}{{Putman} et~al.}{2021}]{Putman2021}
{Putman} M.~E.,  {Zheng} Y.,  {Price-Whelan} A.~M.,  {Grcevich} J.,  {Johnson}
  A.~C.,  {Tollerud} E.,   {Peek} J. E.~G.,  2021, \mn@doi [\apj]
  {10.3847/1538-4357/abe391}, \href
  {https://ui.adsabs.harvard.edu/abs/2021ApJ...913...53P} {913, 53}

\bibitem[\protect\citeauthoryear{{Rey}, {Pontzen}, {Agertz}, {Orkney}, {Read},
  {Saintonge}, {Kim}  \& {Das}}{{Rey} et~al.}{2022}]{Rey_2022}
{Rey} M.~P.,  {Pontzen} A.,  {Agertz} O.,  {Orkney} M. D.~A.,  {Read} J.~I.,
  {Saintonge} A.,  {Kim} S.~Y.,   {Das} P.,  2022, \mn@doi [\mnras]
  {10.1093/mnras/stac502}, \href
  {https://ui.adsabs.harvard.edu/abs/2022MNRAS.511.5672R} {511, 5672}

\bibitem[\protect\citeauthoryear{{Roederer} \& {Kirby}}{{Roederer} \&
  {Kirby}}{2014}]{Roederer-Kirby2014}
{Roederer} I.~U.,  {Kirby} E.~N.,  2014, \mn@doi [\mnras]
  {10.1093/mnras/stu491}, \href
  {https://ui.adsabs.harvard.edu/abs/2014MNRAS.440.2665R} {440, 2665}

\bibitem[\protect\citeauthoryear{{Ry{\'s}}, {Koleva}, {Falc{\'o}n-Barroso},
  {Vazdekis}, {Lisker}, {Peletier}  \& {van de Ven}}{{Ry{\'s}}
  et~al.}{2015a}]{Rys2015_allages}
{Ry{\'s}} A.,  {Koleva} M.,  {Falc{\'o}n-Barroso} J.,  {Vazdekis} A.,  {Lisker}
  T.,  {Peletier} R.,   {van de Ven} G.,  2015a, \mn@doi [\mnras]
  {10.1093/mnras/stv1364}, \href
  {https://ui.adsabs.harvard.edu/abs/2015MNRAS.452.1888R} {452, 1888}

\bibitem[\protect\citeauthoryear{{Ry{\'s}}, {Koleva}, {Falc{\'o}n-Barroso},
  {Vazdekis}, {Lisker}, {Peletier}  \& {van de Ven}}{{Ry{\'s}}
  et~al.}{2015b}]{Rys2015}
{Ry{\'s}} A.,  {Koleva} M.,  {Falc{\'o}n-Barroso} J.,  {Vazdekis} A.,  {Lisker}
  T.,  {Peletier} R.,   {van de Ven} G.,  2015b, \mn@doi [\mnras]
  {10.1093/mnras/stv1364}, \href
  {https://ui.adsabs.harvard.edu/abs/2015MNRAS.452.1888R} {452, 1888}

\bibitem[\protect\citeauthoryear{{Saifollahi} et~al.,}{{Saifollahi}
  et~al.}{2021}]{Saifollahi2021}
{Saifollahi} T.,  et~al., 2021, \mn@doi [\mnras] {10.1093/mnras/stab1118},
  \href {https://ui.adsabs.harvard.edu/abs/2021MNRAS.504.3580S} {504, 3580}

\bibitem[\protect\citeauthoryear{{Sales}, {Navarro}, {Abadi}  \&
  {Steinmetz}}{{Sales} et~al.}{2007}]{Sales2007}
{Sales} L.~V.,  {Navarro} J.~F.,  {Abadi} M.~G.,   {Steinmetz} M.,  2007,
  \mn@doi [\mnras] {10.1111/j.1365-2966.2007.12026.x}, \href
  {https://ui.adsabs.harvard.edu/abs/2007MNRAS.379.1475S} {379, 1475}

\bibitem[\protect\citeauthoryear{{Samuel} et~al.,}{{Samuel}
  et~al.}{2020}]{Samuel_2020}
{Samuel} J.,  et~al., 2020, \mn@doi [\mnras] {10.1093/mnras/stz3054}, \href
  {https://ui.adsabs.harvard.edu/abs/2020MNRAS.491.1471S} {491, 1471}

\bibitem[\protect\citeauthoryear{{Samuel}, {Wetzel}, {Santistevan}, {Tollerud},
  {Moreno}, {Boylan-Kolchin}, {Bailin}  \& {Pardasani}}{{Samuel}
  et~al.}{2022}]{Samuel_2022}
{Samuel} J.,  {Wetzel} A.,  {Santistevan} I.,  {Tollerud} E.,  {Moreno} J.,
  {Boylan-Kolchin} M.,  {Bailin} J.,   {Pardasani} B.,  2022, \mn@doi [\mnras]
  {10.1093/mnras/stac1706}, \href
  {https://ui.adsabs.harvard.edu/abs/2022MNRAS.514.5276S} {514, 5276}

\bibitem[\protect\citeauthoryear{{S{\'a}nchez-Bl{\'a}zquez}
  et~al.,}{{S{\'a}nchez-Bl{\'a}zquez} et~al.}{2006}]{Sanchez-Blazquez2006}
{S{\'a}nchez-Bl{\'a}zquez} P.,  et~al., 2006, \mn@doi [\mnras]
  {10.1111/j.1365-2966.2006.10699.x}, \href
  {https://ui.adsabs.harvard.edu/abs/2006MNRAS.371..703S} {371, 703}

\bibitem[\protect\citeauthoryear{{Sandage} \& {Binggeli}}{{Sandage} \&
  {Binggeli}}{1984}]{Sandage1984}
{Sandage} A.,  {Binggeli} B.,  1984, \mn@doi [\aj] {10.1086/113588}, \href
  {https://ui.adsabs.harvard.edu/abs/1984AJ.....89..919S} {89, 919}

\bibitem[\protect\citeauthoryear{{Schiavon}}{{Schiavon}}{2007}]{Schiavon2007}
{Schiavon} R.~P.,  2007, \mn@doi [\apjs] {10.1086/511753}, \href
  {https://ui.adsabs.harvard.edu/abs/2007ApJS..171..146S} {171, 146}

\bibitem[\protect\citeauthoryear{{Scott} et~al.,}{{Scott}
  et~al.}{2020}]{scott_2020_fornaxI}
{Scott} N.,  et~al., 2020, \mn@doi [\mnras] {10.1093/mnras/staa2042}, \href
  {https://ui.adsabs.harvard.edu/abs/2020MNRAS.497.1571S} {497, 1571}

\bibitem[\protect\citeauthoryear{{Sharp} et~al.,}{{Sharp}
  et~al.}{2006}]{Sharp2006}
{Sharp} R.,  et~al., 2006, in {McLean} I.~S.,  {Iye} M.,  eds,  Society of
  Photo-Optical Instrumentation Engineers (SPIE) Conference Series Vol. 6269,
  Society of Photo-Optical Instrumentation Engineers (SPIE) Conference Series.
  p. 62690G (\mn@eprint {arXiv} {astro-ph/0606137}), \mn@doi{10.1117/12.671022}

\bibitem[\protect\citeauthoryear{{Shetrone}, {C{\^o}t{\'e}}  \&
  {Sargent}}{{Shetrone} et~al.}{2001}]{Shetrone2001}
{Shetrone} M.~D.,  {C{\^o}t{\'e}} P.,   {Sargent} W.~L.~W.,  2001, \mn@doi
  [\apj] {10.1086/319022}, \href
  {https://ui.adsabs.harvard.edu/abs/2001ApJ...548..592S} {548, 592}

\bibitem[\protect\citeauthoryear{{Shetrone}, {Venn}, {Tolstoy}, {Primas},
  {Hill}  \& {Kaufer}}{{Shetrone} et~al.}{2003}]{Shetrone2003}
{Shetrone} M.,  {Venn} K.~A.,  {Tolstoy} E.,  {Primas} F.,  {Hill} V.,
  {Kaufer} A.,  2003, \mn@doi [\aj] {10.1086/345966}, \href
  {https://ui.adsabs.harvard.edu/abs/2003AJ....125..684S} {125, 684}

\bibitem[\protect\citeauthoryear{{Shetrone}, {Siegel}, {Cook}  \&
  {Bosler}}{{Shetrone} et~al.}{2009}]{Shetrone2009}
{Shetrone} M.~D.,  {Siegel} M.~H.,  {Cook} D.~O.,   {Bosler} T.,  2009, \mn@doi
  [\aj] {10.1088/0004-6256/137/1/62}, \href
  {https://ui.adsabs.harvard.edu/abs/2009AJ....137...62S} {137, 62}

\bibitem[\protect\citeauthoryear{{Simon}}{{Simon}}{2019}]{Simon2019}
{Simon} J.~D.,  2019, \mn@doi [\araa] {10.1146/annurev-astro-091918-104453},
  \href {https://ui.adsabs.harvard.edu/abs/2019ARA&A..57..375S} {57, 375}

\bibitem[\protect\citeauthoryear{{Simon}, {Frebel}, {McWilliam}, {Kirby}  \&
  {Thompson}}{{Simon} et~al.}{2010}]{Simon2010}
{Simon} J.~D.,  {Frebel} A.,  {McWilliam} A.,  {Kirby} E.~N.,   {Thompson}
  I.~B.,  2010, \mn@doi [\apj] {10.1088/0004-637X/716/1/446}, \href
  {https://ui.adsabs.harvard.edu/abs/2010ApJ...716..446S} {716, 446}

\bibitem[\protect\citeauthoryear{{Simon}, {Jacobson}, {Frebel}, {Thompson},
  {Adams}  \& {Shectman}}{{Simon} et~al.}{2015}]{Simon2015}
{Simon} J.~D.,  {Jacobson} H.~R.,  {Frebel} A.,  {Thompson} I.~B.,  {Adams}
  J.~J.,   {Shectman} S.~A.,  2015, \mn@doi [\apj]
  {10.1088/0004-637X/802/2/93}, \href
  {https://ui.adsabs.harvard.edu/abs/2015ApJ...802...93S} {802, 93}

\bibitem[\protect\citeauthoryear{{Sitnova} et~al.,}{{Sitnova}
  et~al.}{2021}]{Sitnova2021}
{Sitnova} T.~M.,  et~al., 2021, \mn@doi [\mnras] {10.1093/mnras/stab786}, \href
  {https://ui.adsabs.harvard.edu/abs/2021MNRAS.504.1183S} {504, 1183}

\bibitem[\protect\citeauthoryear{{Smith} et~al.,}{{Smith}
  et~al.}{2008}]{Smith2008}
{Smith} R.~J.,  et~al., 2008, \mn@doi [\mnras]
  {10.1111/j.1745-3933.2008.00469.x}, \href
  {https://ui.adsabs.harvard.edu/abs/2008MNRAS.386L..96S} {386, L96}

\bibitem[\protect\citeauthoryear{{Smith}, {Lucey}, {Hudson}, {Allanson},
  {Bridges}, {Hornschemeier}, {Marzke}  \& {Miller}}{{Smith}
  et~al.}{2009}]{Smith2009_highalpha}
{Smith} R.~J.,  {Lucey} J.~R.,  {Hudson} M.~J.,  {Allanson} S.~P.,  {Bridges}
  T.~J.,  {Hornschemeier} A.~E.,  {Marzke} R.~O.,   {Miller} N.~A.,  2009,
  \mn@doi [\mnras] {10.1111/j.1365-2966.2008.14180.x}, \href
  {https://ui.adsabs.harvard.edu/abs/2009MNRAS.392.1265S} {392, 1265}

\bibitem[\protect\citeauthoryear{{Su} et~al.,}{{Su} et~al.}{2021}]{Su2021}
{Su} A.~H.,  et~al., 2021, \mn@doi [\aap] {10.1051/0004-6361/202039633}, \href
  {https://ui.adsabs.harvard.edu/abs/2021A&A...647A.100S} {647, A100}

\bibitem[\protect\citeauthoryear{{Sybilska} et~al.,}{{Sybilska}
  et~al.}{2017}]{Sybilska2017}
{Sybilska} A.,  et~al., 2017, \mn@doi [\mnras] {10.1093/mnras/stx1138}, \href
  {https://ui.adsabs.harvard.edu/abs/2017MNRAS.470..815S} {470, 815}

\bibitem[\protect\citeauthoryear{{Tafelmeyer} et~al.,}{{Tafelmeyer}
  et~al.}{2010}]{Tafelmeyer2010}
{Tafelmeyer} M.,  et~al., 2010, \mn@doi [\aap] {10.1051/0004-6361/201014733},
  \href {https://ui.adsabs.harvard.edu/abs/2010A&A...524A..58T} {524, A58}

\bibitem[\protect\citeauthoryear{{Thomas}, {Maraston}  \& {Bender}}{{Thomas}
  et~al.}{2003}]{Thomas2003}
{Thomas} D.,  {Maraston} C.,   {Bender} R.,  2003, \mn@doi [\mnras]
  {10.1046/j.1365-8711.2003.06248.x}, \href
  {https://ui.adsabs.harvard.edu/abs/2003MNRAS.339..897T} {339, 897}

\bibitem[\protect\citeauthoryear{{Toloba}, {Boselli}, {Peletier},
  {Falc{\'o}n-Barroso}, {van de Ven}  \& {Gorgas}}{{Toloba}
  et~al.}{2012}]{Toloba2012}
{Toloba} E.,  {Boselli} A.,  {Peletier} R.~F.,  {Falc{\'o}n-Barroso} J.,  {van
  de Ven} G.,   {Gorgas} J.,  2012, \mn@doi [\aap]
  {10.1051/0004-6361/201218944}, \href
  {https://ui.adsabs.harvard.edu/abs/2012A&A...548A..78T} {548, A78}

\bibitem[\protect\citeauthoryear{{Tolstoy}, {Battaglia}  \& {Cole}}{{Tolstoy}
  et~al.}{2008}]{Tolstoy2008}
{Tolstoy} E.,  {Battaglia} G.,   {Cole} A.,  2008, in Low-Metallicity Star
  Formation: From the First Stars to Dwarf Galaxies. pp 310--317,
  \mn@doi{10.1017/S174392130802499X}

\bibitem[\protect\citeauthoryear{{Tolstoy}, {Hill}  \& {Tosi}}{{Tolstoy}
  et~al.}{2009}]{Tolstoy2009}
{Tolstoy} E.,  {Hill} V.,   {Tosi} M.,  2009, \mn@doi [\araa]
  {10.1146/annurev-astro-082708-101650}, \href
  {https://ui.adsabs.harvard.edu/abs/2009ARA&A..47..371T} {47, 371}

\bibitem[\protect\citeauthoryear{{Trager}, {Worthey}, {Faber}, {Burstein}  \&
  {Gonz{\'a}lez}}{{Trager} et~al.}{1998}]{Trager1998}
{Trager} S.~C.,  {Worthey} G.,  {Faber} S.~M.,  {Burstein} D.,   {Gonz{\'a}lez}
  J.~J.,  1998, \mn@doi [\apjs] {10.1086/313099}, \href
  {https://ui.adsabs.harvard.edu/abs/1998ApJS..116....1T} {116, 1}

\bibitem[\protect\citeauthoryear{{Trager}, {Faber}, {Worthey}  \&
  {Gonz{\'a}lez}}{{Trager} et~al.}{2000}]{Traeger2000}
{Trager} S.~C.,  {Faber} S.~M.,  {Worthey} G.,   {Gonz{\'a}lez} J.~J.,  2000,
  \mn@doi [\aj] {10.1086/301442}, \href
  {https://ui.adsabs.harvard.edu/abs/2000AJ....120..165T} {120, 165}

\bibitem[\protect\citeauthoryear{{Ural}, {Cescutti}, {Koch}, {Kleyna},
  {Feltzing}  \& {Wilkinson}}{{Ural} et~al.}{2015}]{Ural2015}
{Ural} U.,  {Cescutti} G.,  {Koch} A.,  {Kleyna} J.,  {Feltzing} S.,
  {Wilkinson} M.~I.,  2015, \mn@doi [\mnras] {10.1093/mnras/stv294}, \href
  {https://ui.adsabs.harvard.edu/abs/2015MNRAS.449..761U} {449, 761}

\bibitem[\protect\citeauthoryear{\VAN{Dokkum}{Van}{van}~Dokkum, {Abraham},
  {Merritt}, {Zhang}, {Geha}  \& {Conroy}}{\VAN{Dokkum}{Van}{van}~Dokkum
  et~al.}{2015}]{Dokkum2015}
\VAN{Dokkum}{Van}{van}~Dokkum P.~G.,  {Abraham} R.,  {Merritt} A.,  {Zhang} J.,
   {Geha} M.,   {Conroy} C.,  2015, \mn@doi [\apjl]
  {10.1088/2041-8205/798/2/L45}, \href
  {https://ui.adsabs.harvard.edu/abs/2015ApJ...798L..45V} {798, L45}

\bibitem[\protect\citeauthoryear{{Vazdekis}}{{Vazdekis}}{1999}]{Vazdekis1999}
{Vazdekis} A.,  1999, \mn@doi [\apj] {10.1086/306843}, \href
  {https://ui.adsabs.harvard.edu/abs/1999ApJ...513..224V} {513, 224}

\bibitem[\protect\citeauthoryear{{Vazdekis}, {S{\'a}nchez-Bl{\'a}zquez},
  {Falc{\'o}n-Barroso}, {Cenarro}, {Beasley}, {Cardiel}, {Gorgas}  \&
  {Peletier}}{{Vazdekis} et~al.}{2010}]{Vazdekis2010}
{Vazdekis} A.,  {S{\'a}nchez-Bl{\'a}zquez} P.,  {Falc{\'o}n-Barroso} J.,
  {Cenarro} A.~J.,  {Beasley} M.~A.,  {Cardiel} N.,  {Gorgas} J.,   {Peletier}
  R.~F.,  2010, \mn@doi [\mnras] {10.1111/j.1365-2966.2010.16407.x}, \href
  {https://ui.adsabs.harvard.edu/abs/2010MNRAS.404.1639V} {404, 1639}

\bibitem[\protect\citeauthoryear{{Vazdekis} et~al.,}{{Vazdekis}
  et~al.}{2015}]{Vazdekis2015}
{Vazdekis} A.,  et~al., 2015, \mn@doi [\mnras] {10.1093/mnras/stv151}, \href
  {https://ui.adsabs.harvard.edu/abs/2015MNRAS.449.1177V} {449, 1177}

\bibitem[\protect\citeauthoryear{{Venhola} et~al.,}{{Venhola}
  et~al.}{2018}]{Venhola_2018_sample_selection}
{Venhola} A.,  et~al., 2018, \mn@doi [\aap] {10.1051/0004-6361/201833933},
  \href {https://ui.adsabs.harvard.edu/abs/2018A&A...620A.165V} {620, A165}

\bibitem[\protect\citeauthoryear{{Venhola} et~al.,}{{Venhola}
  et~al.}{2019}]{Venhola2019}
{Venhola} A.,  et~al., 2019, \mn@doi [\aap] {10.1051/0004-6361/201935231},
  \href {https://ui.adsabs.harvard.edu/abs/2019A&A...625A.143V} {625, A143}

\bibitem[\protect\citeauthoryear{{Venhola} et~al.,}{{Venhola}
  et~al.}{2021}]{Venhola2021}
{Venhola} A.,  et~al., 2021, arXiv e-prints, \href
  {https://ui.adsabs.harvard.edu/abs/2021arXiv211101855V} {p. arXiv:2111.01855}

\bibitem[\protect\citeauthoryear{{Venhola} et~al.,}{{Venhola}
  et~al.}{2022}]{Venhola2022}
{Venhola} A.,  et~al., 2022, \mn@doi [\aap] {10.1051/0004-6361/202141756},
  \href {https://ui.adsabs.harvard.edu/abs/2022A&A...662A..43V} {662, A43}

\bibitem[\protect\citeauthoryear{{Watson} et~al.,}{{Watson}
  et~al.}{2022}]{Watson2022}
{Watson} P.~J.,  et~al., 2022, \mn@doi [\mnras] {10.1093/mnras/stab3477}, \href
  {https://ui.adsabs.harvard.edu/abs/2022MNRAS.510.1541W} {510, 1541}

\bibitem[\protect\citeauthoryear{{Weisz}, {Dolphin}, {Skillman}, {Holtzman},
  {Gilbert}, {Dalcanton}  \& {Williams}}{{Weisz} et~al.}{2014}]{Weisz2014}
{Weisz} D.~R.,  {Dolphin} A.~E.,  {Skillman} E.~D.,  {Holtzman} J.,  {Gilbert}
  K.~M.,  {Dalcanton} J.~J.,   {Williams} B.~F.,  2014, \mn@doi [\apj]
  {10.1088/0004-637X/789/2/147}, \href
  {https://ui.adsabs.harvard.edu/abs/2014ApJ...789..147W} {789, 147}

\bibitem[\protect\citeauthoryear{{Wittmann}, {Lisker}, {Pasquali}, {Hilker}  \&
  {Grebel}}{{Wittmann} et~al.}{2016}]{Wittmann2016}
{Wittmann} C.,  {Lisker} T.,  {Pasquali} A.,  {Hilker} M.,   {Grebel} E.~K.,
  2016, \mn@doi [\mnras] {10.1093/mnras/stw827}, \href
  {https://ui.adsabs.harvard.edu/abs/2016MNRAS.459.4450W} {459, 4450}

\bibitem[\protect\citeauthoryear{{Worthey}, {Faber}  \& {Gonzalez}}{{Worthey}
  et~al.}{1992}]{Worthey1992}
{Worthey} G.,  {Faber} S.~M.,   {Gonzalez} J.~J.,  1992, \mn@doi [\apj]
  {10.1086/171836}, \href
  {https://ui.adsabs.harvard.edu/abs/1992ApJ...398...69W} {398, 69}

\bibitem[\protect\citeauthoryear{{Worthey}, {Faber}, {Gonzalez}  \&
  {Burstein}}{{Worthey} et~al.}{1994}]{Worthey1994}
{Worthey} G.,  {Faber} S.~M.,  {Gonzalez} J.~J.,   {Burstein} D.,  1994,
  \mn@doi [\apjs] {10.1086/192087}, \href
  {https://ui.adsabs.harvard.edu/abs/1994ApJS...94..687W} {94, 687}

\bibitem[\protect\citeauthoryear{{Zheng} et~al.,}{{Zheng}
  et~al.}{2019}]{Zheng2019}
{Zheng} Z.,  et~al., 2019, \mn@doi [\apj] {10.3847/1538-4357/ab03d2}, \href
  {https://ui.adsabs.harvard.edu/abs/2019ApJ...873...63Z} {873, 63}

\bibitem[\protect\citeauthoryear{{{\c{S}}en} et~al.,}{{{\c{S}}en}
  et~al.}{2018}]{Seyda2018solaralpha}
{{\c{S}}en} {\c{S}}.,  et~al., 2018, \mn@doi [\mnras] {10.1093/mnras/stx3254},
  \href {https://ui.adsabs.harvard.edu/abs/2018MNRAS.475.3453S} {475, 3453}

\bibitem[\protect\citeauthoryear{{den Brok} et~al.,}{{den Brok}
  et~al.}{2011}]{denBrok2011}
{den Brok} M.,  et~al., 2011, \mn@doi [\mnras]
  {10.1111/j.1365-2966.2011.18606.x}, \href
  {https://ui.adsabs.harvard.edu/abs/2011MNRAS.414.3052D} {414, 3052}

\makeatother
\end{thebibliography}




\appendix

\section{ATLAS$^{3D}$}\label{appendix_atlas}
For the completeness of this work, we combined our dwarf sample with the whole sample of ATLAS$^{3D}$ galaxies \citep[][]{Cappellari2011-atlas3d}. We downloaded the spectral cubes from the \href{https://www-astro.physics.ox.ac.uk/atlas3d/}{ATLAS$^{3D}$ project website}, and even though these are spatially resolved spectra, in order to apply the same methodology as we did for our dwarfs, we collapsed all data into one single spectrum per galaxy. We used \href{https://www-astro.physics.ox.ac.uk/~mxc/software/}{pPXF} for full spectral fitting, deriving first radial velocity and velocity dispersion applying only additive Legendre polynomials, and then fixed the kinematics to obtain the stellar population properties by fitting the spectra using only multiplicative polynomials.

\begin{figure*}
\centering
\includegraphics[scale=0.8]{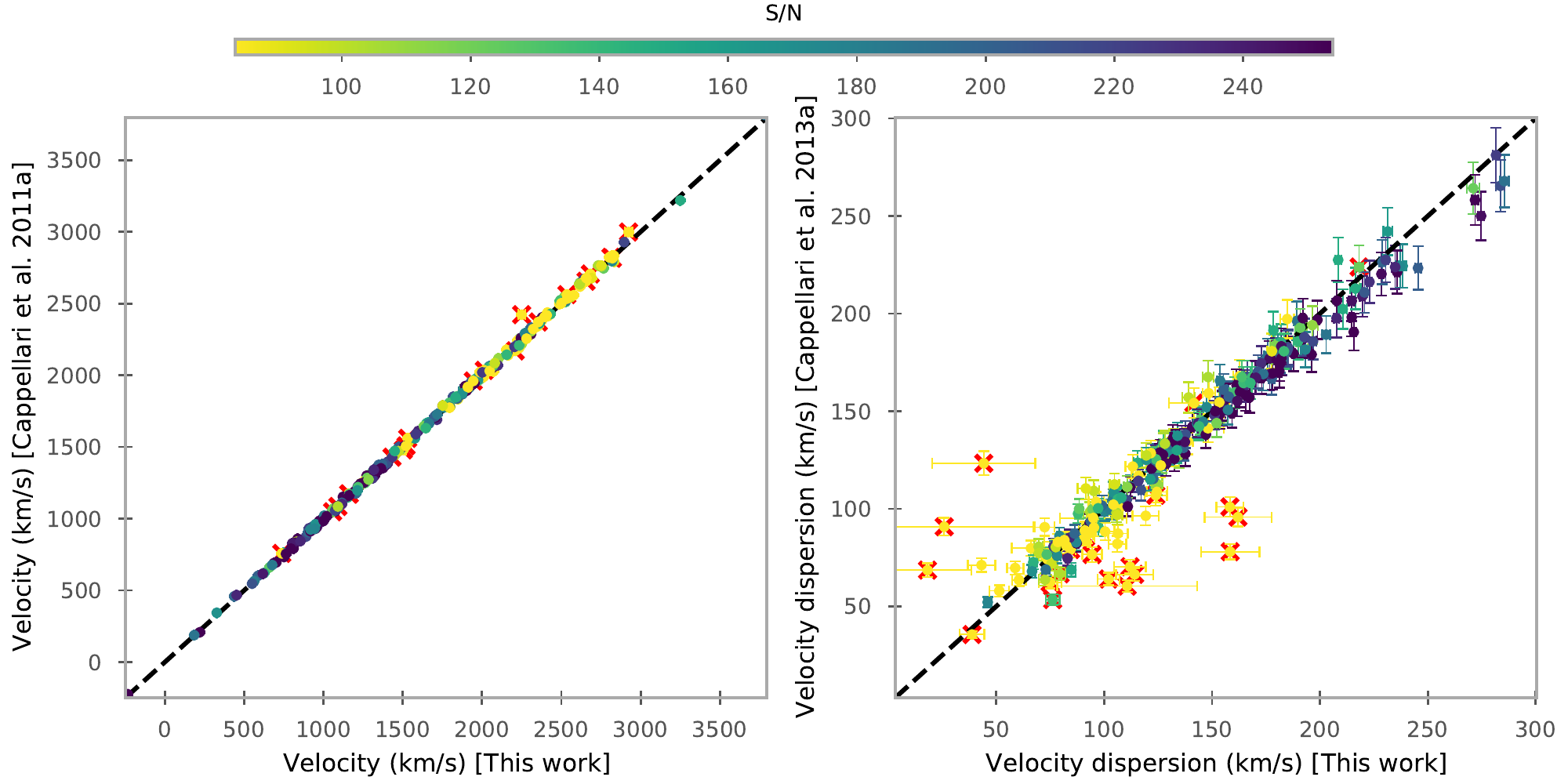}
\caption{Comparison of our kinematics results with those published in the ATLAS$^{3D}$ papers for the ETGs. On the left panel, we show the radial velocity and on the right the velocity dispersion. In both panels, the points are colour coded with the S/N, computed as described in section \ref{kinematics_subsection}, and the red X are those points of bad quality according to \citetalias{McDermid2015}. The black dashed line represents the 1:1 relation.}
\label{fig_atlas_kinematics}
\end{figure*}

In Fig. \ref{fig_atlas_kinematics} we show a comparison between our results for the velocity of each galaxy in the ATLAS$^{3D}$ and the results published in \citet{Cappellari2011-atlas3d}, and also compare velocity dispersion with those in \citet{Cappellari2013a}. We find that our kinematics results are in agreement with the ones published, they fit almost perfectly to the 1:1 relation although there is some scatter for low S/N galaxies in terms of velocity dispersion.

\begin{figure*}
\centering
\includegraphics[scale=0.77]{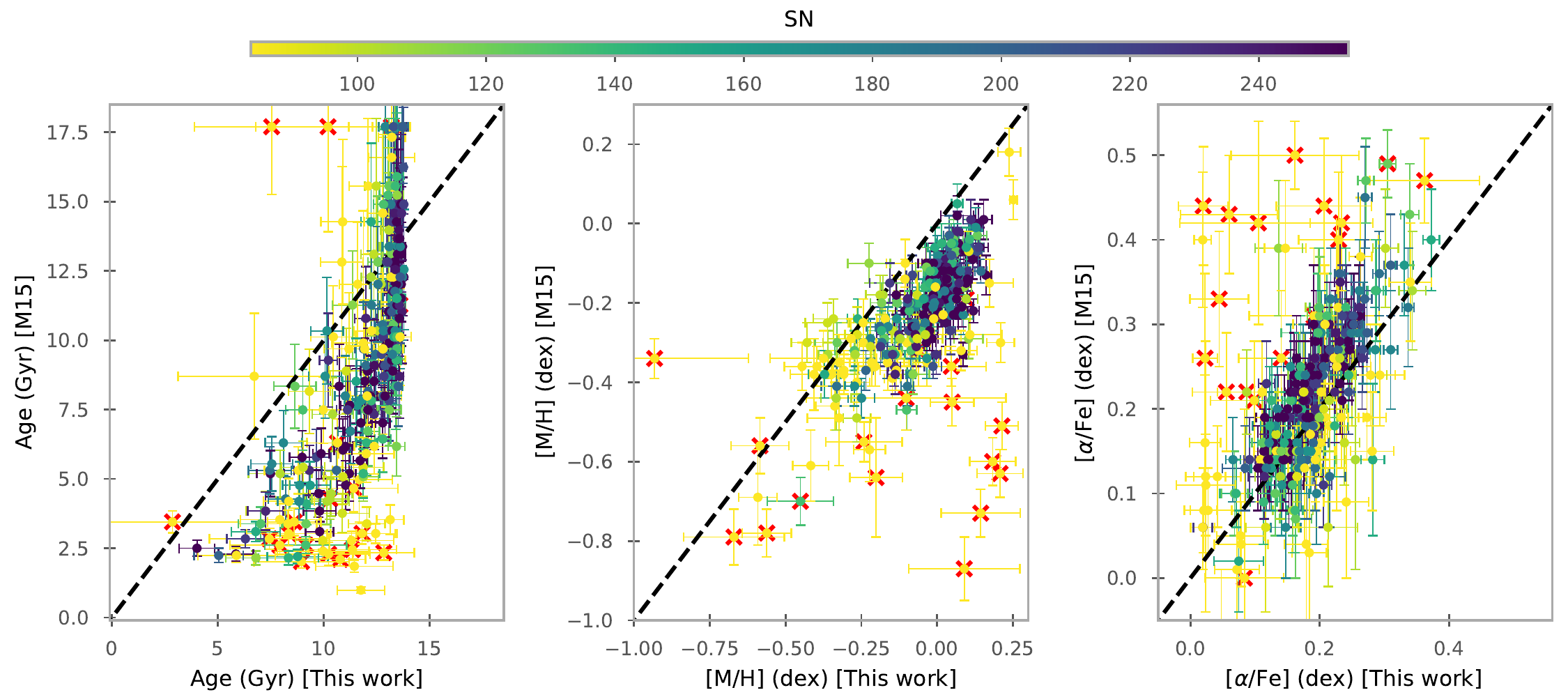}
\caption{Comparison of our stellar population properties integrating all available spectra with those published in \citetalias{McDermid2015} integrating only up to 1.0 R$_e$. On the left panel, the age of each galaxy is shown, on the middle panel the metallicity and on right [$\alpha$/Fe]. Each galaxy is colour coded with the S/N, equal as in Fig. \ref{fig_atlas_kinematics}, and the red X are those points of bad quality according to \citetalias{McDermid2015}. The 1:1 relation is represented by the black dashed line on each panel.}
\label{fig_atlas_populations}
\end{figure*}

In Fig. \ref{fig_atlas_populations} we set side by side our stellar population results with those of \citetalias{McDermid2015}. The ATLAS$^{3D}$ spectra have been published at 0.125, 0.5 and 1.0 effective radii.  We only make comparisons with values at 1R$_e$ because they should resemble more closely to our collapsed spectra. Also, they extracted stellar population properties by using index measurements, in particular, they compute line strength for H$\beta$, Fe5015, Fe5270 and Mgb to retrieve age, metallicity and [$\alpha$/Fe] from  H$\beta$-[MgFe50]' index-index diagrams. As a grid for the diagram they used the models from \citet{Schiavon2007}, and for this reason, their values are not directly comparable to ours with different methodologies and template models. Taking all of this into account, we find some discrepancies between index-index and FSF methods, particularly we notice that index-index results favour younger ages for galaxies reaching in some cases a disagreement of several Gyr. Additionally, \citetalias{McDermid2015} comments that some galaxies give ages older than the universe itself, 13.798$\pm$0.037 Gyr \citep{Planck2014}, but this value is inside the observational uncertainties. With our methodology, the maximum age allowed is fixed by the \href{http://miles.iac.es/}{MILES} models at 14 Gyr. For metallicity and [$\alpha$/Fe] there is a better agreement, nevertheless, for low S/N galaxies, there are also considerable discrepancies. For example, we obtained galaxies with a solar-like metal content while with index-index diagrams those galaxies are very metal-poor. A similar effect happens in $\alpha$-enhancement, high-enhanced galaxies in \citetalias{McDermid2015} have solar-like abundance ratios in our results. Even though these differences are mostly found in low S/N galaxies, we do not discard that the methodology and models play also an important part.
\begin{figure*}
\centering
\includegraphics[scale=0.9]{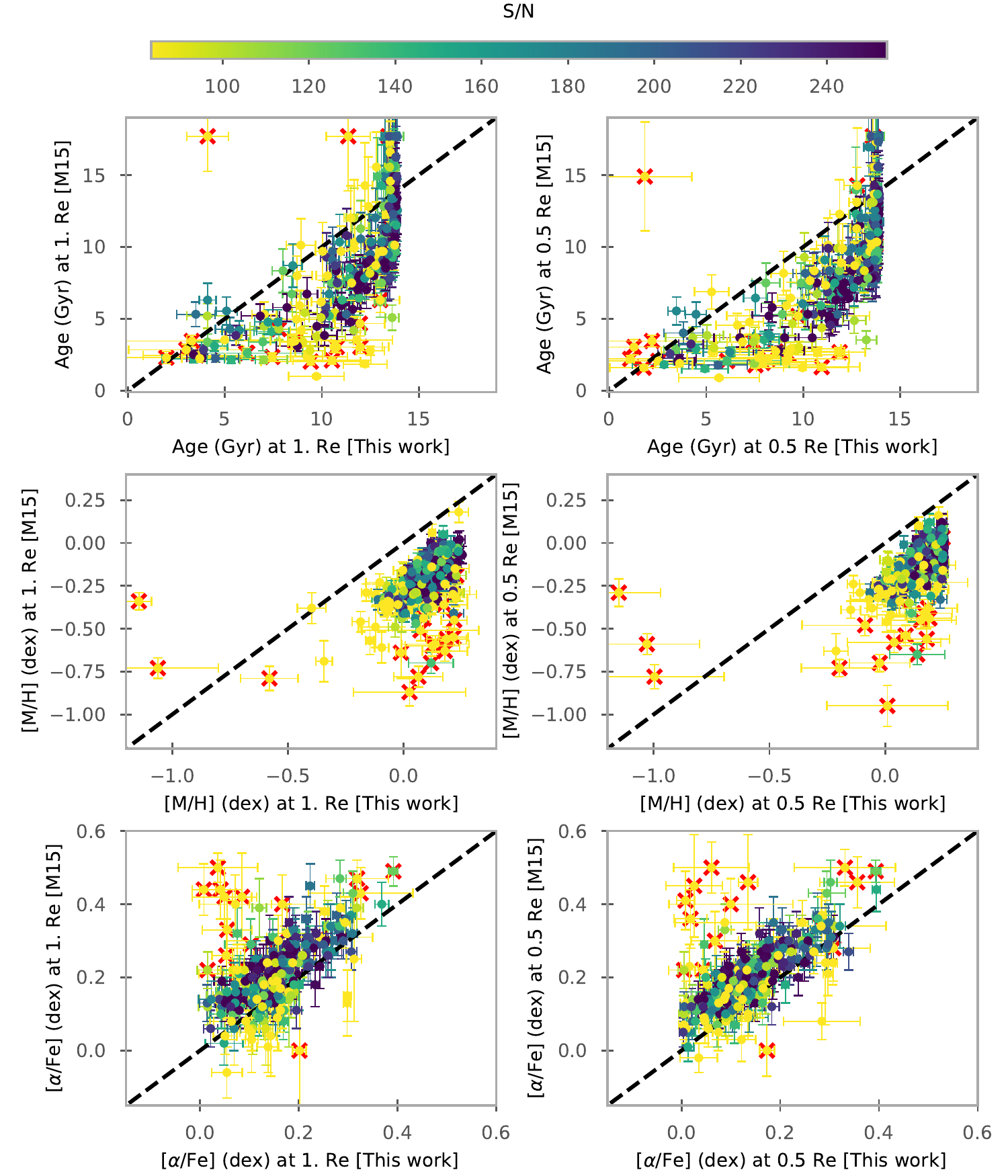}
\caption{Comparison of our stellar population properties with those published in \citetalias{McDermid2015} integrating up to 1.0 and 0.5 R$_e$. On the left column are the values for 1.0 R$_e$, and 0.5 R$_e$ on the right. The top row shows the age, on the middle panel the metallicity and on the bottom [$\alpha$/Fe]. Each galaxy is colour coded with the S/N, equal as in Fig. \ref{fig_atlas_kinematics}, and the red X are those points of bad quality according to \citetalias{McDermid2015}. The 1:1 relation is represented by the black dashed line on each panel.}
\label{fig_atlas_populations_at_re}
\end{figure*}
\section{Local Group dwarfs}\label{appendix_local}
To compare our results to dwarf galaxies in the Local Group we compile abundances for several galaxies. In \citet{Simon2019} and \citet{McConnachie2020} they presented a list of Milky Way satellites with the respective distance, velocity dispersion...etc, and also a list of literature references that have chemical abundances measurements in these dwarfs galaxies. We deeply explore those references and then some more from the literature in order to compile a competent list of measurements in individual stars, all of them in dwarfs galaxies of our Local Group. In order to compare with our FSF results, for each satellite we took as the abundance [Fe/H] or [Mg/Fe] the mean value of all the available stars in our compilation, and as the errors, we use the standard deviation. In Tab. \ref{table_apx_lg_1} we show the compile values from \citet{Simon2019} and \citet{McConnachie2020}, and in Tab. \ref{table_apx_lg_2} we present the statistical results of our compilation, along with all the references used.
\begin{table*}
\caption{Table with general properties of Local Group dwarfs ordered by their distance to the Milky Way. The columns show galaxy name, coordinates and distance modulus from \citet{McConnachie2020}, and distance in kpc, recessional velocity and velocity dispersion are from \citet{Simon2019}. For Aquarius, Leo A and Sag DIG the coordinates and distance modulus come from \citet{McConnachie2012}, and the distance and stellar kinematics from \citet{Kirby2017}.} 
\centering    
\begin{tabular}{ccccccc}     
\hline
Galaxy & RA & DEC & m-M (mag) & D (kpc) & $v_{hel}$ (km/s) & $\sigma$ (km/s) \\

\noalign{\smallskip}
\hline\noalign{\smallskip}

\noalign{\smallskip}
Segue 1 & 151.7667 & 16.0819 & $16.8^{+0.2}_{-0.2}$ & $23.0^{+2.0}_{-2.0}$ & $208.5^{+0.9}_{-0.9}$ & $3.7^{+1.4}_{-1.1}$ \\

\noalign{\smallskip}
Tucana III & 359.15 & -59.6 & $17.01^{+0.16}_{-0.16}$ & $25.0^{+2.0}_{-2.0}$ & $-102.3^{+0.4}_{-0.4}$ & $1.2^{+9.9}_{-9.9}$ \\

\noalign{\smallskip}
Carina3 & 114.63 & -57.8997 & $17.22^{+0.1}_{-0.1}$ & $27.8^{+0.6}_{-0.6}$ & $477.2^{+1.2}_{-1.2}$ & $5.6^{+4.3}_{-2.1}$ \\

\noalign{\smallskip}
Triangulum II & 33.3225 & 36.1783 & $17.4^{+0.1}_{-0.1}$ & $28.4^{+1.6}_{-1.6}$ & $-381.7^{+1.1}_{-1.1}$ & $3.4^{+9.9}_{-9.9}$ \\

\noalign{\smallskip}
Reticulum II & 53.9254 & -54.0492 & $17.4^{+0.2}_{-0.2}$ & $31.6^{+1.5}_{-1.4}$ & $62.8^{+0.5}_{-0.5}$ & $3.3^{+0.7}_{-0.7}$ \\

\noalign{\smallskip}
Ursa Major II & 132.875 & 63.13 & $17.5^{+0.3}_{-0.3}$ & $34.7^{+2.0}_{-1.9}$ & $-116.5^{+1.9}_{-1.9}$ & $5.6^{+1.4}_{-1.4}$ \\

\noalign{\smallskip}
Carina2 & 114.1067 & -57.9992 & $17.79^{+0.05}_{-0.05}$ & $36.2^{+0.6}_{-0.6}$ & $477.2^{+1.2}_{-1.2}$ & $3.4^{+1.2}_{-0.8}$ \\

\noalign{\smallskip}
Segue 2 & 34.8167 & 20.1753 & $17.7^{+0.1}_{-0.1}$ & $37.0^{+3.0}_{-3.0}$ & $-40.2^{+0.9}_{-0.9}$ & $2.2^{+9.9}_{-9.9}$ \\

\noalign{\smallskip}
Coma Berenices & 186.7458 & 23.9042 & $18.2^{+0.2}_{-0.2}$ & $42.0^{+1.6}_{-1.5}$ & $98.1^{+0.9}_{-0.9}$ & $4.6^{+0.8}_{-0.8}$ \\

\noalign{\smallskip}
Bootes II & 209.5 & 12.85 & $18.1^{+0.06}_{-0.06}$ & $42.0^{+1.0}_{-1.0}$ & $-117.0^{+5.2}_{-5.2}$ & $10.5^{+7.4}_{-7.4}$ \\

\noalign{\smallskip}
Tucana II & 342.9796 & -58.5689 & $18.8^{+0.2}_{-0.2}$ & $58.0^{+8.0}_{-8.0}$ & $-129.1^{+3.5}_{-3.5}$ & $8.6^{+4.4}_{-2.7}$ \\

\noalign{\smallskip}
Bootes I & 210.025 & 14.5 & $19.11^{+0.08}_{-0.08}$ & $66.0^{+2.0}_{-2.0}$ & $101.8^{+0.7}_{-0.7}$ & $4.6^{+0.8}_{-0.6}$ \\

\noalign{\smallskip}
Ursa Minor & 227.2854 & 67.2225 & $19.4^{+0.1}_{-0.1}$ & $76.0^{+4.0}_{-4.0}$ & $-247.2^{+0.8}_{-0.8}$ & $9.5^{+1.2}_{-1.2}$ \\

\noalign{\smallskip}
Draco & 260.0517 & 57.9153 & $19.4^{+0.17}_{-0.17}$ & $82.0^{+6.0}_{-6.0}$ & $-290.7^{+0.7}_{-0.8}$ & $9.1^{+1.2}_{-1.2}$ \\

\noalign{\smallskip}
Sculptor & 15.0392 & -33.7092 & $19.67^{+0.14}_{-0.14}$ & $86.0^{+5.0}_{-5.0}$ & $111.4^{+0.1}_{-0.1}$ & $9.2^{+1.1}_{-1.1}$ \\

\noalign{\smallskip}
Horologium I & 43.8821 & -54.1189 & $19.5^{+0.2}_{-0.2}$ & $87.0^{+13.0}_{-11.0}$ & $112.8^{+2.5}_{-2.6}$ & $4.9^{+2.8}_{-0.9}$ \\

\noalign{\smallskip}
Sextans1 & 153.2625 & -1.6147 & $19.67^{+0.1}_{-0.1}$ & $95.0^{+3.0}_{-3.0}$ & $224.3^{+0.1}_{-0.1}$ & $7.9^{+1.3}_{-1.3}$ \\

\noalign{\smallskip}
Carina & 100.4029 & -50.9661 & $20.11^{+0.13}_{-0.13}$ & $106.0^{+5.0}_{-5.0}$ & $222.9^{+0.1}_{-0.1}$ & $6.6^{+1.2}_{-1.2}$ \\

\noalign{\smallskip}
Grus I & 344.1767 & -50.1633 & $20.4^{+0.2}_{-0.2}$ & $120.0^{+12.0}_{-11.0}$ & $-140.5^{+2.4}_{-1.6}$ & $2.9^{+2.1}_{-1.0}$ \\

\noalign{\smallskip}
Hercules & 247.7583 & 12.7917 & $20.6^{+0.2}_{-0.2}$ & $132.0^{+6.0}_{-6.0}$ & $45.0^{+1.1}_{-1.1}$ & $5.1^{+0.9}_{-0.9}$ \\

\noalign{\smallskip}
Fornax & 39.9971 & -34.4492 & $20.84^{+0.18}_{-0.18}$ & $139.0^{+3.0}_{-3.0}$ & $55.2^{+0.1}_{-0.1}$ & $11.7^{+0.9}_{-0.9}$ \\

\noalign{\smallskip}
Leo IV & 173.2375 & -0.5333 & $20.94^{+0.09}_{-0.09}$ & $154.0^{+5.0}_{-5.0}$ & $132.3^{+1.4}_{-1.4}$ & $3.3^{+1.7}_{-1.7}$ \\

\noalign{\smallskip}
Canes Venatici II & 194.2917 & 34.3208 & $21.02^{+0.06}_{-0.06}$ & $160.0^{+4.0}_{-4.0}$ & $-128.9^{+1.2}_{-1.2}$ & $4.6^{+1.0}_{-1.0}$ \\

\noalign{\smallskip}
Canes Venatici I & 202.0146 & 33.5558 & $21.69^{+0.1}_{-0.1}$ & $211.0^{+6.0}_{-6.0}$ & $30.9^{+0.6}_{-0.6}$ & $7.6^{+0.4}_{-0.4}$ \\

\noalign{\smallskip}
Leo II & 168.37 & 22.1517 & $21.84^{+0.13}_{-0.13}$ & $233.0^{+14.0}_{-14.0}$ & $78.3^{+0.6}_{-0.6}$ & $7.4^{+0.4}_{-0.4}$ \\

\noalign{\smallskip}
Leo A & 149.860417 & 30.746389 & $24.51^{+0.12}_{-0.12}$ & $827.0^{+11.0}_{-11.0}$ & $26.2^{+1.0}_{-0.9}$ & $9.0^{+0.8}_{-0.6}$ \\

\noalign{\smallskip}
Aquarius & 311.715833 & -12.848056 & $25.15^{+0.08}_{-0.08}$ & $977.0^{+45.0}_{-45.0}$ & $-141.8^{+1.8}_{-2.0}$ & $7.8^{+1.8}_{-1.1}$ \\

\noalign{\smallskip}
Sag DIG & 292.495833 & -17.678056 & $25.14^{+0.18}_{-0.18}$ & $1047.0^{+53.0}_{-53.0}$ & $-78.4^{+1.6}_{-1.6}$ & $9.4^{+1.5}_{-1.1}$ \\

\hline\noalign{\smallskip}

\end{tabular}
             
\label{table_apx_lg_1}
\end{table*}

\begin{table*}
\caption{Table with the computed [Fe/H] and [Mg/Fe] abundances for Local Group dwarfs. For each galaxy, we show the mean abundance value of all the stars we could find in the literature and the standard deviation as the errors. We also add a column with the number of stars found, the type according to \citet{McConnachie2012} and the references of each consulted paper.} 
\centering
\begin{tabular}{cccccc}
\hline
Galaxy & [Fe/H] (dex) & [Mg/Fe] (dex) & Type & Nº stars & References \\

\noalign{\smallskip}
\hline\noalign{\smallskip}

\noalign{\smallskip}
Segue 1 & -2.63$\pm$0.02 & 0.56$\pm$0.03 & dSph & 14 & \mysplit{\citet{Norris2010b, Sitnova2021} \\ \citet{Frebel2014}} \\

\noalign{\smallskip}
Tucana III & -2.56$\pm$0.08 & 0.48$\pm$0.13 & - & 5 & \citet{Marshall2019, Hansen2017} \\

\noalign{\smallskip}
Carina3 & -3.07$\pm$0.06 & 0.73$\pm$0.02 & - & 2 & \citet{Ji2020} \\

\noalign{\smallskip}
Triangulum II & -2.39$\pm$0.11 & 0.42$\pm$0.27 & - & 8 & \mysplit{\citet{Kirby2017, Sitnova2021} \\ \citet{Ji2019}} \\

\noalign{\smallskip}
Reticulum II & -2.84$\pm$0.0 & 0.33$\pm$0.0 & - & 9 & \citet{Ji2016} \\

\noalign{\smallskip}
Ursa Major II & -2.88$\pm$0.08 & 0.52$\pm$0.08 & dSph & 4 & \citet{Sitnova2021, Frebel2010} \\

\noalign{\smallskip}
Carina2 & -2.67$\pm$0.03 & 0.29$\pm$0.08 & - & 9 & \citet{Ji2020} \\

\noalign{\smallskip}
Segue 2 & -2.96$\pm$0.19 & 0.31$\pm$0.21 & dSph & 1 & \citet{Roederer-Kirby2014} \\

\noalign{\smallskip}
Coma Berenices & -2.47$\pm$0.08 & 0.57$\pm$0.08 & dSph & 6 & \citet{Sitnova2021, Frebel2010} \\

\noalign{\smallskip}
Bootes II & -3.03$\pm$0.0 & 0.36$\pm$0.0 & dSph & 2 & \citet{Francois2016} \\

\noalign{\smallskip}
Tucana II & -2.94$\pm$0.05 & 0.32$\pm$0.08 & - & 7 & \mysplit{\citet{Chiti2018, Marshall2019} \\ \citet{Hansen2017}} \\

\noalign{\smallskip}
Bootes I & -2.53$\pm$0.07 & 0.29$\pm$0.08 & dSph & 95 & \mysplit{\citet{Ishigaki2014, Lai2011} \\ \citet{Gilmore2013, Norris2010a} \\ \citet{Frebel2016, Francois2016} \\ \citet{Gilmore2013}} \\

\noalign{\smallskip}
Ursa Minor & -2.13$\pm$0.08 & 0.17$\pm$0.21 & dSph & 88 & \mysplit{\citet{Shetrone2001, Ural2015} \\ \citet{Kirby2015}} \\

\noalign{\smallskip}
Draco & -2.05$\pm$0.06 & 0.02$\pm$0.21 & dSph & 161 & \citet{Shetrone2001, Kirby2015} \\

\noalign{\smallskip}
Sculptor & -1.77$\pm$0.04 & 0.09$\pm$0.1 & dSph & 211 & \mysplit{\citet{Jablonka2015, Kirby2015} \\ \citet{Tafelmeyer2010, Shetrone2003} \\ \citet{Simon2015}} \\

\noalign{\smallskip}
Horologium I & -2.62$\pm$0.02 & -0.09$\pm$0.03 & - & 3 & \citet{Nagasawa2018} \\

\noalign{\smallskip}
Sextans1 & -2.63$\pm$0.05 & 0.22$\pm$0.09 & dSph & 18 & \mysplit{\citet{Shetrone2001, Mashonkina2022} \\ \citet{Tafelmeyer2010}} \\

\noalign{\smallskip}
Carina & -1.8$\pm$0.06 & 0.3$\pm$0.07 & dSph & 103 & \mysplit{\citet{Ji2020, Shetrone2003} \\ \citet{Norris2017, Fabrizio2015} \\ \citet{}} \\

\noalign{\smallskip}
Grus I & -2.52$\pm$0.01 & 0.32$\pm$0.01 & - & 2 & \citet{Ji2019} \\

\noalign{\smallskip}
Hercules & -2.37$\pm$0.18 & 0.4$\pm$0.0 & dSph & 6 & \citet{Koch2008, Francois2016} \\

\noalign{\smallskip}
Fornax & -1.07$\pm$0.04 & -0.01$\pm$0.12 & dSph & 124 & \mysplit{\citet{Shetrone2003, Kirby2015} \\ \citet{Tafelmeyer2010}} \\

\noalign{\smallskip}
Leo IV & -2.75$\pm$0.16 & 0.23$\pm$0.05 & dSph & 3 & \citet{Francois2016, Simon2010} \\

\noalign{\smallskip}
Canes Venatici II & -2.58$\pm$0.0 & 0.16$\pm$0.0 & dSph & 1 & \citet{Francois2016} \\

\noalign{\smallskip}
Canes Venatici I & -2.35$\pm$0.0 & 0.25$\pm$0.0 & dSph & 2 & \citet{Francois2016} \\

\noalign{\smallskip}
Leo II & -1.85$\pm$0.02 & 0.15$\pm$0.06 & dSph & 25 & \citet{Shetrone2009} \\

\noalign{\smallskip}
Leo A & -1.64$\pm$0.04 & 0.02$\pm$0.1 & dIrr & 59 & \citet{Kirby2017} \\

\noalign{\smallskip}
Aquarius & -1.43$\pm$0.03 & -0.03$\pm$0.09 & dIrr & 17 & \citet{Kirby2017} \\

\noalign{\smallskip}
Sag DIG & -1.89$\pm$0.03 & 0.03$\pm$0.1 & dIrr & 28 & \citet{Kirby2017} \\

\hline\noalign{\smallskip}

\end{tabular}
             
\label{table_apx_lg_2}
\end{table*}
\section{Stellar populations results and fitting relation}\label{appendix_ppxf_table}
In this section, we present a table with the stellar population results from \href{https://www-astro.physics.ox.ac.uk/~mxc/software/}{pPXF} for every dwarf galaxy in our SAMI-Fornax sample, including those with emission lines. With \href{https://www-astro.physics.ox.ac.uk/~mxc/software/}{pPXF} we run 100 MC simulations, and the values and uncertainties in Tab. \ref{table_spp_ppxf} are the mean and standard deviation of those distributions.

Complementary to the relations fit in this paper, in Tab. \ref{table_fit_coef} we show the coefficients of every relation fitted during this work.

\begin{figure}
\centering
\includegraphics[scale=0.06]{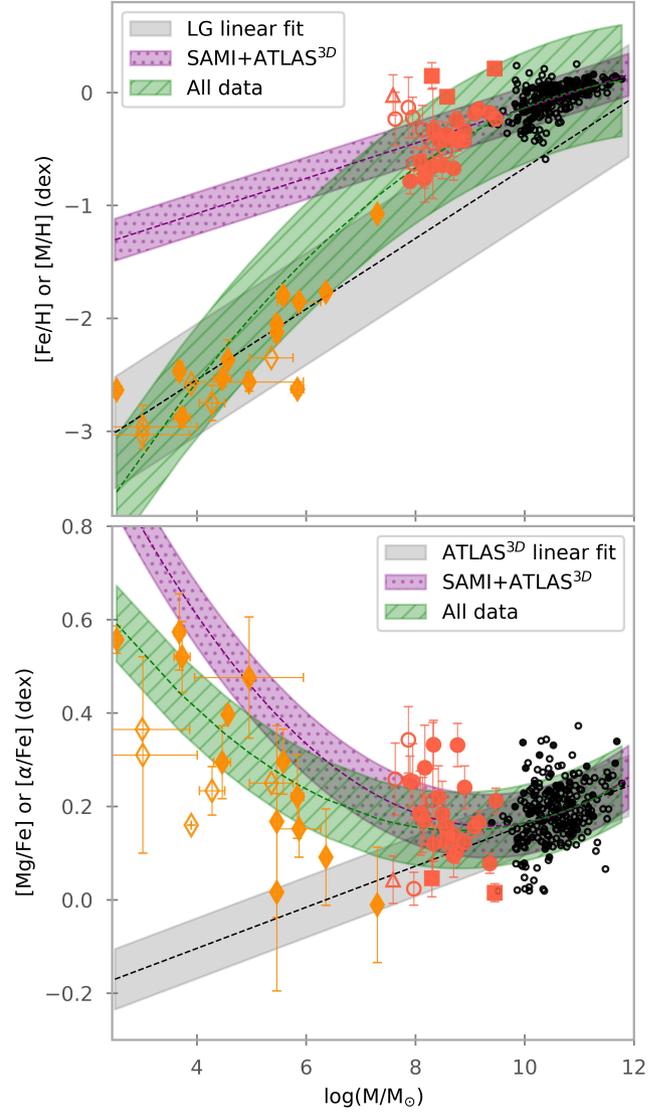}
\caption{Same as Fig. \ref{fig_local_group_alpha-sigma} but with the stellar population properties as a function of the stellar mass. When available, the mass of the Local Group objects has been taken from \citet{Weisz2014}, if not from \citet{Kirby2013}.}
\label{fig_alpha_local_vs_mass}
\end{figure}

Additionally, we include here Fig. \ref{fig_alpha_local_vs_mass}. A plot similar to Fig. \ref{fig_local_group_alpha-sigma}, but representing the metallicity and $\alpha$-enhancement as a function of the stellar mass. Here we only present those galaxies for which we have the value of their stellar mass.


\begin{table}
\caption{Table with the stellar population properties derived from \href{https://www-astro.physics.ox.ac.uk/~mxc/software/}{pPXF} for the SAMI-Fornax dwarfs. Galaxies with * are those with emission lines in their spectra.} 
\centering
\begin{tabular}{cccc}
\hline
FCC name & Age (Gyr) & [M/H] (dex) & [$\alpha$/Fe] (dex) \\

\noalign{\smallskip}
\hline\noalign{\smallskip}

\noalign{\smallskip}
FCC100 & 8.77$\pm$1.55 & -0.67$\pm$0.11 & 0.09$\pm$0.05 \\

\noalign{\smallskip}
FCC106 & 9.74$\pm$0.64 & -0.43$\pm$0.04 & 0.12$\pm$0.02 \\

\noalign{\smallskip}
FCC113* & 8.25$\pm$1.57 & -0.92$\pm$0.11 & 0.07$\pm$0.05 \\

\noalign{\smallskip}
FCC134 & 2.94$\pm$1.42 & -0.24$\pm$0.24 & 0.26$\pm$0.08 \\

\noalign{\smallskip}
FCC135 & 8.76$\pm$0.97 & -0.41$\pm$0.07 & 0.14$\pm$0.03 \\

\noalign{\smallskip}
FCC136 & 10.14$\pm$0.97 & -0.19$\pm$0.04 & 0.16$\pm$0.02 \\

\noalign{\smallskip}
FCC143 & 13.41$\pm$0.56 & 0.21$\pm$0.06 & 0.02$\pm$0.02 \\

\noalign{\smallskip}
FCC164 & 7.56$\pm$1.42 & -0.33$\pm$0.08 & 0.12$\pm$0.04 \\

\noalign{\smallskip}
FCC178 & 6.72$\pm$2.27 & -0.22$\pm$0.18 & 0.25$\pm$0.08 \\

\noalign{\smallskip}
FCC181 & 10.51$\pm$2.77 & -0.13$\pm$0.27 & 0.34$\pm$0.07 \\

\noalign{\smallskip}
FCC182 & 10.56$\pm$0.74 & -0.15$\pm$0.05 & 0.17$\pm$0.02 \\

\noalign{\smallskip}
FCC188 & 9.34$\pm$2.5 & -0.41$\pm$0.16 & 0.22$\pm$0.07 \\

\noalign{\smallskip}
FCC195 & 5.69$\pm$2.18 & -0.33$\pm$0.22 & 0.18$\pm$0.08 \\

\noalign{\smallskip}
FCC202 & 7.04$\pm$1.37 & -0.34$\pm$0.09 & 0.24$\pm$0.05 \\

\noalign{\smallskip}
FCC203 & 6.85$\pm$1.16 & -0.24$\pm$0.07 & 0.1$\pm$0.02 \\

\noalign{\smallskip}
FCC207* & 6.51$\pm$1.62 & -0.54$\pm$0.15 & 0.16$\pm$0.07 \\

\noalign{\smallskip}
FCC211 & 9.65$\pm$1.39 & -0.65$\pm$0.09 & 0.33$\pm$0.05 \\

\noalign{\smallskip}
FCC222 & 9.39$\pm$1.28 & -0.43$\pm$0.09 & 0.33$\pm$0.05 \\

\noalign{\smallskip}
FCC223 & 6.47$\pm$4.23 & -0.45$\pm$0.49 & 0.21$\pm$0.17 \\

\noalign{\smallskip}
FCC235* & 4.39$\pm$1.85 & -0.49$\pm$0.23 & 0.12$\pm$0.08 \\

\noalign{\smallskip}
FCC245 & 6.26$\pm$1.51 & -0.37$\pm$0.11 & 0.15$\pm$0.04 \\

\noalign{\smallskip}
FCC250 & 6.14$\pm$1.92 & -0.63$\pm$0.15 & 0.02$\pm$0.04 \\

\noalign{\smallskip}
FCC252 & 12.93$\pm$0.8 & -0.04$\pm$0.06 & 0.13$\pm$0.02 \\

\noalign{\smallskip}
FCC253 & 12.64$\pm$1.29 & 0.15$\pm$0.12 & 0.05$\pm$0.04 \\

\noalign{\smallskip}
FCC263* & 3.39$\pm$0.74 & -0.37$\pm$0.05 & 0.15$\pm$0.03 \\

\noalign{\smallskip}
FCC264 & 9.62$\pm$2.01 & -0.61$\pm$0.11 & 0.19$\pm$0.09 \\

\noalign{\smallskip}
FCC266 & 10.02$\pm$1.82 & -0.64$\pm$0.12 & 0.18$\pm$0.07 \\

\noalign{\smallskip}
FCC274 & 10.02$\pm$2.45 & -0.73$\pm$0.24 & 0.28$\pm$0.09 \\

\noalign{\smallskip}
FCC277 & 5.79$\pm$0.69 & -0.23$\pm$0.06 & 0.21$\pm$0.03 \\

\noalign{\smallskip}
FCC285* & 4.92$\pm$1.19 & -0.74$\pm$0.1 & 0.24$\pm$0.05 \\

\noalign{\smallskip}
FCC298 & 11.3$\pm$1.33 & -0.76$\pm$0.08 & 0.17$\pm$0.05 \\

\noalign{\smallskip}
FCC300 & 10.35$\pm$1.79 & -0.52$\pm$0.11 & 0.16$\pm$0.06 \\

\noalign{\smallskip}
FCC301 & 8.96$\pm$0.61 & -0.18$\pm$0.04 & 0.08$\pm$0.02 \\

\noalign{\smallskip}
FCC306* & 1.58$\pm$0.69 & -0.45$\pm$0.12 & 0.12$\pm$0.06 \\

\noalign{\smallskip}
FCC033* & 6.23$\pm$0.84 & -0.43$\pm$0.06 & 0.08$\pm$0.02 \\

\noalign{\smallskip}
FCC037* & 7.79$\pm$2.12 & -0.62$\pm$0.2 & 0.26$\pm$0.08 \\

\noalign{\smallskip}
FCC046* & 3.95$\pm$1.08 & -0.38$\pm$0.11 & 0.12$\pm$0.04 \\

\noalign{\smallskip}
FCCB442 & 1.82$\pm$1.3 & -0.02$\pm$0.18 & 0.04$\pm$0.05 \\

\noalign{\smallskip}
FCCB904 & 10.8$\pm$1.87 & -0.79$\pm$0.11 & 0.25$\pm$0.05 \\

\hline\noalign{\smallskip}

\end{tabular}
             
\label{table_spp_ppxf}
\end{table}
\begin{table*}
\caption{Table with the coefficients of all the fits shown in this paper.} 
\centering    
\begin{tabular}{cccccc}     
\hline
Relation & a & b & c & Fitted data & Figure \\

\hline\noalign{\smallskip}

\noalign{\smallskip}
[$\alpha$/Fe] = a*log($M_{\star}/M_{\odot}$) + b & 0.04$\pm$0.01 & -0.28$\pm$0.09 &  & ATLAS$^{3D}$ & \ref{fig_alpha_vs_sigma} \\

\noalign{\smallskip}
[$\alpha$/Fe] = a*log($\sigma$)$^2$ + b*log($\sigma$) + c & 0.24$\pm$0.04 & -0.82$\pm$0.15 & 0.87$\pm$0.14 & SAMI-Fornax+ATLAS$^{3D}$ & \ref{fig_alpha_vs_sigma}, \ref{fig_local_group_alpha-sigma} \\

\noalign{\smallskip}
[M/H] = a*log($M_{\star}/M_{\odot}$) + b & 0.14$\pm$0.01 & -1.52$\pm$0.09 &  & SAMI-Fornax+ATLAS$^{3D}$ & \ref{fig_all_vs_all} \\

\noalign{\smallskip}
[M/H] = a*log($\sigma$) + b & 0.48$\pm$0.03 & -1.07$\pm$0.07 &  & SAMI-Fornax+ATLAS$^{3D}$ & \ref{fig_local_group_alpha-sigma} \\

\noalign{\smallskip}
[M/H] = a*log($\sigma$) + b & 0.86$\pm$0.37 & -3.07$\pm$0.28 &  & Local Group & \ref{fig_local_group_alpha-sigma} \\

\noalign{\smallskip}
[M/H] = a*log($\sigma$)$^2$ + b*log($\sigma$) + c & -0.75$\pm$0.06 & 3.62$\pm$0.19 & -4.35$\pm$0.14 & Local Group, SAMI-Fornax, Coma, ATLAS$^{3D}$ & \ref{fig_local_group_alpha-sigma} \\

\noalign{\smallskip}
[$\alpha$/Fe] = a*log($\sigma$) + b & 0.18$\pm$0.02 & -0.19$\pm$0.05 &  & ATLAS$^{3D}$ & \ref{fig_local_group_alpha-sigma} \\

\noalign{\smallskip}
[$\alpha$/Fe] = a*log($\sigma$)$^2$ + b*log($\sigma$) + c & 0.16$\pm$0.02 & -0.52$\pm$0.06 & 0.58$\pm$0.04 & Local Group, SAMI-Fornax, Coma, ATLAS$^{3D}$ & \ref{fig_local_group_alpha-sigma} \\

\noalign{\smallskip}
[$\alpha$/Fe] = a*(r/R$_{200}$) + b & -0.34$\pm$0.12 & 0.44$\pm$0.06 &  & Local Group & \ref{fig_local_group_alpha-distance} \\

\noalign{\smallskip}
[$\alpha$/Fe] = a*(r/R$_{200}$) + b & -0.19$\pm$0.05 & 0.26$\pm$0.02 &  & SAMI-Fornax & \ref{fig_local_group_alpha-distance} \\

\noalign{\smallskip}
[$\alpha$/Fe] = a*(r/R$_{200}$) + b & -0.08$\pm$0.07 & 0.20$\pm$0.05 &  & Virgo & \ref{fig_local_group_alpha-distance} \\

\noalign{\smallskip}
[$\alpha$/Fe] = a*(r/R$_{200}$) + b & -0.15$\pm$0.04 & 0.15$\pm$0.02&  & Coma & \ref{fig_local_group_alpha-distance} \\

\end{tabular}
             
\label{table_fit_coef}
\end{table*}

\bsp	
\label{lastpage}
\end{document}